\documentclass[epsfig,showpacs,floatfix,superscriptaddress]{revtex4}
\usepackage{epsfig}
\usepackage{color}
\usepackage{graphicx}
\usepackage{graphics}
\usepackage[section] {placeins}

\newcommand{\Fritiof}{\textsc{Fritiof}}

 \def\be{\begin{equation}}
 \def\ee{\end{equation}}
 \def\bea{\begin{eqnarray}}
 \def\eea{\end{eqnarray}}
 \def\bean{\begin{eqnarray*}}
 \def\eean{\end{eqnarray*}}

\begin{document}

\title{System Size and Energy Dependence of Dilepton Production in
Heavy-Ion Collisions at 1-2 $A$GeV Energies}

\author{E.L. Bratkovskaya}
\affiliation{Institute for Theoretical Physics, Frankfurt University,
         60438 Frankfurt-am-Main, Germany } 
\affiliation{ Frankfurt Institut for Advanced Studies,
         Frankfurt University, 60438 Frankfurt-am-Main, Germany}
\author{J. Aichelin}
\affiliation{SUBATECH, Laboratoire de Physique Subatomique et des
Technologies Associ\'ees, \\
Universit\'e de Nantes - IN2P3/CNRS - Ecole des Mines de Nantes \\
4 rue Alfred Kastler, F-44072 Nantes, Cedex 03, France}
\author{M. Thomere}
\affiliation{SUBATECH, Laboratoire de Physique Subatomique et des
Technologies Associ\'ees, \\
Universit\'e de Nantes - IN2P3/CNRS - Ecole des Mines de Nantes \\
4 rue Alfred Kastler, F-44072 Nantes, Cedex 03, France}
\author{S. Vogel}
\affiliation{Institute for Theoretical Physics, Frankfurt University,
         60438 Frankfurt-am-Main, Germany }
\author{M. Bleicher}
\affiliation{Institute for Theoretical Physics, Frankfurt University,
         60438 Frankfurt-am-Main, Germany } 
\affiliation{ Frankfurt Institut for Advanced Studies,
         Frankfurt University, 60438 Frankfurt-am-Main, Germany}

\begin{abstract}
We study the dilepton production in heavy-ion collisions at energies of 1-2 $A$GeV
as well as in proton induced $pp, pn, pd$ and $p+A$ reactions from 1 GeV up to 3.5 GeV
where data have been taken by the HADES collaboration.
For the analysis we employ three different transport
models - the microscopic off-shell Hadron-String-Dynamics (HSD)
transport approach, the Isospin Quantum Molecular Dynamics (IQMD)
approach as well as the Ultra-relativistic Quantum Molecular Dynamics
(UrQMD) approach. We find that the HSD and IQMD models
describe very reasonably the elementary $pp$, $pn$ and $\pi N$ reactions
despite of different assumptions on quantities like the excitation function
of the $\Delta$ multiplicity, where solid experimental constraints
are not available. Taking these data on elementary collisions as input,
the three models provide a good description of the presently available heavy ion data.
In particular, we confirm the experimentally observed
enhancement of the dilepton yield (normalized to the multiplicity of
neutral pions $N_{\pi^0}$) in heavy-ion collisions with respect to that measured in
 $NN = (pp+pn)/2$ collisions. We identify two contributions to this enhancement:
a) the $pN$ bremsstrahlung which scales with the number
of collisions and not with the number of participants, i.e. pions;
b) the dilepton emission from intermediate $\Delta$'s which are part
of the reaction  cycles $\Delta \to \pi N ; \pi N \to \Delta$ and $NN\to N\Delta;
N\Delta \to NN$.  With increasing system size more generations of intermediate
$\Delta$'s are created. If such $\Delta$ decays into a pion, the pion can be
reabsorbed, however, if it decays into a dilepton, the dilepton escapes from
the system.  Thus, experimentally one observes only one pion (from the last
produced $\Delta$) whereas the dilepton yield accumulates
the contributions from all $\Delta$'s of the cycle.
We show as well that the Fermi motion enhances the production of pions and
dileptons in the same way. Furthermore, employing the off-shell HSD approach,
we explore the influence of in-medium effects like the modification of
self-energies and spectral functions of the vector mesons due to their interactions
with the hadronic environment. We find only a modest influence of the in-medium effects
on the dilepton spectra in the invariant mass range where data with small error bars exist.
\end{abstract}

\date{\today}
\maketitle
\section{Introduction}

According to the theory of strong interactions, the Quantum Chromo
Dynamics (QCD), hadrons are bound objects of quarks and gluons.  The
properties of hadrons in vacuum are well known and confirmed by lattice
QCD calculations \cite{Fodor:2012gf} while the properties of hadrons in
a strongly interacting environment are subject of intensive research.
QCD inspired approaches as well as phenomenological models based on
phase shifts and SU(3) symmetry
\cite{Fodor:2012gf,Brown:1991kk,Brown:1993yv,Hatsuda:1995dy,asakawa,Herrmann,Chanfray,Klingl96,Rapp:1995zy,Rapp:1997fs,Peters,Leupold,lauran,Tolos:2011ge}
predict significant changes of hadron properties in a strongly interacting medium. The results of the different models vary substantially.  It is
therefore one of the challenges of novel experimental heavy-ion physics
to study these in-medium modifications of hadrons. Besides the in-medium
properties of the antikaon, interesting also for astrophysical reasons,
the vector mesons and especially the $\rho$ meson have been in the
focus of the theoretical interest because the $\rho$ has
the quantum numbers of a photon and can therefore disintegrate into
an electron-positron pair.  Having only electromagnetic interactions this
pair may easily leave the reaction region without further collisions.
This allows to reconstruct the invariant mass of the decaying $\rho^0$.
Thus there is a hope that by measuring the dilepton invariant mass
spectra the in-medium mass and width of the $\rho$ meson become
experimentally accessible.  For this it is necessary to separate the
background from known dilepton sources  -- Dalitz decays
of baryonic and mesonic resonances as well as $pN$ and $\pi N$ bremsstrahlung.  At SPS
energies of 40 and 158 $A$GeV such an enhancement above the known background has been
measured by the CERES \cite{Agakichiev:2005ai} and the NA60
\cite{Seixas:2007ua} collaborations. The experimental results are compatible with the
assumption that in the medium the peak position of the $\rho$ meson mass distribution
remains rather unchanged while the width increases considerably (cf.
\cite{vanHees:2007yi,Gale:2006hg,Renk:2006ax,Brat:2009PLB,Santini:2011zw,Linnyk:2011vx}).

At much lower energies, at energies around 1 $A$GeV dileptons have been measured in heavy-ion collisions at the
BEVALAC  in Berkeley by the DLS Collaboration
\cite{Matis:1994tg,Wilson:1997sr,Wilson:1993mp,Porter:1997rc}.  These
data led to the so called 'DLS puzzle' because the DLS dilepton
yield   in C+C and Ca+Ca collisions in the invariant mass range from 0.2 to 0.5
GeV  \cite{Porter:1997rc}  was about five
times higher than the results from different transport models at that time
using the 'conventional' dilepton sources such as bremsstrahlung, $\pi^0,
\eta, \omega$ and $\Delta$ Dalitz decays and direct decay of vector
mesons ($\rho, \omega, \phi$)
\cite{Wolf:1992gg,Bratkovskaya:1996bv,Xiong:1990bg}.  This discrepancy
remained even after including in the transport calculations the different scenarios for the in-medium
modifications of vector meson properties,
as dropping mass or collisional broadening of the $\rho$ and  $\omega$
spectral functions
\cite{Ernst:1997yy,BratRapp98,BratKo99,Fuchs:2005zga}.  To solve this
puzzle was one of the main motivations to build  the HADES (High
Acceptance Dilepton Spectrometer) detector at  GSI in Germany
\cite{Agakishiev:2007ts,Pachmayer:2008yn,Sudol:2008zz,Agakishiev:2007ts,Agakishiev:2009yf,Lapidus:2009aa,Agakishiev:2011vf}.
In 2008 the HADES collaboration confirmed the DLS data
\cite{Agakishiev:2007ts,Pachmayer:2008yn} for  C+C at 1.0 $A$GeV.  In the mean time also
the theoretical transport approaches as well as effective models for
the elementary $NN$ reactions have been further developed. As it has
been suggested in Ref.  \cite{Bratkovskaya:2007jk}, the DLS puzzle can
be solved when incorporating stronger $pn$ and $pp$ bremsstrahlung
contributions in line with the  updated One-Boson-Exchange (OBE) model
calculations from \cite{Kaptari:2005qz}.  The previous OBE approaches
\cite{Schafer:1989dm} used in the old transport calculations  for the
analysis of the DLS data, gave results close to the soft photon
approximation.  As shown in Ref. \cite{Bratkovskaya:2007jk}  the
results of the HSD model (off-shell microscopic Hadron-String-Dynamics
(HSD) transport approach)  with  'enhanced' bremsstrahlung cross
sections agree very well  with the HADES experimental data for C+C
at 1 and 2 $A$GeV as well as with the DLS data for C+C and
Ca+Ca at 1 $A$GeV,  especially when one includes
a collisional broadening in the vector-meson spectral functions.
Similar results have been obtained by other, independent transport approaches
-- IQMD \cite{Thomere:2007cj} and the Rossendorf BUU \cite{Barz:2009yz}.

Despite of the fact that theory predicts that the vector meson
properties are modified substantially  already at energies as low as 1-2 $A$GeV,
it is quite difficult to observe these
changes experimentally. The production yield of $\rho^0$ and $ \omega$ mesons is small
at these energies  and the background from other dilepton sources like
$\Delta$-Dalitz decay and $pN$ bremsstrahlung is large in  the mass
range of interest, $M>0.4$ GeV. Therefore the presently available data
do not allow for a detailed investigation of  the in-medium properties
of vector mesons.

This focuses the interest of the present studies to the question
whether the invariant mass spectrum below the $\rho/ \omega$ peak depends on
the system size and on the beam energy in a non-trivial way, i.e.
whether the dilepton invariant mass spectra can be understood as a
superposition of individual $pp$ and $pn$ interactions.
In a first publication \cite{Agakishiev:2009yf} the HADES collaboration
found that the invariant mass spectra of dileptons, observed in 1
and 2 $A$GeV C+C collisions, are practically coincident below $M =0.4$ GeV
if divided by the total number of observed $\pi^0$ and after
subtracting the $\eta$ Dalitz decay contribution. It is strongly
suppressed at 1 $A$GeV  but becomes essential at 2 $A$GeV due to the
kinematical threshold for the $\eta$ production in $NN$ collisions.
This scaling with the $\pi^0$ number can be interpreted as a scaling
with the number of participants $N_{part}$. The HADES collaboration,
comparing the dilepton yield from the light C+C systems with the
elementary $pp$ and $pd$ interactions, albeit taken at different energies,
has concluded that the dilepton invariant mass spectra in these light systems
can be considered as a mere superposition of $pn$ and $pp$ collisions without any 'in-medium'
enhancement.

In a more recent publication \cite{Agakishiev:2011vf} a heavier system,
Ar+KCl at 1.75 $A$GeV,  was investigated and the collaboration came to
the conclusion that in this reaction the dilepton invariant mass yield between 0.2 and 0.6 GeV
is about 2-3 times larger than expected
from a mere superposition of $pp$ and $pn$ collisions
\cite{Agakishiev:2011vf}.
 From the analysis
of the excess in the transverse-mass slope and from the angular anisotropies
the HADES collaboration concluded that the excess of dileptons in the low invariant
mass region scales with the system size
very differently than the freeze-out yield of pions and $\eta$ and that the data
are compatible with the assumption that they originate from $\Delta$ Dalitz
decays, being suggestive of resonance matter \cite{Agakishiev:2011vf}.

It is the purpose of the present study to investigate this
enhancement within the presently available transport codes -- HSD,  IQMD
and UrQMD --  in order to explore whether the dilepton production in these
systems can be reproduced by the theoretical approaches and to identify
eventually the origin of this in-medium enhancement. All these codes
have been successfully employed to investigate a multitude of
experimental observables. They are, however, not of the same
sophistication as far as the dileptons are concerned. The dilepton part
of the UrQMD program is still under development and up to now
bremsstrahlung is not included. This limits the predictive power to
parts of the spectra where bremsstrahlung is not essential. For the
study of the in-medium enhancement we limit ourselves to HSD and IQMD
calculations. We start out in section II with a short description of
the HSD model and of the improvements made compared to the
'standard' HSD 2.5 version used for the extended dilepton study
in Ref. \cite{Bratkovskaya:2007jk}. Then we come to a brief description of dilepton production
in IQMD.  In section III the dilepton production in elementary
reactions, measured by  the HADES and and DLS collaboration, is compared with HSD
and IQMD calculations. The forth section is devoted to the study of dilepton
production in heavy-ion collisions. We discuss our calculations for
all systems which have been measured by the HADES collaboration and present also our
predictions for Au+Au collision at 1.25 $A$GeV which is presently
analysed.  After checking that the invariant mass spectra of
dileptons in heavy-ion collisions are well reproduced by the HSD as well as by the IQMD
approach we study in section V the enhancement of the dilepton
production in heavy-ion collisions as compared to the elementary
reactions and identify its origin. In section VI we present our conclusions.

\section{ The transport models}
\subsection{ The HSD model}

Our analysis of the experimental results is carried out within the
off-shell HSD transport model
\cite{Bratkovskaya:1996qe,Cassing:1999es,Ehehalt:1996uq,Bratkovskaya:2007jk} -
based on covariant self energies
for the baryons \cite{Weber:1992qc}. It has been used for the
description of $pA$ and $AA$ collisions from SIS to RHIC energies.
We recall that in the HSD approach nucleons, $\Delta$'s,
N$^*$(1440), N$^*$(1535), $\Lambda$, $\Sigma$ and $\Sigma^*$
hyperons, $\Xi$'s, $\Xi^*$'s and $\Omega$'s as well as their
antiparticles are included on the baryonic side whereas the $0^-$
and $1^-$ octet states are incorporated in the mesonic sector.
Inelastic baryon--baryon (and meson-baryon) collisions with energies
above $\sqrt{s_{th}}\simeq 2.6$ GeV (and  $\sqrt s_{th}\simeq 2.3$ GeV) are
described by the \Fritiof{} string model \cite{Andersson:1992iq} whereas low
energy hadron--hadron collisions are modelled using experimental
cross sections.

The dilepton production by the decay of a (baryonic or mesonic) resonance $R$
can be schematically presented in the following way:
\begin{eqnarray}
 BB &\to&R X   \label{chBBR} \\
 mB &\to&R X \label{chmBR} \\
      && R \to  e^+e^- X, \label{chRd} \\
      && R \to  m X, \ m\to e^+e^- X, \label{chRMd} \\
      && R \to  R^\prime X, \ R^\prime \to e^+e^- X. \label{chRprd}
\end{eqnarray}
In a first step a resonance $R$ might be produced in baryon-baryon
($BB$) or meson-baryon ($mB$) collisions (\ref{chBBR}), (\ref{chmBR}).
Then this resonance can either couple directly to dileptons
(\ref{chRd}) (e.g. Dalitz decay of the $\Delta$ resonance: $\Delta \to
e^+e^-N$) or produces mesonic (\ref{chRMd}) or baryonic  (\ref{chRprd})
resonances which then produce dileptons via direct decays ($\rho,
\omega$) or Dalitz decays ($\pi^0, \eta, \omega$).  With increasing
energy hadrons are created by non-resonant mechanisms or string decay.
This is also true for those which disintegrate into dileptons.
The electromagnetic part of all conventional dilepton sources  --
$\pi^0, \eta, \omega, \Delta$  Dalitz decays as well as direct decay of vector
mesons $\rho, \omega$ and $\phi$ -- are calculated as described in detail
in Ref.~\cite{Bratkovskaya:2000mb}.
We note that we use here again (as in early HSD dilepton studies
\cite{Bratkovskaya:1996qe,Cassing:1999es}) the "Wolf" model for the
differential electromagnetic decay width of the $\Delta$ resonance
\cite{Wolf90} instead of the "Ernst" description
\cite{Ernst:1997yy} adopted at that time in  \cite{Bratkovskaya:2000mb}.

The treatment of the 'enhanced' bremsstrahlung contribution from $pp$, $pn$
as well as $\pi N$ 'quasi-elastic' scattering, based on the OBE calculations by Kaptari
and K\"ampfer  \cite{Kaptari:2005qz}, is discussed in detail in Ref.
\cite{Bratkovskaya:2007jk} (cf. Section 2.6 there) where also a discussion of the
different models \cite{Schafer:1989dm,Shyam03,deJong97},
which formulate bremsstrahlung in the elementary reactions, can be found.
We note here that the  OBE models mentioned above
\cite{Kaptari:2005qz,Schafer:1989dm,Shyam03,deJong97} provide
different results not only for the $pN$ bremsstrahlung
contribution (which might be attributed to the different way to
realize the gauge invariance) but for the $\Delta$-Dalitz decay,
due to the different form factors.
In our transport analysis we use only the bremsstrahlung contribution from \cite{Kaptari:2005qz}
avoiding the uncertainties in the $\Delta$ channel in the OBE  models
and neglecting the quantum mechanical interference between individual contributions
which can not be treated consistently in transport approaches.
Also  we stress here again that in order to separate the bremsstrahlung ($pp\to ppe^+e^-$)
from a vector-dominance like dilepton production via the $\rho$-meson
($pp\to pp\rho, \rho\to e^+e^-$), we do not employ a vector-dominance
form factor when calculating the bremsstrahlung. Thus, the dilepton
radiation via the decay of the virtual photon ($pp\to pp\gamma^*,
\gamma^*\to e^+e^-$) and the direct $\rho$ decay to $e^+e^-$ are
distinguished explicitly in the calculations.
In the Section VI we discuss the model uncertainties concerning
the treatment of $\Delta$'s and bremsstrahlung.

The off-shell HSD transport approach incorporates the {\em off-shell
propagation} for vector mesons {as described in}
Ref.~\cite{Cassing_off} in extension of early BUU transport models \cite{Cass90,Cassing:1999es}.
In the off-shell transport description, the hadron
spectral functions change dynamically during the propagation through
the medium and evolve towards the on-shell spectral function{s} in the
vacuum.   As demonstrated in
Refs.~\cite{Bratkovskaya:2007jk,Brat:2009PLB}, the off-shell dynamics
is important for resonances with a rather long lifetime in {the} vacuum
but strongly decreasing lifetime in the nuclear medium (especially
$\omega$ and $\phi$ mesons) and also be proven to be vital for the correct
description of the dilepton decay  of $\rho$ mesons with masses close to
the two pion decay threshold. For a detailed description of the
off-shell dynamics and the implementation of in-medium scenarios (as a
collisional broadening and/or dropping mass scenario) in HSD as well as for an
extension of the LUND string model to include 'modified' spectral
functions we refer the reader to
Refs.~\cite{Cassing_off,Bratkovskaya:2007jk,Brat:2009PLB}.

For the present study we consider the scenario of a 'collisional
broadening' of the vector meson spectral functions. This is
also supported by experimental data in contrast to the 'dropping mass' scenario (cf.
\cite{vanHees:2007yi,Gale:2006hg,Renk:2006ax,Brat:2009PLB,Santini:2011zw,Linnyk:2011vx}).
We incorporate the effect of collisional broadening of the
vector-meson spectral functions by using for the vector meson width
\begin{eqnarray}
\Gamma^*_V(M,|\vec p|,\rho_N)=\Gamma_V(M) + \Gamma_{coll}(M,|\vec
p|,\rho_N) . \label{gammas}
\end{eqnarray}
Here $\Gamma_V(M)$ is the total width of the vector mesons
($V=\rho,\omega$) in the vacuum.
The collisional width in (\ref{gammas}) is approximated as
\begin{eqnarray}
\Gamma_{coll}(M,|\vec p|,\rho_N) = \gamma \ \rho_N < v \
\sigma_{VN}^{tot} > \approx  \ \alpha_{coll} \ \frac{\rho_N}{\rho_0}
. \label{dgamma}
\end{eqnarray}
Here $v=|{\vec p}|/E; \ {\vec p}, \ E$ are the velocity, 3-momentum
and energy of the vector meson in the rest frame of the nucleon
current and $\gamma^2=1/(1-v^2)$; $\rho_N$ is the
nuclear density and $\sigma_{VN}^{tot}$ the meson-nucleon total
cross section. We use the 'broadening coefficients'
$\alpha_{coll} \approx 150$~MeV for the $\rho$ and $\alpha_{coll} \approx 70$~MeV
for $\omega$ mesons as obtained in \cite{Bratkovskaya:2007jk}.
For the further details we refer the reader to Ref. \cite{Bratkovskaya:2007jk}.

We use the time integration method to calculate dilepton spectra which
means that vector mesons and resonances can emit dileptons from their
production ('birth') up to their absorption ('death'). This is
especially important for the study of in-medium effects because this
method takes the full in-medium dynamics into account.

We note that it is very important to have an adequate description of
the elementary reactions, especially near the threshold  where the cross
sections grow very rapidly.  This rise has a big
impact on the description of the experimental data. The comparison of
the latest experimental data from the HADES collaboration on $pp$
collisions at 3.5 GeV \cite{HADES_pp35} with HSD calculations shows
that the previous parametrizations of $\eta$-meson and of vector mesons
($\rho,\omega$) production cross sections from Ref.
\cite{Bratkovskaya:2007jk} overestimate the data. The over-prediction
of the dilepton yield at the $\rho$-peak has been already realized in
Ref. \cite{Bratkovskaya:2007jk} from the comparison to the DLS data for
$pp$ at 2.09 GeV. Thus, we have modified the HSD model accordingly in
order to obtain a better description of the existing experimental data
on elementary reactions (cf. the discussions in the next subsection).

We note that the HSD model is well tested with respect to the bulk
observables at low energy and in light systems, relevant for present
study. The pion and eta production from C+C collisions at
the energies considered here are shown in Section 3 of Ref. \cite{Bratkovskaya:2007jk}.

\subsubsection{Particle production from elementary reactions}

\begin{figure}[h!]
\phantom{a}\vspace*{5mm}
\includegraphics[width=8cm]{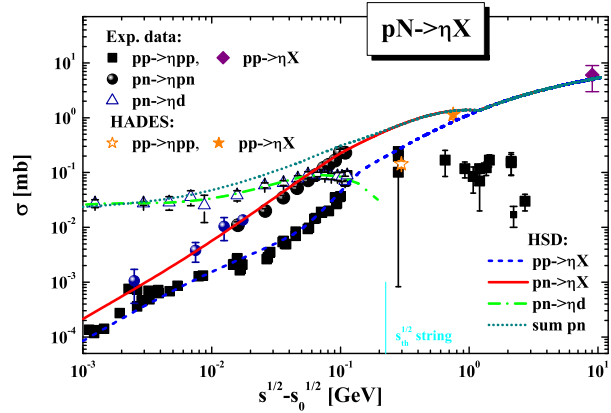}
\caption{(Color online) The $\eta$ production cross section in $pp$ and $pn$
reactions as a function of the invariant energy above threshold
$(\sqrt{s}-\sqrt{s_0})$.  The solid  and dashed  lines represent the
inclusive  $pp \to\eta X$ and $pn \to\eta X$ cross sections from the
HSD model, the dashed-dotted line indicates the $pn \to\eta d$ channel
and the dotted line the  $pn \to\eta X$ cross section including $pn
\to\eta d$.
The experimental data are collected from Refs.
\cite{etanew,Moskal02,Moskal05,HADES_pion125,HADES_pp35}:  the full
squares and open star stand  for the exclusive $pp\to\eta pp$
experimental data, the dots for  $pn\to\eta pn$ and open triangles
for  $pn\to\eta d$ experimental data; the solid diamond and solid star
show inclusive experimental data for $pp\to\eta X$.
The open and full stars indicate the exclusive and inclusive
cross sections extrapolated from the dilepton data by
the HADES collaboration \cite{HADES_pp35}. }
\label{Fig_xsEta}
\end{figure}

Here we describe the major changes/improvements  made in the HSD model
used here compared to the basic HSD version 2.5 used for the dilepton analysis in Ref.
\cite{Bratkovskaya:2007jk}:

\vspace*{3mm}
{\bf 1)} The high energy part of the $pp\to \eta X$ and the $pn \to
\eta X$ cross sections have newly been parametrized. The new
parametrization, compared to the experimental data, is shown in Fig.
\ref{Fig_xsEta} for $pp$ and $pn$ reactions as a function of the
centre of mass energy above threshold $(\sqrt{s}-\sqrt{s_0})$.  The solid
and dashed  lines represent the inclusive  $pp \to\eta X$ and $pn
\to\eta X$ cross sections from the HSD model.  The experimental data
are collected from Refs.
\cite{etanew,Moskal02,Moskal05,HADES_pion125,HADES_pp35}:  the full
squares and the open star stand  for the exclusive $pp\to\eta pp$ data,
the dots  for  $pn\to\eta pn$ data. The full diamond and the full star
show inclusive data for $pp\to\eta X$.
The open and full stars indicate the exclusive and inclusive
cross sections extrapolated from the dilepton data by
the HADES collaboration \cite{HADES_pp35}.
It is important to note that
in elementary reactions also a deuteron can be produced in the final state
via $pn \to\eta d$.  The cross section for
this channel is indicated by the dashed-dotted line in Fig.
\ref{Fig_xsEta}, whereas the dotted line shows the  $pn \to\eta X$
cross section including the $pn \to\eta d$ contribution.  The open
triangles show the  experimental data for $pn\to\eta d$ from Refs.
\cite{etanew,Moskal02,Moskal05}. The channel $pn\to\eta d$ is not
considered in the HSD calculations for $A+A$ and $p+A$ reactions
because the probability for  deuteron formation in the  baryonic
medium is negligibly small.

\begin{figure}[h!]
\phantom{a}\hspace*{-40mm}
\includegraphics[width=11cm]{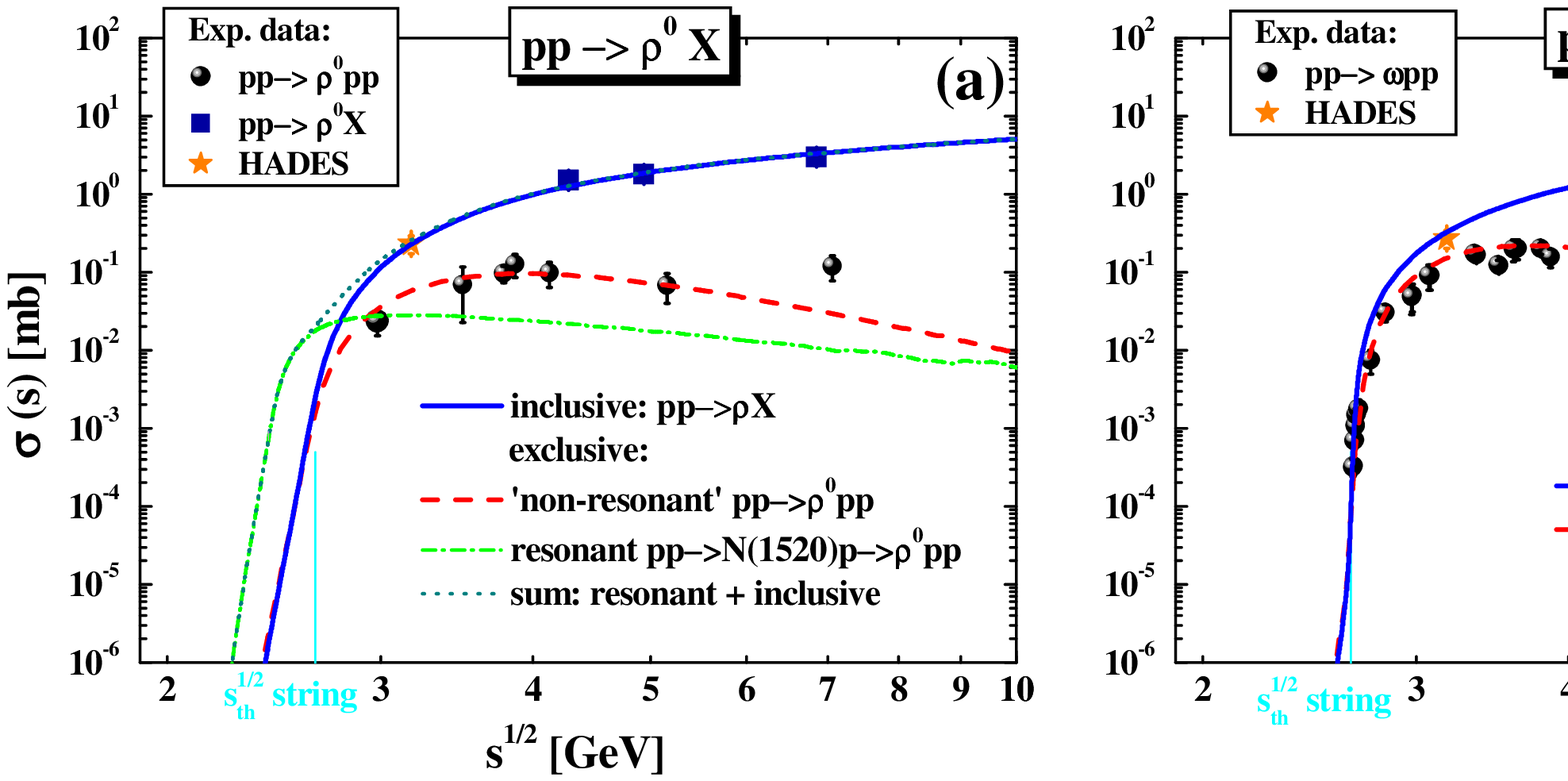}
\caption{(Color online) The production cross sections for the channels $pp \rightarrow
\rho X$ (left plot (a)) and $pp \rightarrow \omega X$ (right plot (b)) as a
 function of the centre of mass energy $\sqrt{s}$.  The solid lines
 represent the parametrizations of the inclusive $p p \rightarrow V X \
 (V=\rho,\omega)$ cross sections while the dashed lines stand for the
 exclusive  'non-resonant' cross sections. The dashed-dotted line on the
 left plot shows the contribution from the $N(1520)$ resonance to the
$\rho^0$ production via the process $pp\to N(1520)p\to \rho^0 pp$ and the
dotted line indicates the sum of the inclusive non-resonant and exclusive
resonant contributions.  The experimental data
\protect\cite{expVprod,DISTO02,Moskal02} are shown for exclusive $pp
\rightarrow V pp$ (dots) and inclusive $pp \rightarrow V X$ (squares)
vector meson production.
The stars indicate the inclusive cross sections extrapolated from
the dilepton data by  the HADES collaboration \cite{HADES_pp35}.
The vertical light blue lines show the threshold for meson production
by string formation and decay ($\sqrt{s}_{th}=2.6$ GeV) as
implemented in HSD for baryon-baryon channels.  }
 \label{Fig_xsVM}
\end{figure}

\vspace*{3mm}
{\bf 2)} The HSD model has also been improved  concerning the isospin
separation of the vector meson production in baryon-baryon ($BB$) and
secondary meson-baryon ($mB$) reactions.  The  isospin averaged cross
sections $ BB \to V BB$ ($V= \rho,\omega,\phi$) and $ mB \to V B $ have
been replaced by cross sections which take explicitly  the
isospin for each channel into account.
The new parametrization of the cross section as a function of
the centre of mass energy, $\sqrt{s}$  for the $pp$ reaction is compared in Fig.  \ref{Fig_xsVM}
to the experimental data. The solid lines represent the parametrizations of the inclusive $p p
\rightarrow V X \ (V=\rho,\omega)$ cross sections  while the dashed
lines stand for the exclusive cross sections.  We denote these
exclusive cross sections for the $\rho$-meson production as
'non-resonant' since in this study we consider explicitly the possible
contribution of the baryonic resonance $N(1520)$ to the sub-threshold
$\rho^0$ production ($pp\to N(1520)p\to \rho^0 pp$). It is indicated as
the dashed-dotted line on the left plot. The dotted line shows the sum
of the inclusive 'non-resonant' and exclusive 'resonant' contribution.
The experimental data  \cite{expVprod,DISTO02,Moskal02} are shown for
exclusive $pp \rightarrow V pp$ (dots) and inclusive $pp \rightarrow V X$
(squares) vector meson production.
The stars indicate the inclusive cross sections extrapolated from
the dilepton data by  the HADES collaboration \cite{HADES_pp35}.

We note that we do not propagate explicitly the $N(1520)$  resonance in
the HSD approach, rather we consider it as an excitation in the amplitude which
enhances the $\rho$-meson production in $NN$ and $\pi N$ reactions at
sub-threshold energies.  The modelling of the $N(1520)$ production in $NN$
collisions is based on a phase space model with a constant matrix
element adopted from Ref. \cite{TeisZP97}. The contribution  of
the $N(1520)$ to the $\rho$ cross section is included in line with Ref.
\cite{Peters} which has been used in our previous work
\cite{BratKo99}.  The decay channels of the $N(1520)$  resonance are not well established.
Especially the disintegration into a $\rho$ is estimated in between 15 -25\%.
Including this contribution from the $N(1520)$ resonance decay presents an upper estimate for the $\rho$-meson
production in $NN$ and $\pi N$ reactions at sub-threshold energies.
This model assumption can be checked experimentally via an observation
of an enhancement of the dilepton yield near the $\rho$-peak in the
elementary reactions at sub-threshold energies.  In the case of
heavy-ion collisions at  low bombarding energies the contribution of
the $N(1520)$ resonance to the dilepton spectra can hardly be seen
especially not in reactions of the light nuclei as C+C as
measured by the HADES collaboration at 1.0 $A$GeV.  The Fermi motion
modifies  the available energy for meson production and due to the
rapid rise of the cross section at threshold the inclusive $\rho$-meson
production mechanism starts to dominate, see Fig. \ref{Fig_xsVM}, even
if the nominal energy is below threshold. On the other hand, the
$N(1520)$ resonance can be excited by pion-baryon collisions and
contribute to the $\rho$-meson production via the process $\pi N\to
N(1520) \to \rho N$.  The probability of such processes is larger for
heavy nuclei collisions but the pion density is relatively small at
sub-threshold energies where the possible contribution of $N(1520)$
plays a role. Consequently, the enhancement of the $\rho$-meson
production by accounting for the $N(1520)$ channel is relatively small in
the HSD model.  This differs from e.g. the UrQMD model
\cite{UrQMD1,UrQMD2} where a much larger cross sections for the
$N(1520)$ production is used. We will come back to this discussion in
Section IV.\\

\vspace*{3mm}
{\bf 3)} We improved also the description of multi-meson production between
the two pion production threshold and $\sqrt{s}=2.6$ GeV where we match
the standard HSD description of particle production via strings.

Close to the two-pion threshold the two pions are dominantly produced by
the decay of 2 Deltas created in NN collisions. With increasing energy
the available phase space is sufficient for multi-meson production and
a lot of extra channels become open. However, it is unknown whether the
light mesons are produced by the decay of heavy baryonic resonances
or directly from the excitation and decay of the strings.
Since there is very little experimental information on the exclusive
channel decomposition in this 'intermediate' energy range we used
the FRITIOF LUND string model as an 'event generator' for the
production of such 'multi-meson' channels by adding them to the
exclusive channels which are modeled in the HSD explicitly, such
that we obtain the inelastic $NN$ cross section, i.e.
$\sigma_{inel}=\sigma_{excl} +\Delta\sigma_{incl}$,
where $\sigma_{excl}$ stands for the exclusive channels
such as $NN\to \Delta N$, $NN\to \Delta N m,  \ m \equiv \pi, \rho, \omega, \phi, ...$
and channels with strangeness production such as $NN\to YNK \ (Y\equiv \Lambda, \Sigma)$
and $NN\to NN K\bar K$.  Here $\Delta\sigma_{incl}$
corresponds to the sum of the 2 pion production channels as $NN\to \Delta\Delta$
and 'multi-meson' production $NN\to NN(\Delta) + n\times m$ $(n=2,3,4,...)$ and
channels with the final hyperons and strange mesons.
We note that since close to the threshold the FRITIOF
model doesn't provide the correct isospin decomposition for
$\Delta$ production, since e.g. an exclusive channel $NN\to \Delta^{++}n$ is
missing, we have adjusted the FRITIOF model to correct for the isospin decomposition
of produced $\Delta$'s:  for the exclusive channel $pp \to N\Delta$ we assume
now that 3/4 of the produced $\Delta$ are in the $\Delta^{++}$ state and
only 1/4 in the dilepton producing $\Delta^{+}$ state. This leads
to a reduction of $\Delta^+$ production and an enhancement of
$\Delta^{++}$ production, respectively.

\begin{figure}[t!]
\phantom{a}\hspace*{-25mm}
\includegraphics[width=6.cm]{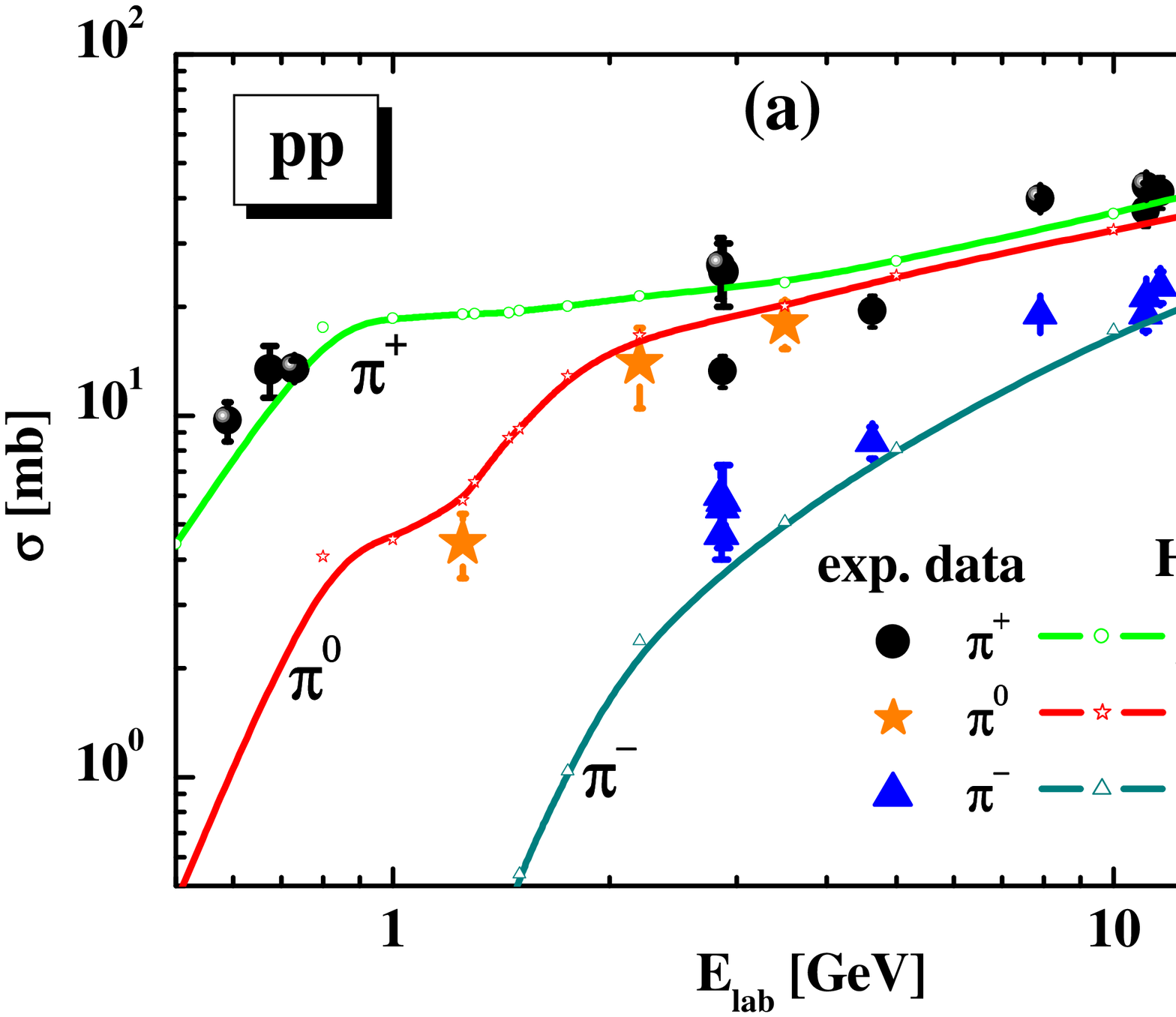}
\hspace*{19mm}
\includegraphics[width=6.15cm]{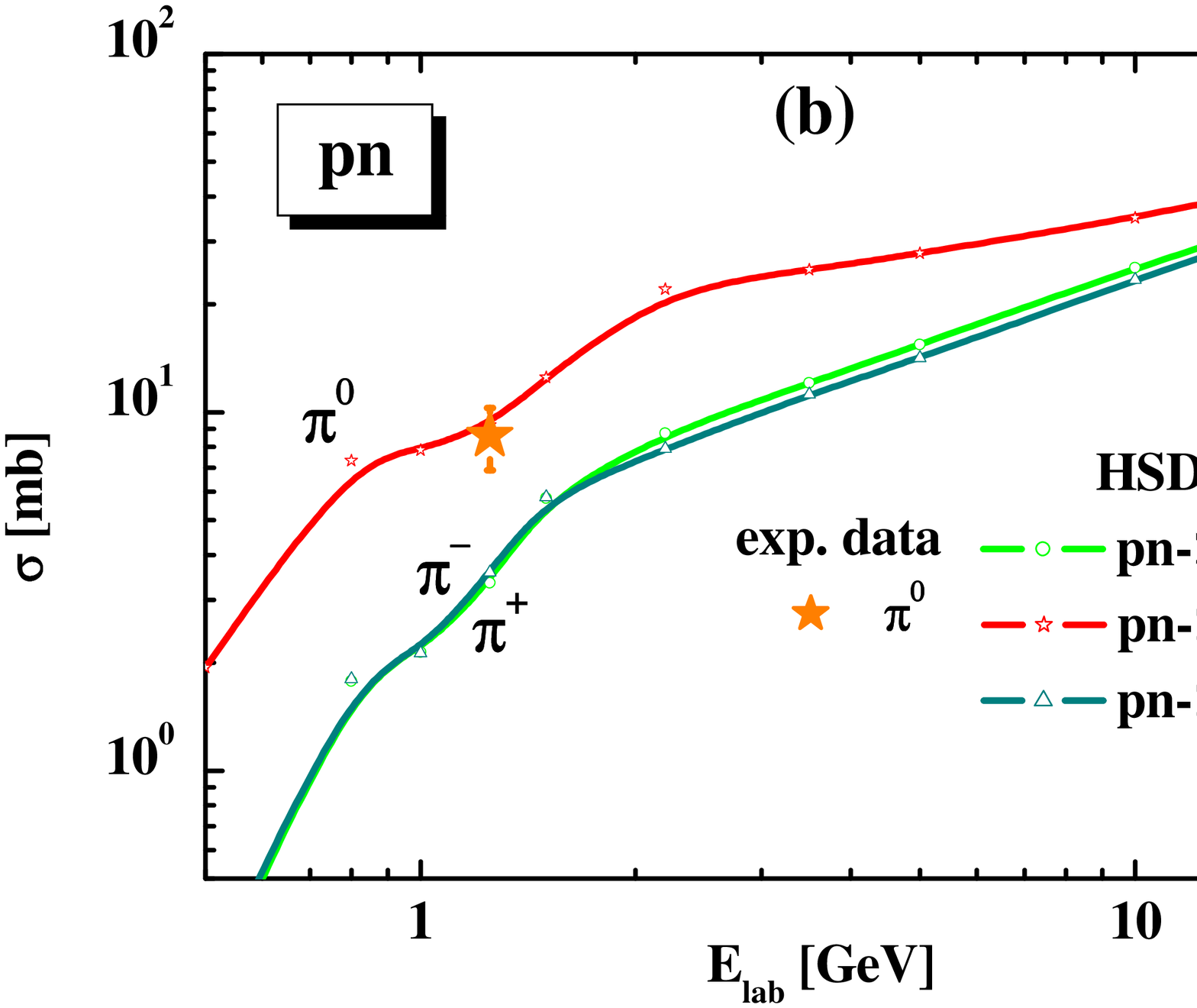}
%
\caption{(Color online) Left (a): The inclusive pion  production cross sections as a
function of the proton bombarding energy $E_{lab}$.  The HSD results
are shown in terms of lines with open symbols whereas the experimental
data are indicated by the corresponding solid symbols, i.e. for $pp
\rightarrow \pi^+ + X$: HSD - the solid line with open dots,
experimental data - solid dots from Refs.
\protect\cite{expVprod,expPiprod}; for $pp \rightarrow \pi^0 + X$: HSD
- the solid line with open stars, the HADES data - full stars from
Refs.  \protect\cite{HADES_pp22,HADES_pp35}; for $pp \rightarrow \pi^-
+ X$: HSD - the solid line with open triangles, experimental data -
full triangles from  Refs.  \protect\cite{expVprod,expPiprod}.
Right (b):  The production cross sections for $pn \rightarrow \pi X, \
\pi=\pi^+, \pi^0, \pi^-$ from the HSD model, the HADES data
-- full star from the extrapolation in Ref.  \protect\cite{Agakishiev:2009yf}.}
\label{Fig_xspi}
\end{figure}

\begin{figure}[h]
\includegraphics[width=9.5cm]{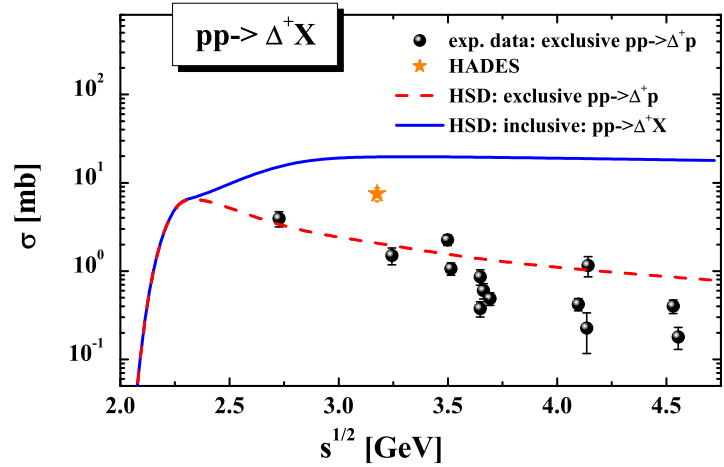}
\caption{(Color online) The production cross sections for the inclusive channels
$pp \rightarrow \Delta^+ X$ (solid line) and the exclusive channel
$pp \rightarrow \Delta^+ p$ (dashed line)
from HSD  as a function of the invariant energy $\sqrt{s}$.
The experimental data  \protect\cite{expVprod} are shown for
exclusive $pp \rightarrow \Delta^+ p$ production.
The star indicates the inclusive cross section extrapolated from
the dilepton data by  the HADES collaboration \cite{HADES_pp35}.}
\label{Fig_xsDel}
\end{figure}

The excitation function of the multiplicity of the different pions in HSD is shown
in Fig.\ref{Fig_xspi}, on the left hand side for $pp$ collisions,
on the right hand side for $pn$ reactions. These multiplicities
are compared with the available data which are very scarce for $pn$ reactions.
Additionally to the total pion multiplicity, the multiplicity of
$\Delta$'s themself is very important for the dilepton study
because the $\Delta$ resonances decay into pions as well as into
dileptons whereas other sources of pions do not contribute to the
dilepton yield. The $\Delta$ production in $pp$ collisions in the
HSD approach is shown in Fig. \ref{Fig_xsDel} and compared with
the available experimental data. Here the production cross
sections for the inclusive channels $pp \rightarrow \Delta^+ X$
(solid line) and for the exclusive channel $pp \rightarrow \Delta^+ p$
(dashed line) from HSD are presented  as a function of
the invariant energy $\sqrt{s}$. The experimental data
\cite{expVprod} are shown for exclusive $pp \rightarrow \Delta^+ p$ production.
The star indicates the extrapolation for the $\Delta$ inclusive cross section
from the dilepton spectra by the
HADES collaboration based on the PLUTO simulation program \cite{PLUTO} from
Ref. \cite{HADES_pp35}. One can see from  Fig. \ref{Fig_xsDel}
that the inclusive $\Delta$ production dominates the exclusive one
already at relatively low $\sqrt{s}$.
However, due to the lack of inclusive experimental data on $\Delta$
production it is hard to justify the modeling of $\Delta$ dynamics
beyond the exclusive channels which are relatively well known
experimentally and accurately modeled in transport approaches.


As said, above a kinetic energy of 1.5 GeV there is no experimental
information available on whether resonance are involved in the
production. This is the reason way different parameterizations
have been advanced. For example in the resonance based GiBUU model \cite{GiBUU} a  lower
inclusive $\Delta^+$ production cross section is used as compared to HSD.
Also the isospin relations are different which leads to a lower dilepton
contribution from $\Delta$ Dalitz decay. Additionally the
different parametrization for the differential dilepton decay
width is employed which lowering the $\Delta$ Dalitz decay channal
substantially compare to the HSD (cf. discussions in Section VI).

\subsection{Open questions related to the elementary reactions in transport models}

In nucleon-nucleon collisions at low energies, i.e.  below $\sqrt{s}< 2.2$ GeV,
very seldom more than one meson is  produced. The cross sections for these reactions have been
measured experimentally (cf. \cite{Moskal02}) and are used in the transport approach.
Above $\sqrt{s}\approx  2.2\ GeV $ the multi-meson production starts to dominate,
but the experimental information on inclusive as well as exclusive multi-meson production
channels are very poor. Also it is not known whether
the mesons are directly produced or whether they are decay products of
intermediate resonances or strings.  The theoretical analysis of these data has
not produced yet a consistent knowledge on the channel decomposition
\cite{WASAatCOSY:2011aa,AlvarezRuso:1997mx,Oset:1998bt,Cao:2010km}.
This introduces large uncertainties for the prediction of the dilepton yield
in transport theories because it depends on the formation of specific
intermediate resonances.
We note that in the UrQMD model the production of mesons at intermediate
energies is realized exclusively via excitation and decay of heavy baryonic resonances
which are explicitly propagated  in the transport model \cite{UrQMD1,UrQMD2},
whereas in HSD the string mechanism is used (as discussed above) for the description of the same final
meson spectra. Thus, one needs more exclusive experimental information in order
to differentiate between the models.

\subsection{The IQMD model}

\begin{figure}[h]
\includegraphics[width=9.5cm]{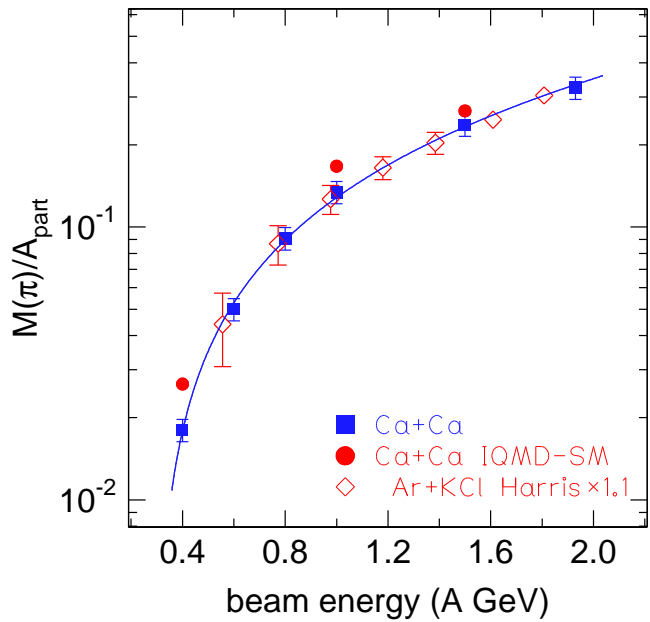}
\caption{(Color online) The excitation function of the $\pi$ multiplicity per participating nucleon
($({N(\pi^+)+N(\pi^-))}/{A_{part}}$) for Ca+Ca collisions using $A_{part}$ as $0.9 A$.
Data of the FOPI collaboration are compared with data of Harris et al. \cite{Harris:1987md}
and  predictions of the IQMD model \cite{Reisdorf:2006ie,reis}.}
\label{Fig_piiqmd}
\end{figure}

The IQMD model used for the calculations in this study is the same
as introduced in the first IQMD paper on dilepton production
\cite{Thomere:2007cj}. In this model all pions are produced by the
decay of $\Delta$ resonances. Because no higher mass resonances
are included we limit the prediction to beam energies up to 2 $A$GeV.
The excitation function of the pion yield for the Ca+Ca
system, compared with the available data, is shown in Fig.
\ref{Fig_piiqmd}. We see that the pion multiplicity, the result of a
complicated interplay between $\Delta$ creation, absorption and
decay, is quite reasonable reproduced by the IQMD approach
\cite{reis}. This is also the case for heavier systems
\cite{Reisdorf:2006ie}. Thus both, the IQMD as well as the HSD
approach, describe the available pion data quite well, a
prerequisite for an analysis of the dilepton spectra which are not
only normalized to the pion yield but have an important
contribution from the $\Delta$ decay. For other models which are
used to describe the dilepton production, like \cite{Buss:2011mx},
it remains to be seen whether they reproduce heavy-ion pion data.

For the calculations of the dilepton spectra the standard IQMD
program \cite{Hartnack:1997ez,Hartnack:2011cn} has been supplemented
with all elementary cross sections which are important for this process\cite{Thomere:2007cj}.
For that we have used the parametrizations of available
experimental data, but for many channels, $pp$ data are only available for low $\sqrt{s}$ values
and $np$ data are very scarce. Consequently, in heavy-ion collisions at beam energies
larger than 1.5 $A$GeV most of the particles which emit dileptons are produced
using theoretically calculated cross sections.  In Ref. \cite{Thomere:2007cj} we have
studied how the uncertainties of the cross sections from elementary
reactions influence the dilepton spectra in heavy-ion collisions.
For these studies we use the set up in which the $pn \to \omega pn$
cross section is 5 times higher than the $pp\to \omega pp$ cross section.
This explains the difference between HSD and IQMD at dileption
invariant masses close to the $\rho,\omega$ peak.

In the IQMD approach the dileptons are calculated perturbatively
using the 'spontaneous decay' method - contrary to the time
integration method in HSD and UrQMD.
It is based on the assumption that all hadrons which decay into dileptons
and which are produced in the heavy-ion collision contribute to the dilepton
yield as if they were produced in free space. This implies that a possible later
reabsorption of the hadrons is not taken into account.
Because in reality some of the $\Delta$'s and of the other dilepton producing
hadrons are reabsorbed, the IQMD calculations give an upper limit for the dilepton
production in heavy-ion collisions. Consequently, the spontaneous decay method limits
the approach to small systems contrary to the time integration method which
follows
the in-medium dynamics of all dilepton sources exactly. However, for the systems
studied here the 'spontaneous decay' method is still acceptable.
For the details of the cross sections for the creation of dilepton
producing particles we refer to Ref. \cite{Thomere:2007cj}.

\section{Dilepton production in elementary $pp, pd$ and $p+A$ reactions}

The first reaction considered here is the dilepton production in elementary reactions like
$p+p$, quasi-free $p+n(d)$ and $p+Nb$ reactions.

\subsection{Dilepton production in pp and pd at energies around 1.25 GeV}

\begin{figure}[h!]
\phantom{a}\hspace*{-50mm}
\includegraphics[width=13cm]{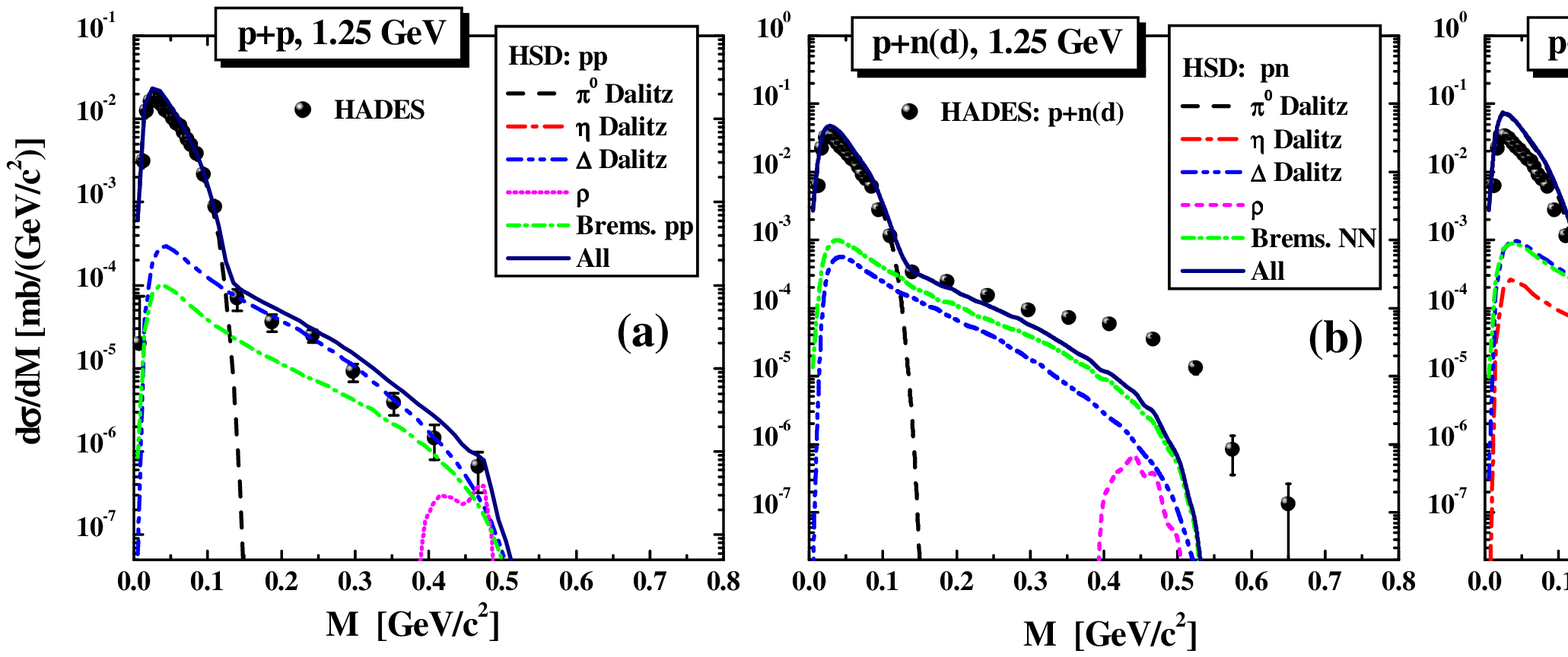}
\caption{(Color online) The HSD results for the dilepton differential cross section
$d\sigma/dM$ for $pp$ (left plot (a)), $pn$ (middle plot (b)) and $pd$ (right
plot (c)) reactions at 1.25 GeV  in comparison to the experimental data for
$pp$ (left) and quasi-free $pn$ (middle and right plots) reactions from
the HADES collaboration
\cite{Lapidus:2009aa,Agakishiev:2009yf}.  The individual colored lines
display the contributions from the various channels in the HSD
calculations (see color coding in the legend). The theoretical
calculations passed through the corresponding HADES acceptance filters
and mass/momentum resolutions. }
\label{Fig_pp125}
\end{figure}

We start our discussion with the HADES and DLS data a 1.25 GeV.
Fig. \ref{Fig_pp125} shows the differential cross section $d\sigma/dM$
for dileptons as a function of the invariant mass M for $pp$ (left),
$pn$ (middle)  and
$pd$ (right) reactions at 1.25 GeV. The HSD results are presented  in
comparison to the experimental data  from the HADES collaboration
\cite{Lapidus:2009aa,Agakishiev:2009yf}.  The different lines display
the contributions from the various channels in the HSD calculations
(for the colour coding we refer to the legend). We note here (and that
applies to all further plots) that the theoretical calculations passed
through the appropriate experimental acceptance filters and that the
mass/momentum resolution is taken into account.

As seen in the left part of Fig. \ref{Fig_pp125} the $pp$ dilepton
yield is dominated by the $\Delta$-Dalitz decay while bremsstrahlung is
sub-leading due to the destructive interference  between initial and
final state amplitudes in case of equal charges due to a different sign
in the acceleration. Thus these HADES data provide a solid constraint
on the $\Delta$ production whose control will be very important for a
robust interpretation of the heavy-ion data. In $pn$ collisions,
however, bremsstrahlung is dominating as can be seen from the middle
part of Fig. \ref{Fig_pp125}. Because the form of the dilepton
invariant mass spectrum  from $\Delta$ decay and from
bremsstrahlung is not completely the same, the form of the $pp$ and the $pn$ spectra is not
identical and we see in $np$ a slight enhancement close to the
kinematic limit.  In the right part of Fig. \ref{Fig_pp125} we
compare the HSD results for $pd$ collisions with the so called quasi-free $pn$ HADES
data, used later as the 'reference' spectra $NN=(pn+pp)/2$ for the
interpretation of the heavy-ion data. Experimentally the quasi-free $pn$ events have been
separated  by measuring the proton spectator in the $pd$ reactions at
$1.6 < p_{lab}<2.6$~$GeV/c$.

The comparison of the $pp$ and the $pn(d)$ data shows clearly that in $pn(d)$ the
proton does not scatter on a quasi free neutron. The kinematical limit for the invariant mass of the
dilepton which is $M_{max}=\sqrt{s_{NN}}-2m_N= 0.545$ GeV in $pp$ and $np$ collisions
is well exceeded in the $pd$ collisions. The largest invariant mass observed
($M \approx 0.66$  GeV) corresponds to the maximal invariant mass which is kinematically
allowed in the {\it  three} body $pd$
system under the condition that the outgoing proton has at least a momentum of 1.6 $GeV/c$.
Therefore at the upper end of the invariant mass spectra we have a collision of the proton with
the deuteron with a center of mass energy of $\sqrt{s_{pd}}=\sqrt{(p_p+p_d)^2}$.
This observation one has to keep in mind for the interpretation of dilepton production
in heavy-ion collisions, when the $pd$ results are used as a reference to discuss
the in-medium enhancement of the dilepton yield.

In semiclassical transport calculations, like HSD, one simulates the deuteron as a bound
system of  a proton and a neutron which are redistributed in coordinate
and momentum space according to the wave function of the Paris
potential \cite{Paris_wf}.  The energy of each nucleon (in the deuteron
rest frame) is taken as $E_N=m_N+\varepsilon/2$, where
$\varepsilon=-2.2$~MeV is the binding energy of the deuteron.  We use the
energy-momentum relation for free particles to determine the effective
mass of the nucleon and then the energy-momentum 4-vector to describe
the collision. An incoming nucleon scatters with one or subsequently with both nucleons
of the deuteron but never with the two at the same time.
This gives another kinematics as compared to a true three-body collision and therefore
HSD calculations underpredict the dilepton production close to the
kinematical limit of $pd$ collisions.

Another problem with the quasi-free $pn$ scattering is related to the
possibility of deuteron formation in the final state. This is not probable in
heavy-ion collisions (cf. Ref. \cite{Danielew91}) and not included in HSD.
However, as seen from the Fig. \ref{Fig_xsEta}, the process $pn\to\eta d$
might be important for the $\eta$-production at threshold energies.
Thus, we include this contribution as an enhanced cross section for $\eta$
production in $pn$  (this was not included in our previous work
\cite{Bratkovskaya:2007jk}) but we do not treat the deuteron formation
explicitly in the code.  As seen from the right part of Fig.
\ref{Fig_pp125} in np collisions around $M=0.4$ GeV the $\eta$
contribution turns out to be of the same order of importance as
$\Delta$-Dalitz decays and bremsstrahlung.

Figure \ref{Fig_pp125} (right) shows that in $pd$ collisions the HSD model
underestimates the dilepton yield between  $0.35<M<0.5$ \ GeV, a region which
is accessible in two-body collisions at this energy. A possible
candidate to explain this enhancement is the contribution of
sub-threshold $\rho$-meson production via excitation and decay of the
$N(1520)$ resonance shown as the dashed-dotted line in Fig. \ref{Fig_xsVM}.  A very small
contribution of this resonant $\rho$ production channel is even seen in
$pp$ collisions (dotted line on the left plot).  However, this
contribution is not sufficient to describe the experimental data.  This
is in line with a recent study by the GiBUU group \cite{GiBUU}.  Also IQMD
calculations fail to describe this part of the spectrum.

\begin{figure}[h!]
\phantom{a}\hspace*{-30mm}
\includegraphics[width=11cm]{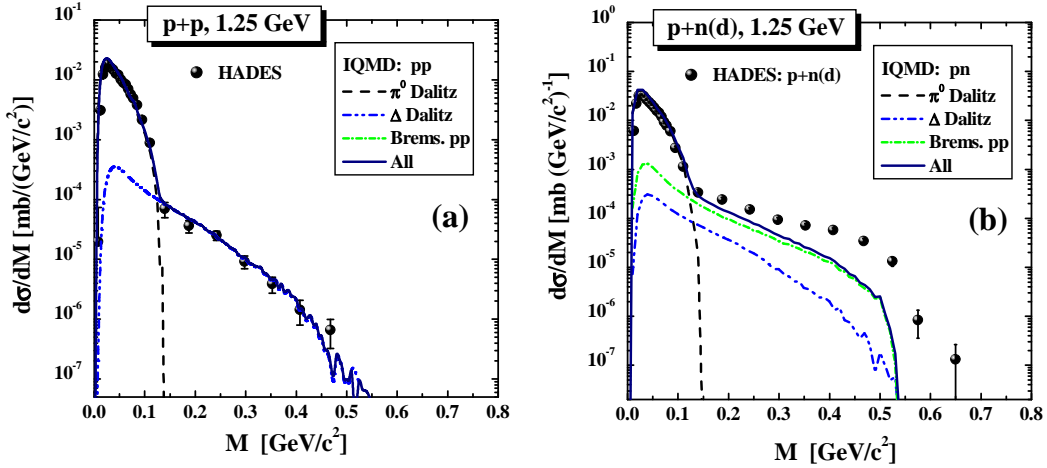}
\caption{(Color online) The IQMD results for the dilepton differential cross section
$d\sigma/dM$ for $pp$ (left (a)) and $pn$ (right (b)) reactions at
1.25 GeV  in comparison to the experimental data for $pp$ (left) and
quasi-free $pn(d)$ (right) reactions from the HADES collaboration
\cite{Lapidus:2009aa,Agakishiev:2009yf}.  The individual
colored lines display the contributions from the various channels in
the IQMD calculations (see color coding in the legend). The theoretical
calculations passed through the corresponding HADES acceptance filters
and mass/momentum resolutions. }
\label{Fig_IQNN125}
\end{figure}

Fig. \ref{Fig_IQNN125} shows the IQMD predictions for $pp$ and $np$
collisions as well compared to $pp$ and $pd$ HADES data. We see a very good
agreement between HSD and IQMD predictions
for the elementary $pp$ and $pn$ reactions.

\begin{figure}[h]
\includegraphics[width=15cm]{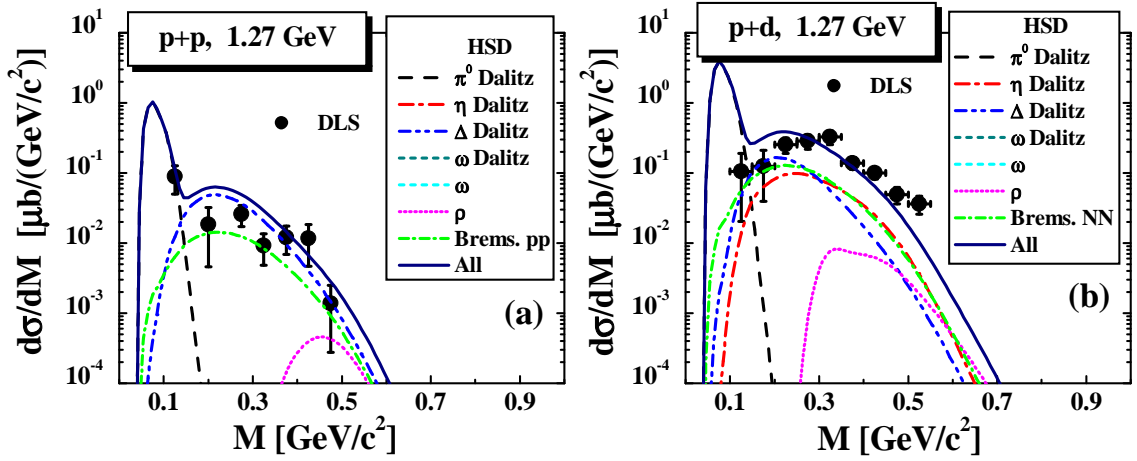}
\caption{(Color online) The dilepton differential cross section $d\sigma/dM$ for
$pp$ (left plot (a)) and $pd$ (right plot (b)) at 1.27 GeV in comparison to the
DLS data \cite{Wilson:1997sr}.  The HSD calculations passed
through the corresponding DLS acceptance filter and mass resolution.
}
\label{Fig_pp125DLS}
\end{figure}

The cross section $d\sigma/dM$ at 1.27 GeV, calculated in the HSD model, is compared
in Fig. \ref{Fig_pp125DLS} to the $pp$ (left) and $pd$ (right) DLS data \cite{Wilson:1997sr} .
The theoretical calculations passed
through the corresponding DLS acceptance filter and mass resolution.
While the agreement between HSD and the data looks reasonable,
one has to keep in mind that due to the very broad mass resolution
the spectra are strongly distorted at large invariant masses.
There seems to be an underestimation of the last experimental
point for $pd$, however, the quality of the data does not allow for robust
conclusions.

\subsection{Dilepton production in pp and pd at energies around 2.2 GeV}

\begin{figure}[h]
\includegraphics[width=8cm]{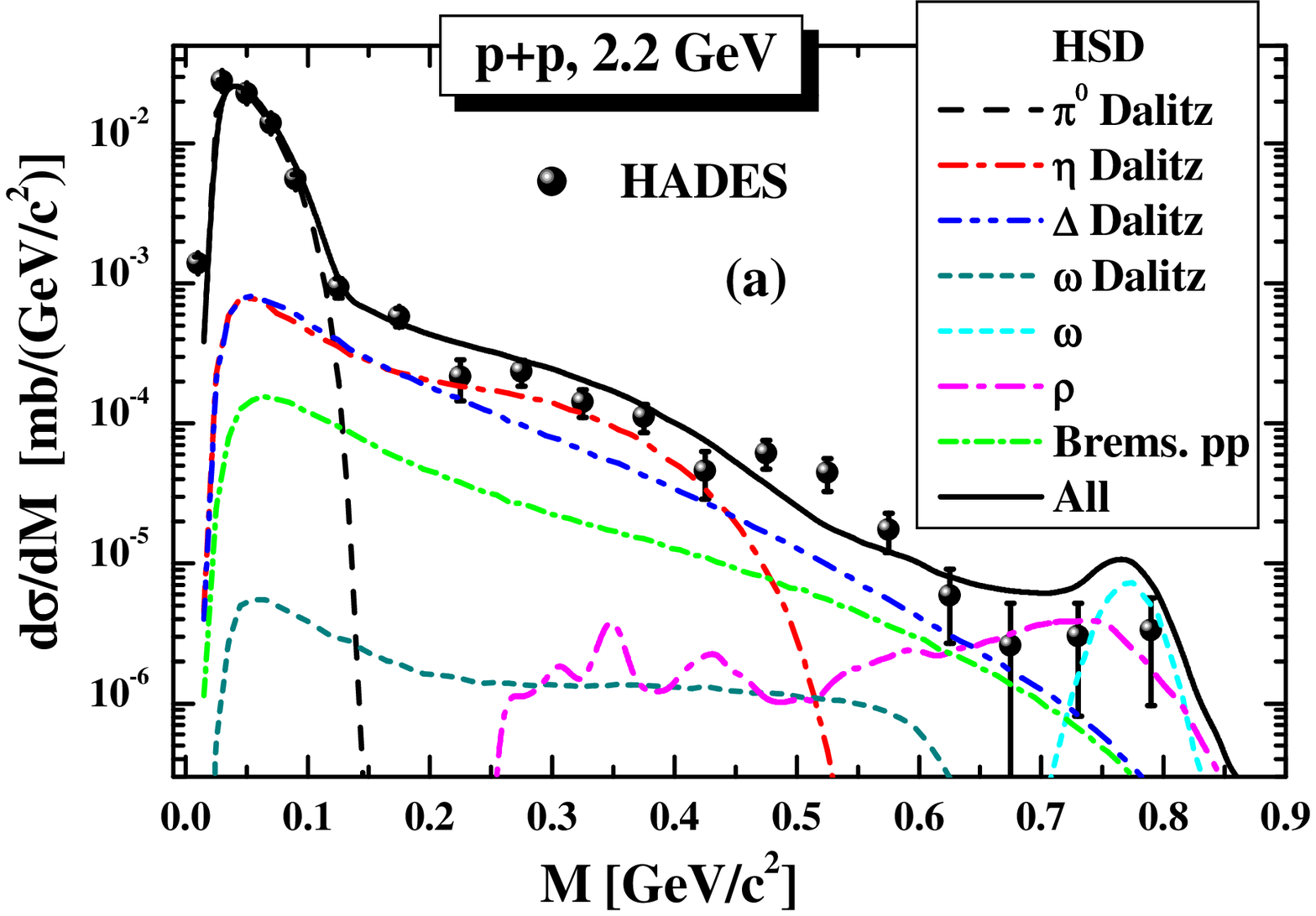} \hspace*{5mm}
\includegraphics[width=8cm]{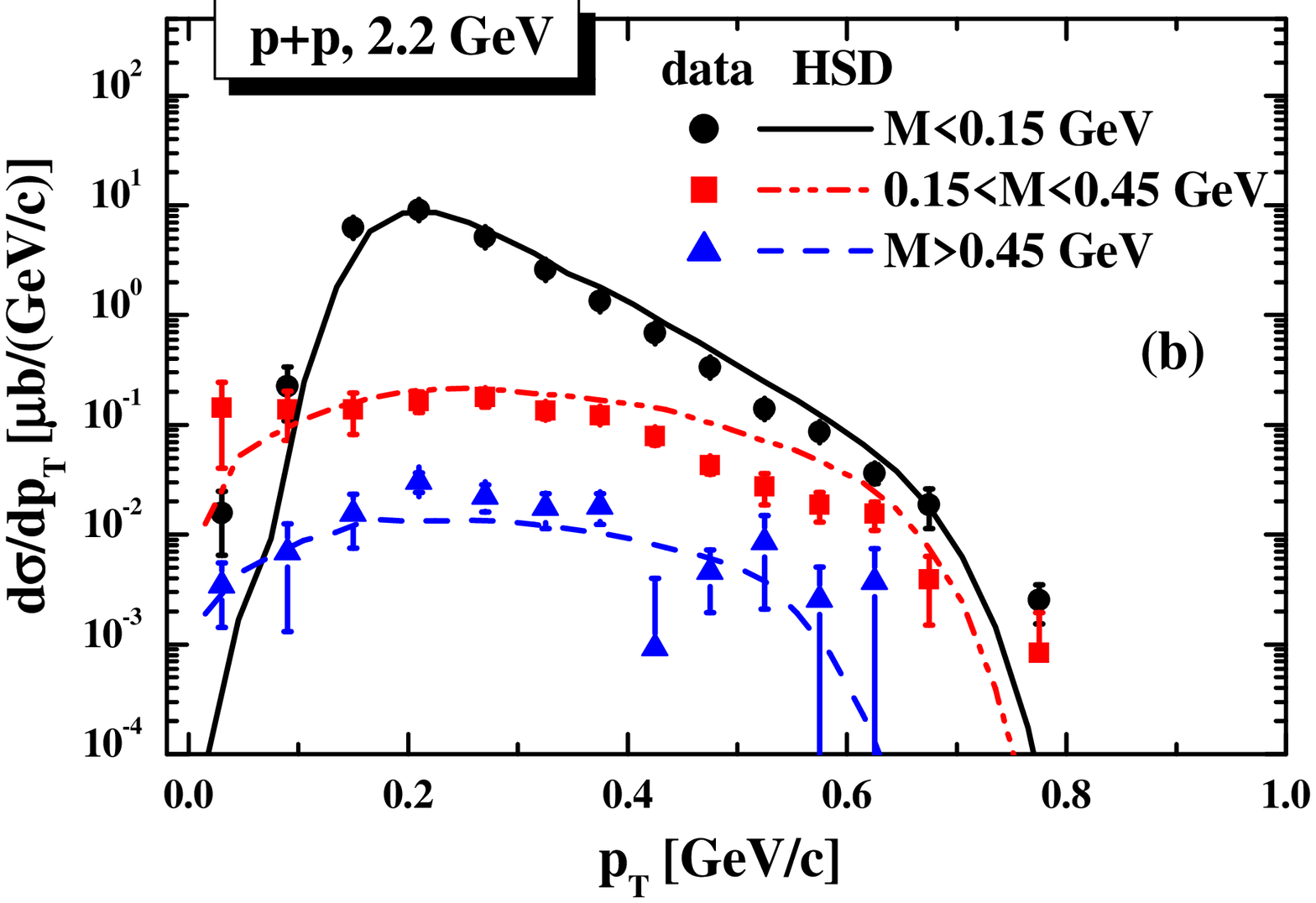}
\caption{(Color online) Left (a): the differential  cross section $d\sigma/dM$ from HSD calculations
 for $e^+e^-$ production in $pp$ reactions at a bombarding energy of 2.2 GeV
 in comparison to the HADES data \cite{HADES_pp22}.
The individual coloured lines display the
contributions from the various channels in the HSD calculations
(for the colour coding see legend).
Right (b): HSD results for the differential dilepton transverse momentum
cross section for $pp$ at 2.2 GeV and for different mass bins:
$M \leq$ 0.15 GeV,  0.15 $\leq M \leq$ 0.55 GeV and  $M \geq$ 0.55 GeV.
The theoretical calculations passed through the corresponding HADES
acceptance filter and mass/momentum resolution.
}
\label{Fig_Mpp22}
\end{figure}

The differential  cross section $d\sigma/dM$ from HSD calculations   for $e^+e^-$
production in $pp$ reactions at bombarding energies of 2.2 GeV in
comparison to the HADES data \cite{HADES_pp22} is presented in Fig.
\ref{Fig_Mpp22} (left).  The right part of Fig.  \ref{Fig_Mpp22} shows
for the same reaction the HSD results for the differential
transverse momentum cross sections for $pp$ at 2.2 GeV separated for
different invariant mass bins:  $M \leq$ 0.15 GeV,  0.15 $\leq M \leq$
0.55 GeV and  $M \geq$ 0.55 GeV. Also at an energy of 2.2 GeV we
see a quite satisfying agreement between theory and experiment.

\begin{figure}[h]
\includegraphics[width=8cm]{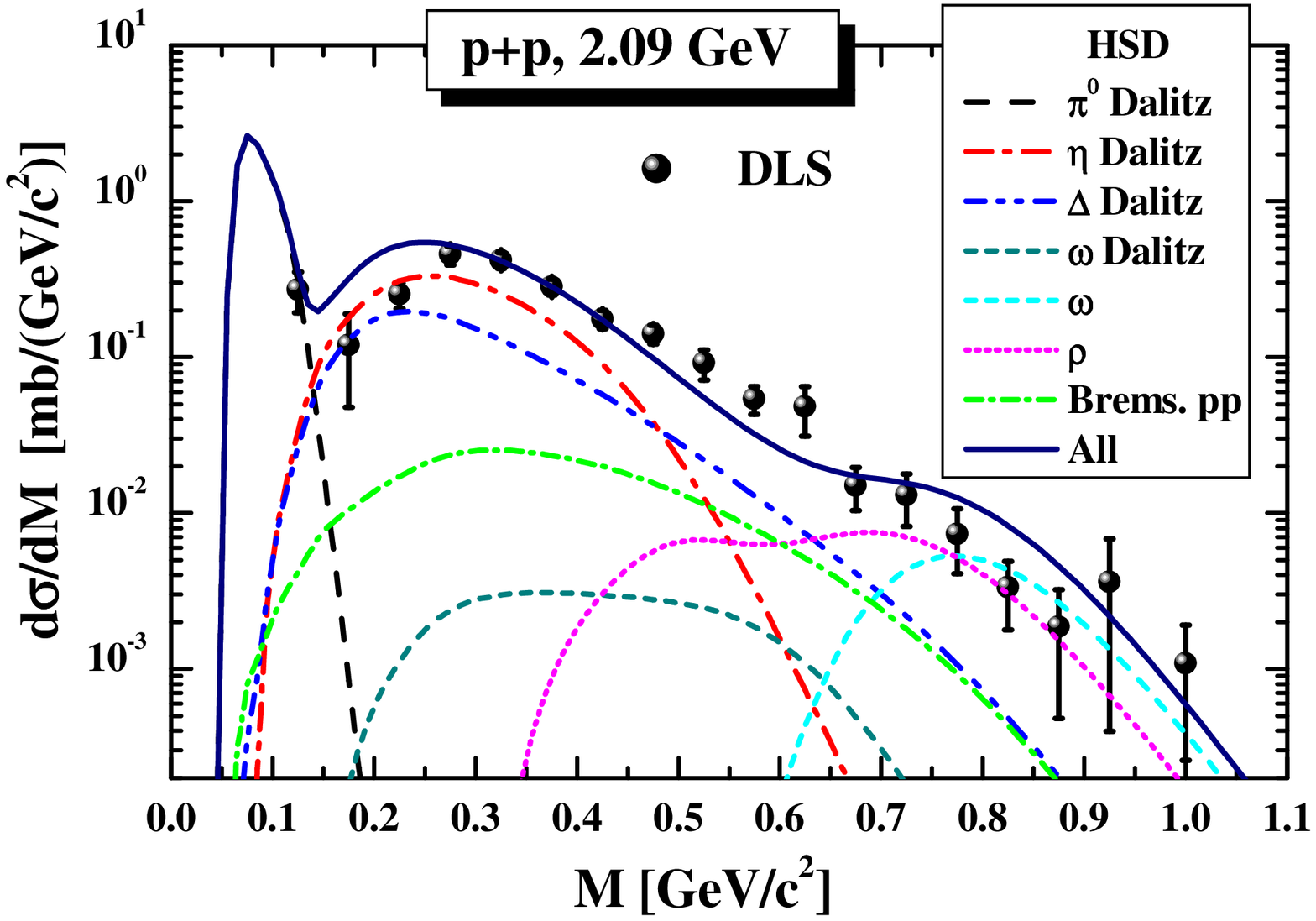}\hspace*{5mm}
\includegraphics[width=8cm]{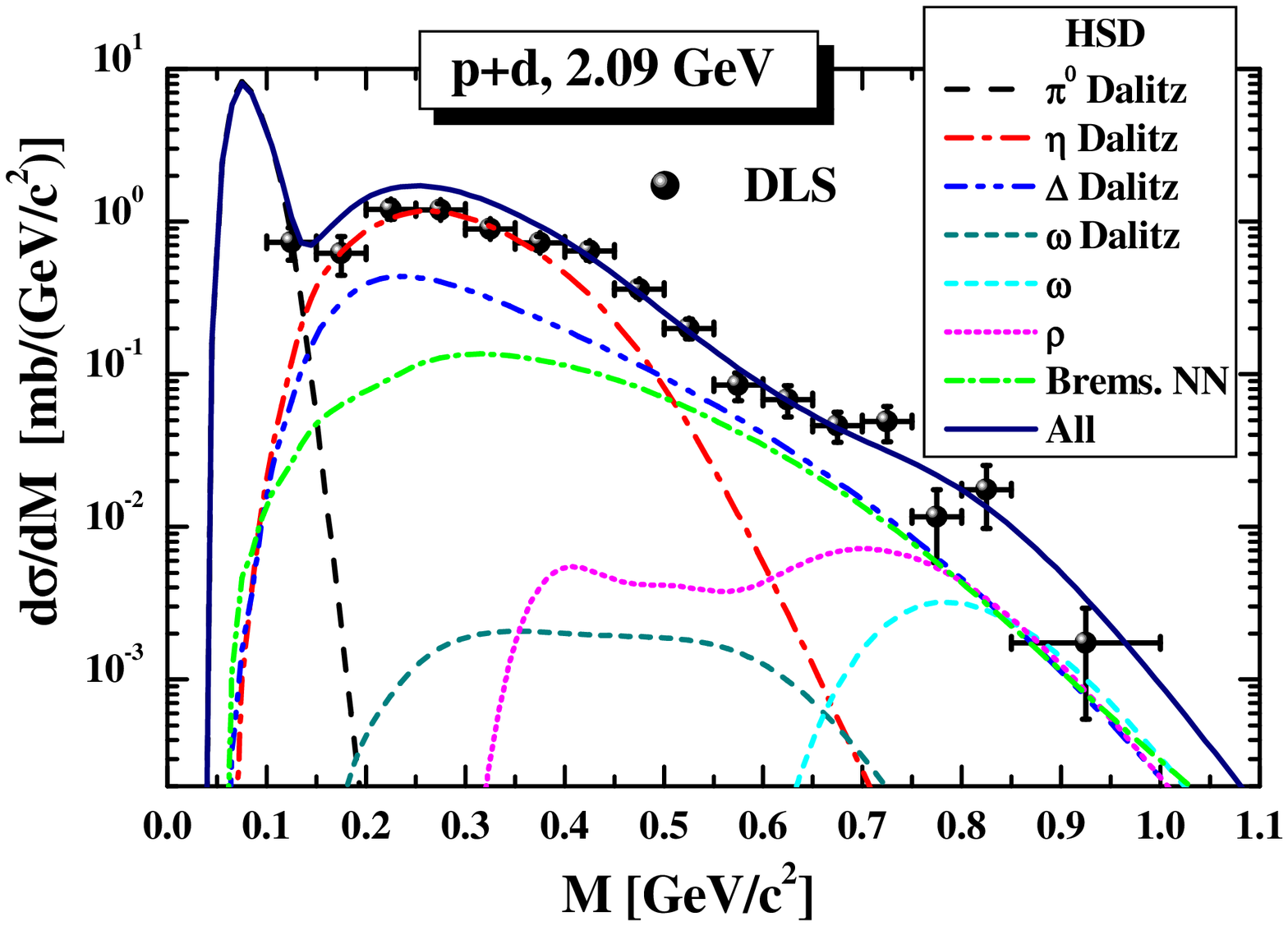}
\caption{(Color online) The dilepton differential cross section $d\sigma/dM$ for the
$pp$ (left plot (a)) and $pd$ (right plot (b)) at 2.09 GeV in comparison to the
DLS data \cite{Wilson:1997sr}.  The theoretical calculations passed
through the corresponding acceptance filters and mass resolutions. }
\label{Fig_pp2DLD}
\end{figure}

Fig. \ref{Fig_pp2DLD} shows the dilepton differential cross section
$d\sigma/dM$ for the $pp$ (left plot) and $pd$ (right plot) at 2.09 GeV
from HSD calculations in comparison to the DLS data
\cite{Wilson:1997sr}. We see also here a good agreement and the fact
that the DLS as well as the HADES data are reproduced with the same
theory underlines the consistency of both data sets which have quite
different acceptance cuts.

\subsection{Dilepton production in pp  at 3.5 GeV}

\begin{figure}[t]
\phantom{a}\vspace*{5mm}
\includegraphics[width=7.5cm]{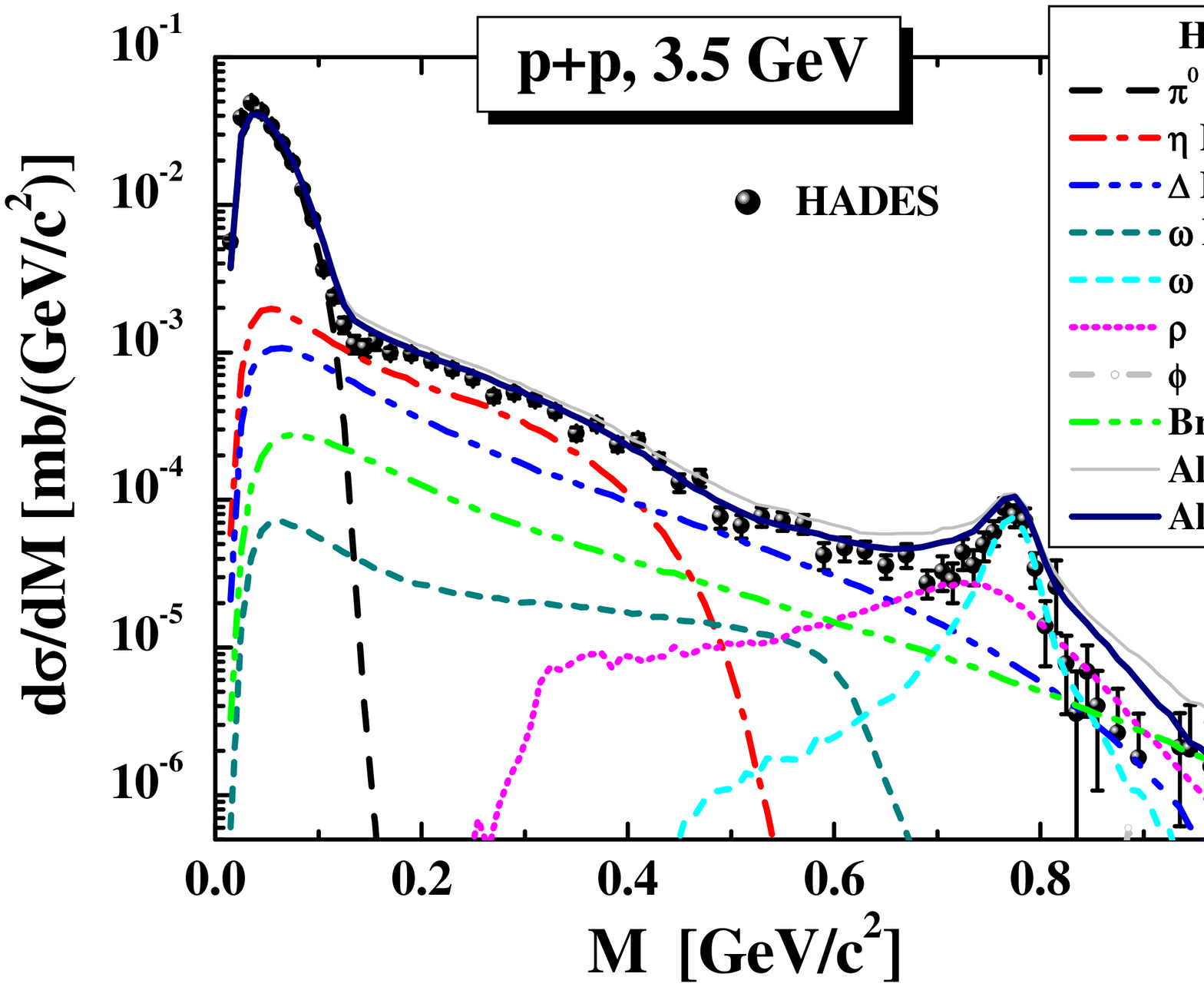}
\caption{(Color online) The differential  cross section $d\sigma/dM$ from HSD calculations for
$e^+e^-$ production in $pp$ reactions at a bombarding energy of 3.5 GeV
in comparison to the HADES data \cite{HADES_pp35}.  The individual
colored lines display the contributions from the various channels in
the HSD calculations (see color coding in the legend). The thick line,
labeled as "All wo Brems", shows the total sum of all
channels (labeled as "All") without $pp$ Bremsstrahlung.  The
theoretical calculations passed through the corresponding HADES
acceptance filters and mass/momentum resolutions.
 }
\label{Fig_M35}
\end{figure}

Finally we come to the HADES pp data at 3.5 GeV. Although HADES
has not measured heavy ion collisions at this energy we include
these results for completeness. Fig. \ref{Fig_M35}
shows the differential  cross section $d\sigma/dM$ from HSD
calculations for dilepton  production in $pp$ reactions at a bombarding
energy of 3.5 GeV in comparison to the HADES data \cite{HADES_pp35}.
We present the results including and excluding the bremsstrahlung
contribution because at this energy there exist no solid bremsstrahlung
calculations. The validity of our approach, to take the Kaptari and Kaempfer matrix
element and to adjust only the phase space, as described in detail in
Ref. \cite{Bratkovskaya:2007jk}, becomes questionable at such a high energy.
The thick lines, labelled in the legend as "All wo Brems", show the sum of
all channels (labelled as "All") without $pp$ bremsstrahlung. For the
distribution of the invariant masses of the dileptons,  bremsstrahlung
does not play a major role at this energy in $pp$, as expected.

\begin{figure}[h]
\phantom{a}\hspace*{-30mm}
\includegraphics[width=9.5cm]{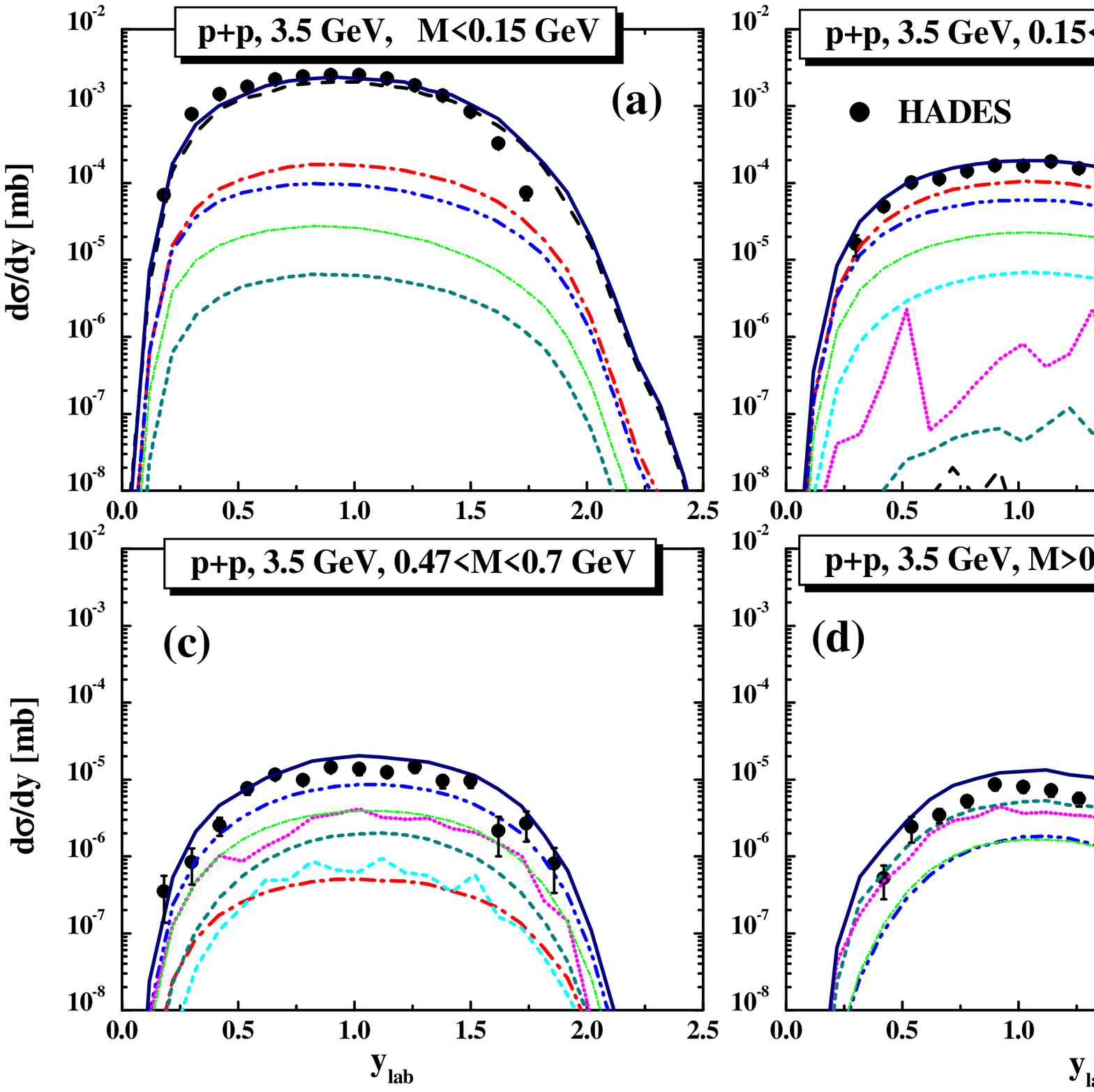}\\[3mm]
\hspace*{-27mm}\includegraphics[width=9.5cm]{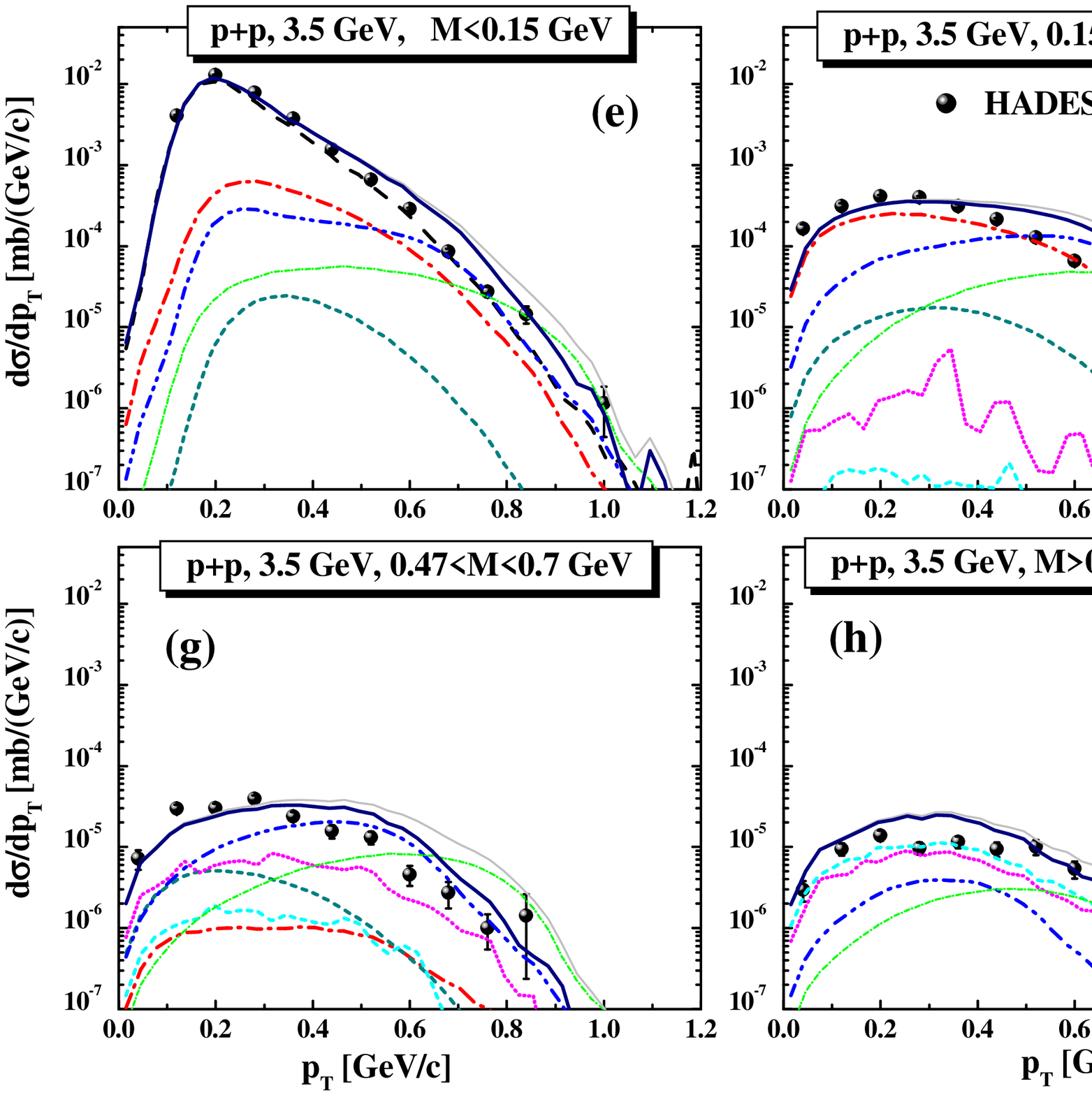}
\caption{(Color online) The HSD results for the rapidity distribution (upper 4 plot (a-d))
and the transverse momentum spectra (lower 4 plots (c-h)) for $pp$ at 3.5 GeV and for 4
different mass bins:  $M \leq$ 0.15 GeV,  0.15 $\leq M \leq$ 0.47 GeV,
0.47 $\leq M \leq$ 0.7 GeV and  $M \geq$ 0.7 GeV in comparison to the
 HADES data \cite{HADES_pp35}.  The individual coloured lines display
the contributions from the various channels in the HSD calculations
(see colour coding in the legend).  The tick lines, labelled in the legend as
"All wo Brems", show the sum of all channels (labelled as "All")
without $pp$ Bremsstrahlung.  The theoretical calculations passed
through the corresponding HADES acceptance filters and mass/momentum
resolutions.
}
\label{Fig_y35}
\end{figure}

In Fig. \ref{Fig_y35}  we compare the HSD results for $pp$ at 3.5 GeV
and for 4 different mass bins:  $M \leq$ 0.15 GeV,  0.15 $\leq M \leq$
0.47 GeV, 0.47 $\leq M \leq$ 0.7 GeV and $M \geq$ 0.7 GeV to the HADES
data \cite{HADES_pp35}.  The upper 4 plots show the rapidity
distribution and the lower 4 plots the transverse momentum spectra.  As
in Fig. \ref{Fig_M35} the thick lines, labelled in the legend as "All
wo Brems", show the sum of all channels (labelled as "All") without
$pp$ bremsstrahlung.  We observe that the rapidity distribution is well
described  except  for invariant masses around the $\rho$ peak where we
overpredict the data by a constant factor. Also the transverse
momentum distribution is well described by theory with the exception of a region
around $M\approx 0.6$ GeV where our calculations overpredict the data.

We note that the present result is in a better agreement with the
HADES $p_T$ data as compared to the early HSD predictions
\cite{Bratkovskaya:2007jk,HADES_pp35} due to the following
reasons: a lowering of the $\eta$ Dalitz dilepton contribution due
to the reduction of the $\eta$ production cross section in line
with the new HADES data (cf. discussion in Section II.A.1(1)); a
lowering of the direct $\rho,\omega$ dilepton decay contributions
due to the modification of the vector meson production cross
section (cf. discussion in Section II.A.1(2)); a lowering of the
$\Delta$ Dalitz dilepton contribution due to the adjustment of the
isospin decomposition in the exclusive channel $NN\to
\Delta^{++}n$ from FRITIOF (cf. discussion in Section II.A.1 (3)).
The latter reduces the total (inclusive) $\Delta^+$ production by
a factor up to 1.4 at 3.5 GeV and correspondingly the dilepton
yield. This reduction is even larger (more then a factor of 3) for
dileptons with high invariant masses and high $p_T$ since they
stem dominantly from the Dalitz decay of exclusive $\Delta$'s
simply due to kinematical reasons - a lower amount of associated
particles leaves more energy for the generation of high mass
$\Delta$'s. An addition reduction of the $\Delta$ dilepton yield
stems from the different parametrizations used  for the
differential electromagnetic decay width of the $\Delta$ resonance
(cf. discussion in Section VI): presently - "Wolf" \cite{Wolf90}
instead of the original "Ernst" description \cite{Ernst:1997yy}
with a coupling constant $g=3$ instead of $g=2.7$ which is
consistent with the 'photon' ($M\to 0$) limit. Without these
modification the present HSD version reproduces the results of
\cite{Bratkovskaya:2007jk,HADES_pp35}.

We speculate that HSD produces slightly too many $\Delta$ at 3.5
GeV. Since the elementary cross section for inclusive $\Delta$
production in $pp$ reactions at this energy is not available, the
repartition of the pion yield between $\Delta$ resonances (which
produce dileptons) and other resonances (which do not produce
dileptons) is not well known and may be the origin of the
deviation obtained in the $p_T$ spectra. For the mass bin $
0.47<M<0.7$ GeV we see that above $p_T=0.7$ GeV/$c$ bremsstrahlung
is the dominating source of dilepton production. We plot the sum
of all contributions without bremsstrahlung as well.

\subsection{Dilepton production in pA collisions at 3.5 GeV}

\begin{figure}[h]
\phantom{a}\vspace*{5mm}
\includegraphics[width=7.5cm]{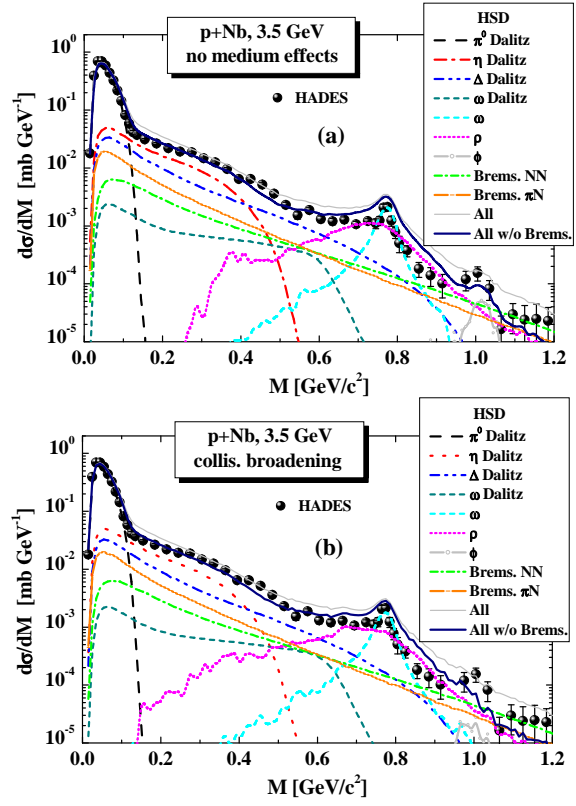}
\caption{(Color online) The differential  cross section $d\sigma/dM$ from HSD calculations for
 $e^+e^-$ production in the $p+Nb$ reaction at a bombarding energy of 3.5 $A$GeV
 in comparison to the HADES data \cite{HADES_pNb35}.  The upper
 part (a) shows the case of 'free' vector-meson spectral functions while
the lower part (b)  gives the result for the 'collisional broadening'
scenario.  The individual coloured lines display the contributions from
the various channels in the HSD calculations (see colour coding in the
legend).  The tick lines, labelled in legend as "All wo Brems", show
the sum of all channels (labelled as "All") without $pp$
bremsstrahlung.  The theoretical calculations passed through the
corresponding HADES acceptance filters and mass/momentum resolutions.
 }
\label{Fig_MpNb35}
\end{figure}

We are coming now to $p+A$ reactions. Fig. \ref{Fig_MpNb35} compares the
differential  cross section $d\sigma/dM$ from HSD calculations for $e^+e^-$
production in $p+Nb$ reaction at a bombarding energy of 3.5 GeV to
the HADES data \cite{HADES_pNb35}.  The upper part shows the case of the
'free' vector-meson spectral functions while the lower part gives the
result for the 'collisional broadening' scenario.  Again the thick
lines, labeled in the legend as "All wo Brems", show the sum of all
channels (labeled as "All") without $NN$ bremsstrahlung. We display both cases
since the treatment of bremsstrahlung using the extrapolation of the OBE
model to such high energy is questionable, as discussed above. For
the same reason the $\pi N$ bremsstrahlung presented Fig. in \ref{Fig_MpNb35}
has to be considered with care.
The collisional broadening scenario comes closer to the experimental
results in the region around the $\rho$ peak.
We thus find a nice agreement between theory and experiment also for proton-nucleus collisions.

\section{Dilepton production in heavy-ion collisions}

\subsection {Dileptons from the HSD and IQMD  models}

\begin{figure}[!h]
\centerline{\psfig{figure=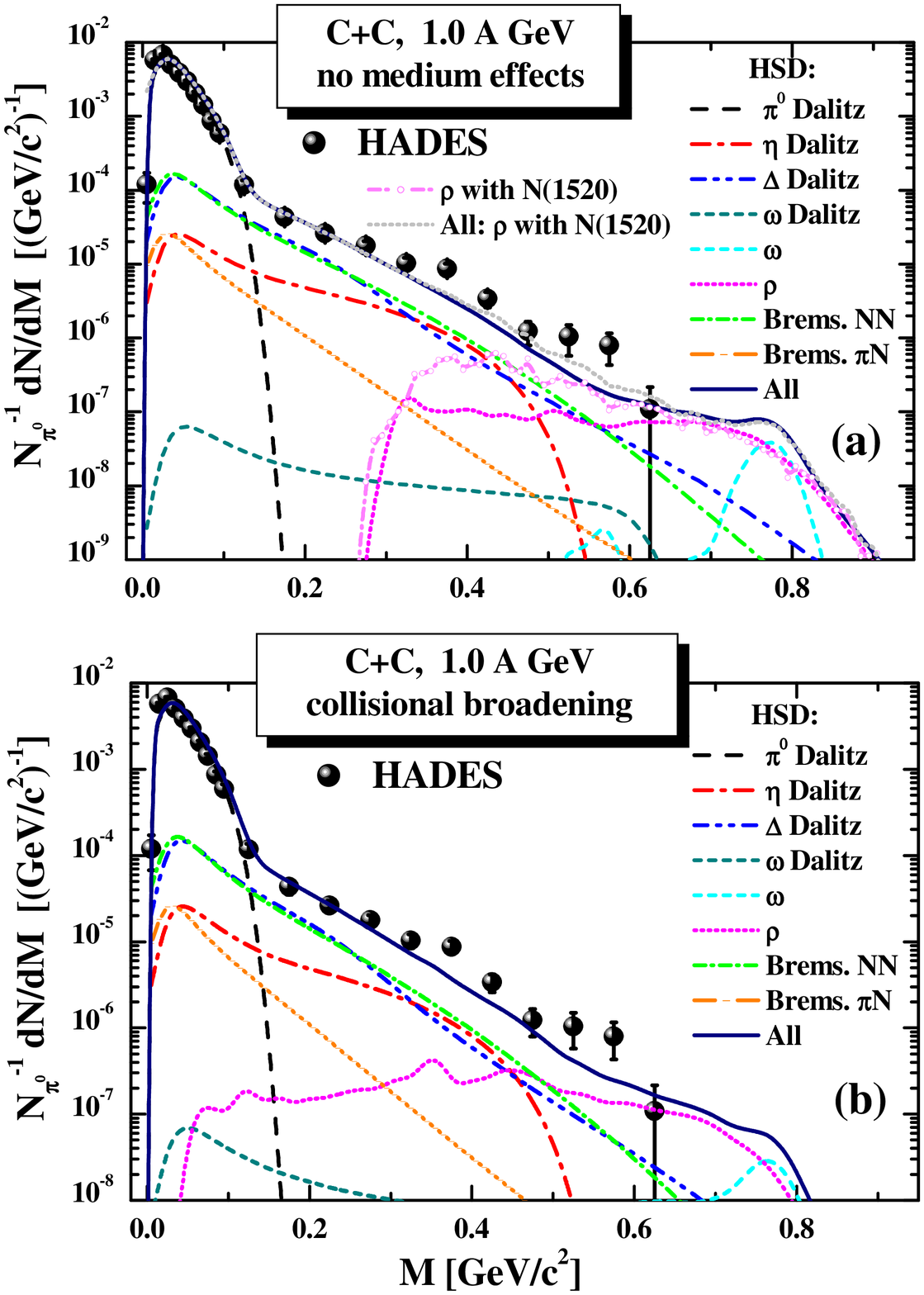,width=9.cm}}
\caption{(Color online) The results of the HSD transport calculation  for the mass
differential dilepton spectra - normalized to the
$\pi^0$ multiplicity  - for C+C at 1.0 $A$GeV in
comparison to the HADES data \cite{Agakishiev:2007ts}. The upper part (a) shows the case
of 'free' vector-meson spectral functions while the lower part (b) gives
the result for the 'collisional broadening'
scenario. In both scenarios the HADES acceptance filter and
mass/momentum resolution have been incorporated. The different color lines
display individual channels in the transport calculation (see
legend). } \label{Fig_CC10}
\phantom{a}\vspace*{5mm}
\centerline{\psfig{figure=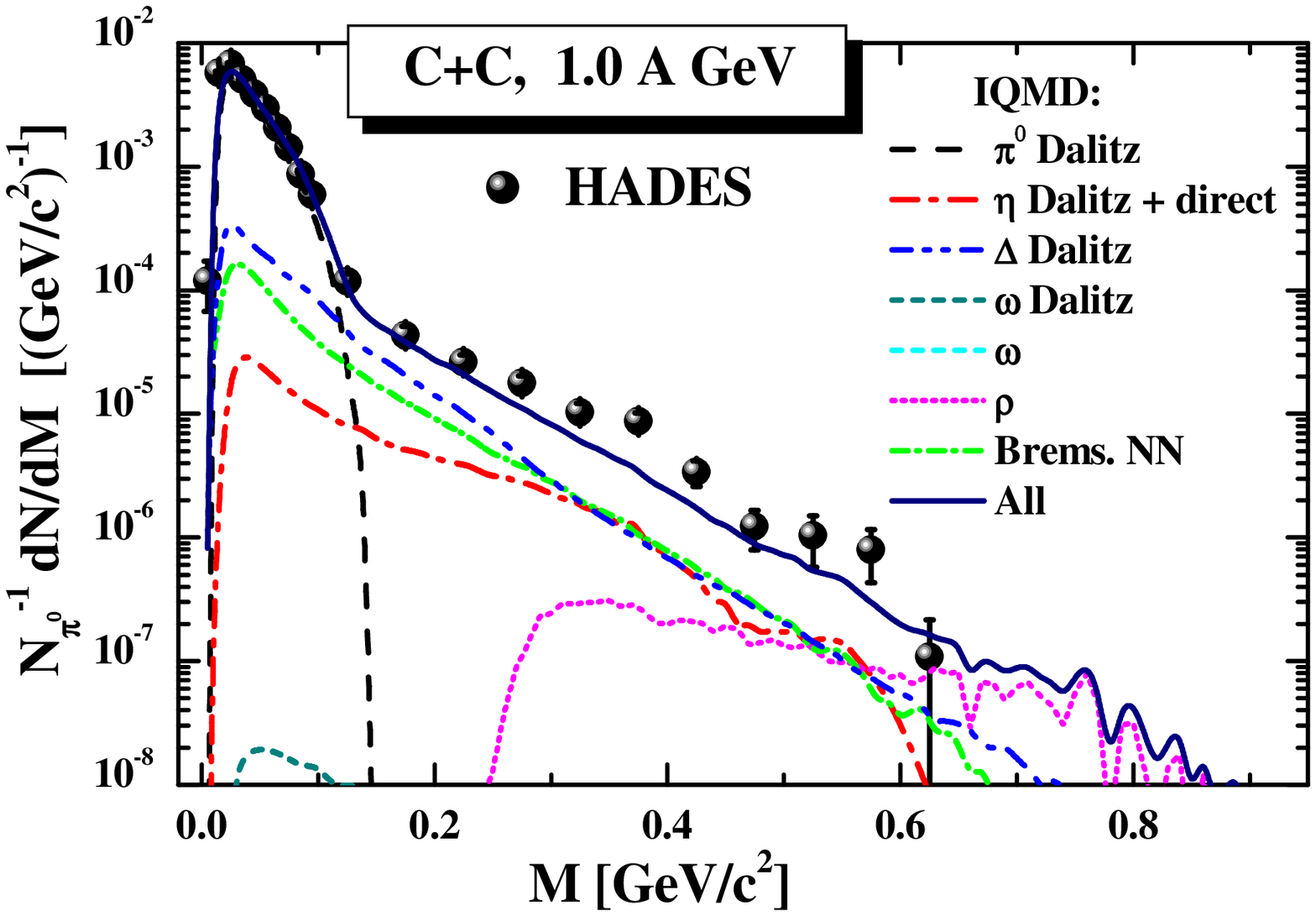,width=9.cm}}
\caption{(Color online) The mass differential dilepton spectra - normalized to the  $\pi^0$
multiplicity - from IQMD calculations for C+C - at 1 $A$GeV in
comparison to the HADES data \cite{Agakishiev:2007ts}.
The different colour lines display individual channels in the transport
calculation (see legend).  The theoretical calculations passed through
the corresponding HADES acceptance filter and mass/momentum
resolutions.
} \label{Fig_IQCC10}
\end{figure}

Now we come to the heavy-ion results and start with showing in Fig.
\ref{Fig_CC10} the mass differential dilepton spectra - normalized
to the $\pi^0$ multiplicity  - of HSD calculations for C+C at 1.0 $A$GeV
in comparison to the HADES data \cite{Agakishiev:2007ts}.  The HADES
collaboration has obtained the $\pi^0$ multiplicity by the average of the
multiplicity of charged  pions \cite{Agakishiev:2009zv} and we apply the same method for the
theoretical calculations. The upper part
displays the results for  'free' vector-meson spectral functions while
the lower part shows the result for the 'collisional broadening'
scenario.  We note here, and this holds for all dilepton spectra
normalized to the number of $\pi^0$'s,  that the normalization is done
by the total number of $\pi^0$'s in $4\pi$, i.e. without applying an
experimental acceptance. This allows for a direct comparison with the
published HADES results.

The $\Delta$ Dalitz decay and bremsstrahlung contributions are
the dominant channels and contribute with about the same weight to the
invariant mass spectra. For invariant masses $M>0.3$~GeV also the
subthreshold $\eta$ channel contributes in an important way. The
different descriptions of the $\rho$ meson become important only at
large invariant masses where no experimental data are available. The figure shows as well
the contribution from direct $\rho$ decays when including the $N^*(1520)$
resonance which may enhance the $\rho$ meson production at sub-threshold
energies as discussed in Section II.B.
As seen in the figure, there is indeed a small contribution but not larger
than the experimental error bars. At higher energies other channels dominate. Therefore the
$N^*(1520)$ resonance is not an important source for dilepton production
in heavy-ion reactions.
Also the 'in-medium' effects due to the collisional broadening of
the spectral functions for  $\rho$ and $\omega$ mesons  is not visible in the final
spectra due to the strong contributions from other dilepton sources
at low invariant masses where this effect is most pronounced and partly
due to the limited experimental mass resolution at high invariant masses
which smears out the spectra.

Fig. \ref{Fig_IQCC10} shows the results of IQMD calculations,
including acceptance in the same way as the HSD calculations. It is
remarkable that the two quite sophisticated transport theories predict
results which are that similar.  Even the channel decomposition is very
similar what is all but trivial because the invariant mass spectra depend on
many details of the reaction. They include the $\Delta$ dynamics in a nucleus,
which we will discuss in section V in more detail, the number of collisions and hence of the
spatial distributions of the nucleons in the colliding nuclei,
the Fermi momentum and the Pauli blocking
of reactions if final state nucleons would be placed in already
occupied phase space regions.

\begin{figure}[h!]
\includegraphics[width=9.cm]{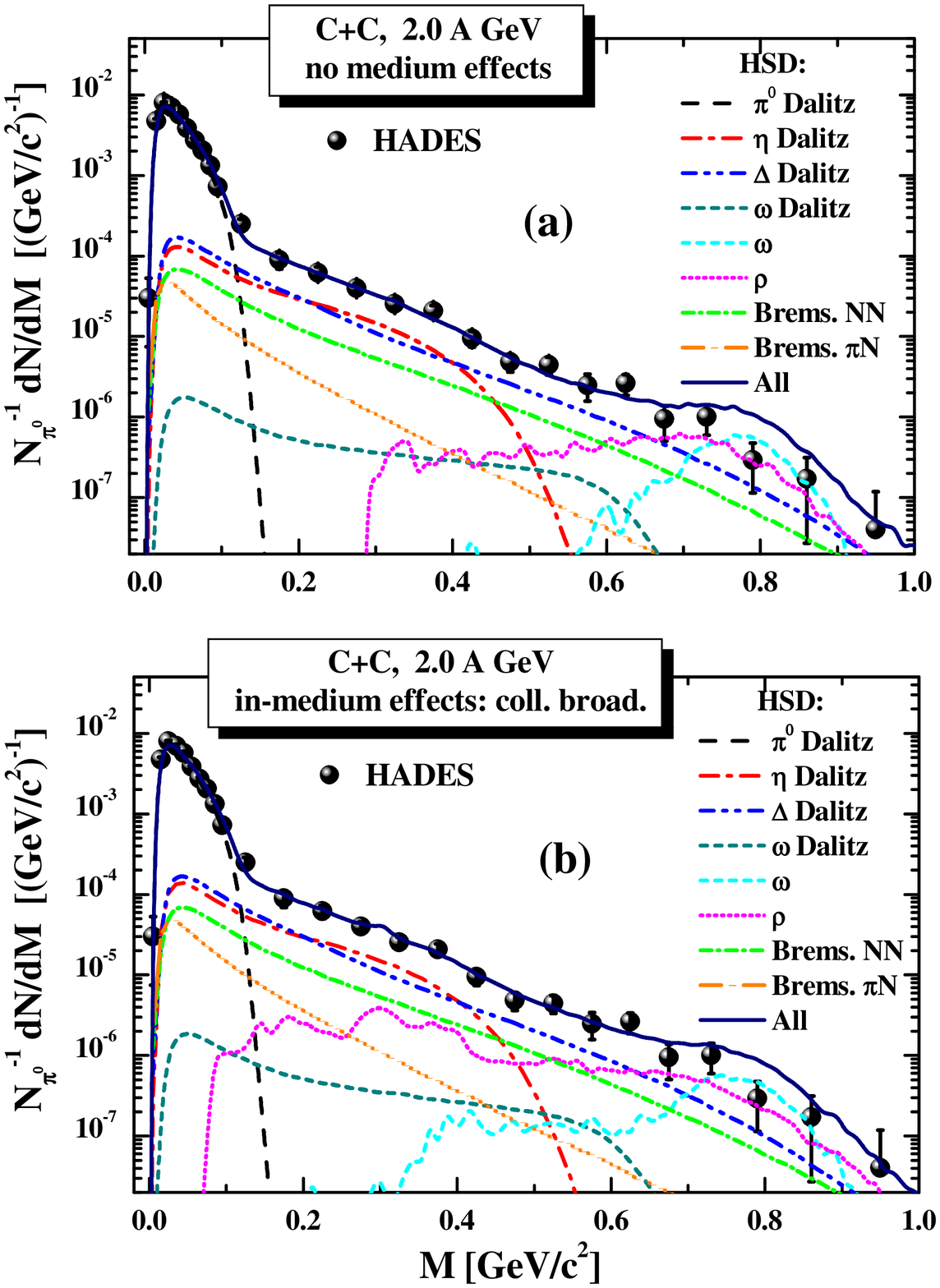}
\caption{(Color online) The mass differential dilepton spectra - normalized to the $\pi^0$ multiplicity
 - from HSD calculations for C+C  at 2 $A$GeV in
comparison to the HADES data \cite{Agakishiev:2009yf}.  The upper
part (a) shows the case of 'free' vector-meson spectral functions while the
lower part (b) gives the result for the 'collisional broadening' scenario.
The different colour lines display individual channels in the transport
calculation (see legend).  The theoretical calculations passed through
the corresponding HADES acceptance filter and mass/momentum
resolutions.
} \label{Fig_CC20}
\phantom{a}\vspace*{5mm}
\includegraphics[width=9.cm]{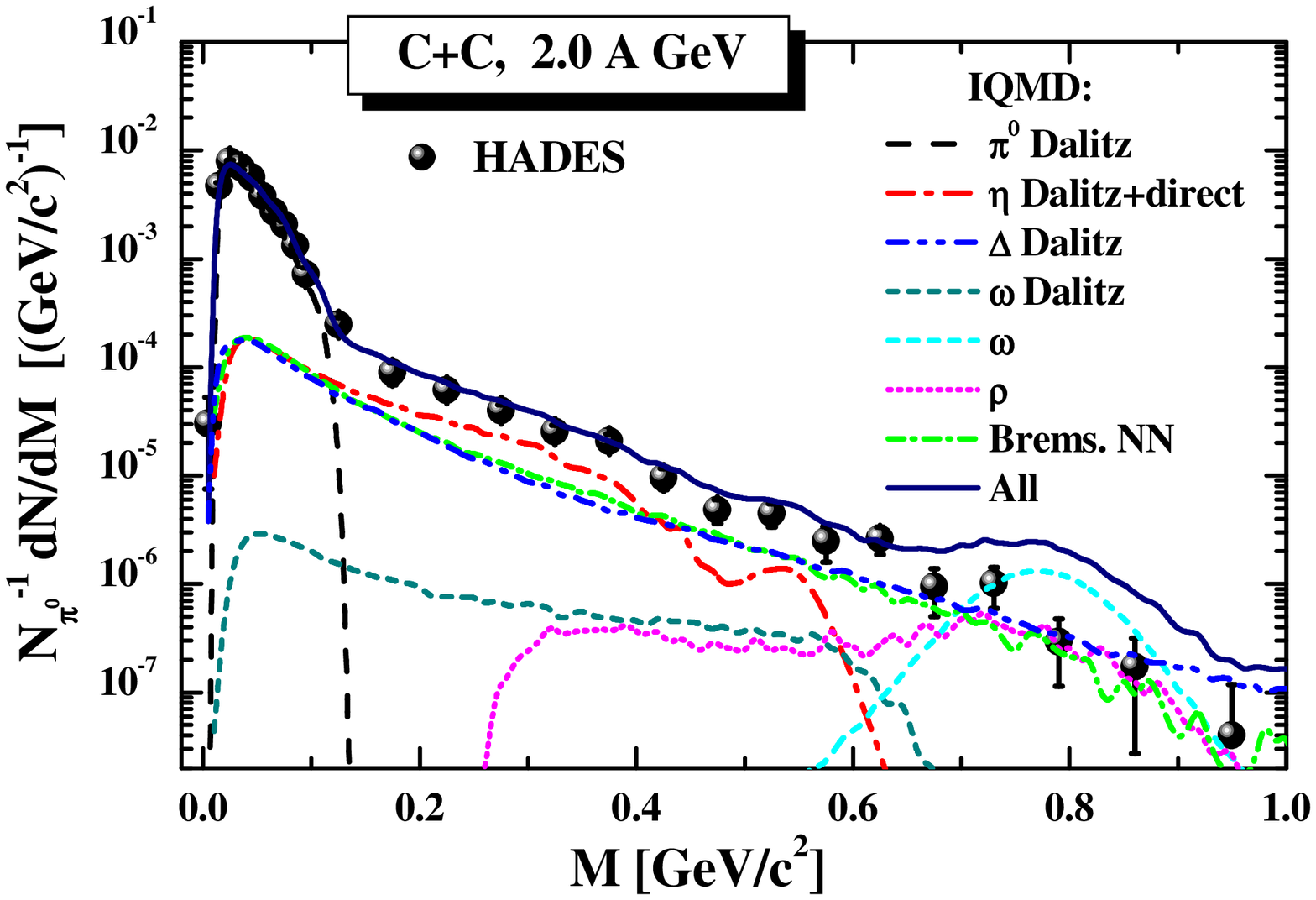}
\caption{(Color online) The mass differential dilepton spectra - normalized to the
$\pi^0$ multiplicity - from IQMD for C+C at 2 $A$GeV in
comparison to the HADES data \cite{Agakishiev:2009yf}.
The different colour lines display individual channels in the transport
calculation (see legend).  The theoretical calculations passed through
the corresponding HADES acceptance filter and mass/momentum
resolutions.
} \label{Fig_IQCC20}
\end{figure}

Fig. \ref{Fig_CC20} shows the mass differential dilepton spectra -
normalized to the $\pi^0$ multiplicity - from HSD calculations for C+C
- at 2 $A$GeV in comparison to the HADES data \cite{Agakishiev:2009yf}.
 The theoretical calculations passed through
the corresponding HADES acceptance filters and mass/momentum
resolutions  which leads to a smearing of the spectra at high invariant mass
and particularly in the $\omega$ peak region.
The upper part shows again the case of 'free' vector-meson spectral
functions while the lower part presents the result for the 'collisional
broadening' scenario. Also here the difference between the
in-medium scenarios is of minor importance, partly due to the limited mass resolution
which smears out the spectra. Nevertheless,
one can conclude that the 'free' calculations predict an enhancement
in the region of the $\rho$ mass which is not seen in the
experimental data, which are more in favor to the collisional
broadening scenario.

Fig. \ref{Fig_IQCC20} compares the same data with the results from  IQMD calculations for
C+C - at 2 $A$GeV which have been acceptance corrected in the same way
as the HSD data. Again we see a very good agreement between the two
theoretical approaches. Only the different parametrizations of the
$\omega$ cross section yield deviations at invariant masses around 0.77 GeV.

\begin{figure}[h]
\phantom{a}\vspace*{5mm}
\includegraphics[width=9.cm]{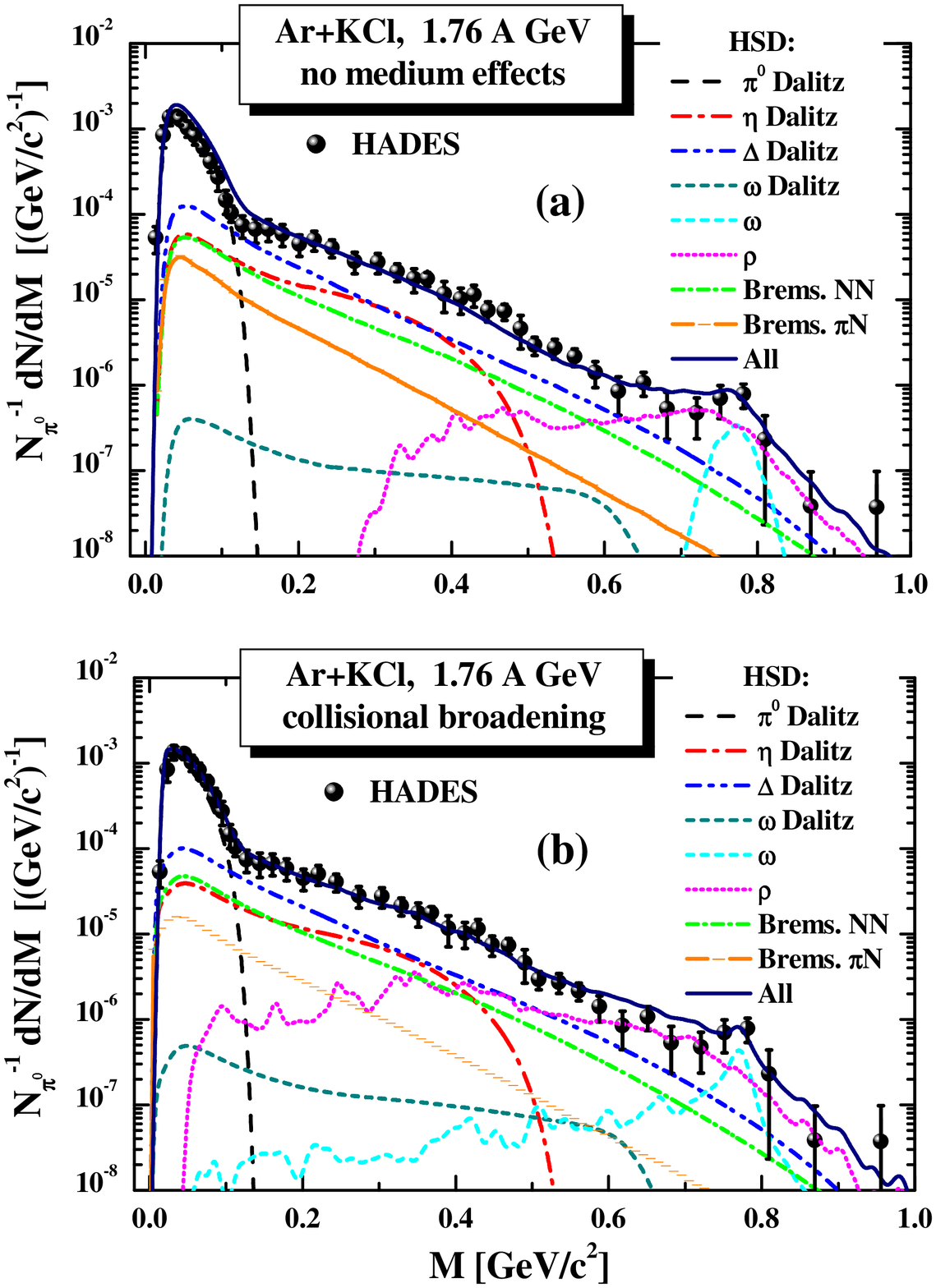}
\caption{(Color online) The mass differential dilepton spectra - normalized to the
$\pi^0$ multiplicity - from HSD for Ar+KCl at 1.76 $A$GeV  in
comparison to the HADES data \cite{Agakishiev:2011vf}.  The upper
part(a) shows the case of 'free' vector-meson spectral functions while the
lower part (b) gives the result for the 'collisional broadening' scenario.
The individual colored lines display the contributions from the various
channels in the HSD calculations (see color coding in the legend).  The
theoretical calculations passed through the corresponding HADES
acceptance filter and mass/momentum resolutions.
}
\label{Fig_MArKCl}
\phantom{a}\vspace*{5mm}
\includegraphics[width=9.cm]{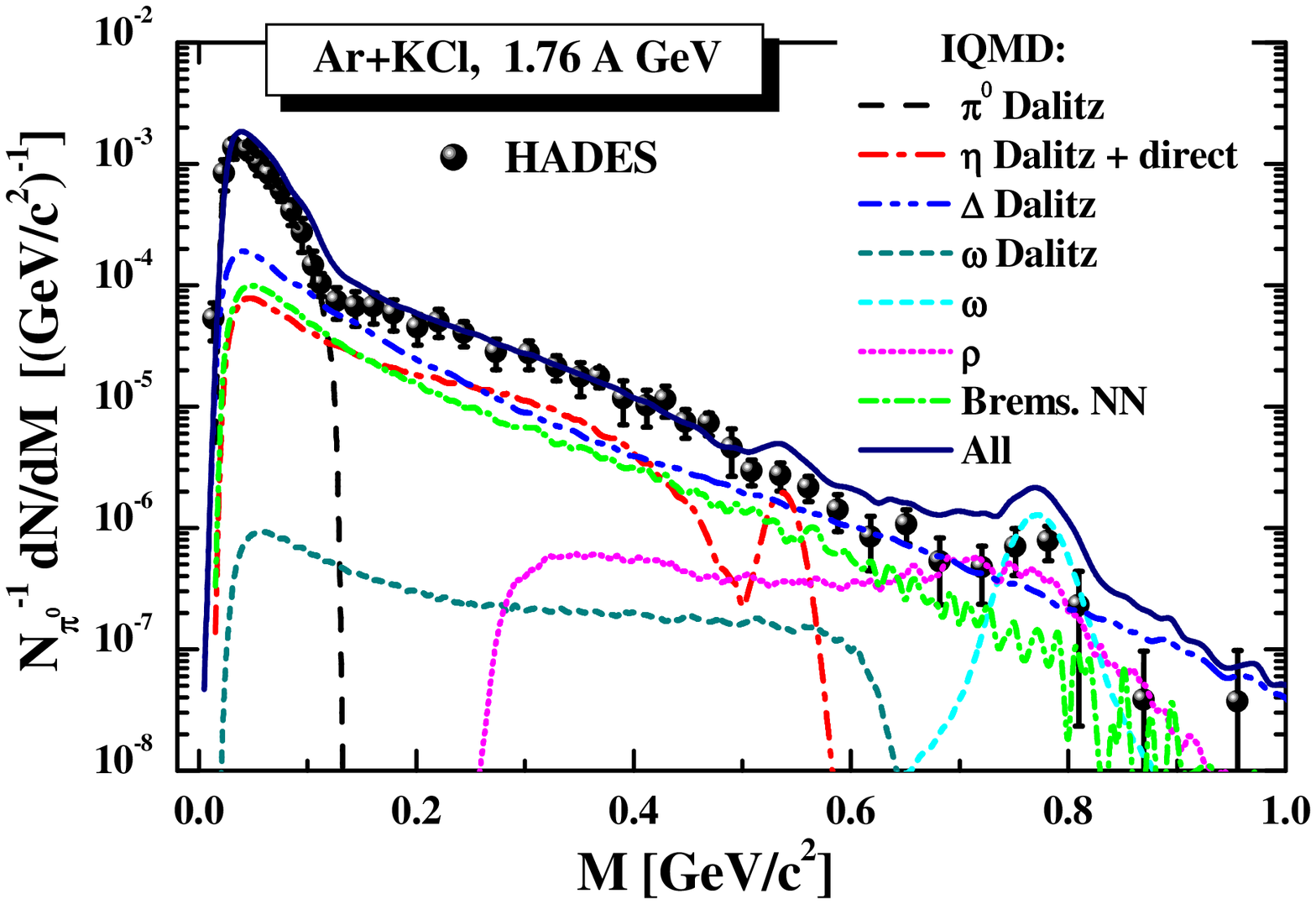}
\caption{(Color online) The mass differential dilepton spectra - normalized to the
$\pi^0$ multiplicity - from IQMD for Ar+KCl at 1.76 $A$GeV  in
comparison to the HADES data \cite{Agakishiev:2011vf}.
The individual colored lines display the contributions from the various
channels in the IQMD calculations (see color coding in the legend).  The
theoretical calculations passed through the corresponding HADES
acceptance filter and mass/momentum resolutions.
}
\label{Fig_IQArKCl}
\end{figure}

Fig. \ref{Fig_MArKCl} displays the mass differential dilepton spectra -
normalized to the $\pi^0$multiplicity - from HSD calculations for
Ar+KCl at 1.76 $A$GeV  in comparison to the HADES data
\cite{Agakishiev:2011vf}.  The upper part shows again the case of
'free' vector-meson spectral functions while the lower part gives the
result for the 'collisional broadening' scenario. Also in this data set
the enhancement around the $\rho$ mass is clearly visible. For this
heavier system the 'collisional broadening' scenario shows  a slightly
better agreement with experiment than the 'free' result and we expect
that for larger systems the difference between the two approaches increases.

Fig. \ref{Fig_IQArKCl}, which presents the IQMD results for this reaction,
shows that the agreement between both theories
continues also for heavier systems.  Again up to invariant masses of
0.7 GeV both invariant mass spectra are almost identical and agree with
data.  Also the channel decomposition is rather similar.
Here one can see again the overestimation of the dilepton yield
by IQMD at $\rho/omega$ peak which is related to the enhance $\omega$
production cross section in elementary $pn$ collisions relative to $pp$
collisions due to the isospin model used in IQMD (cf. Section II.C).

\begin{figure}[h!]
\includegraphics[width=9.cm]{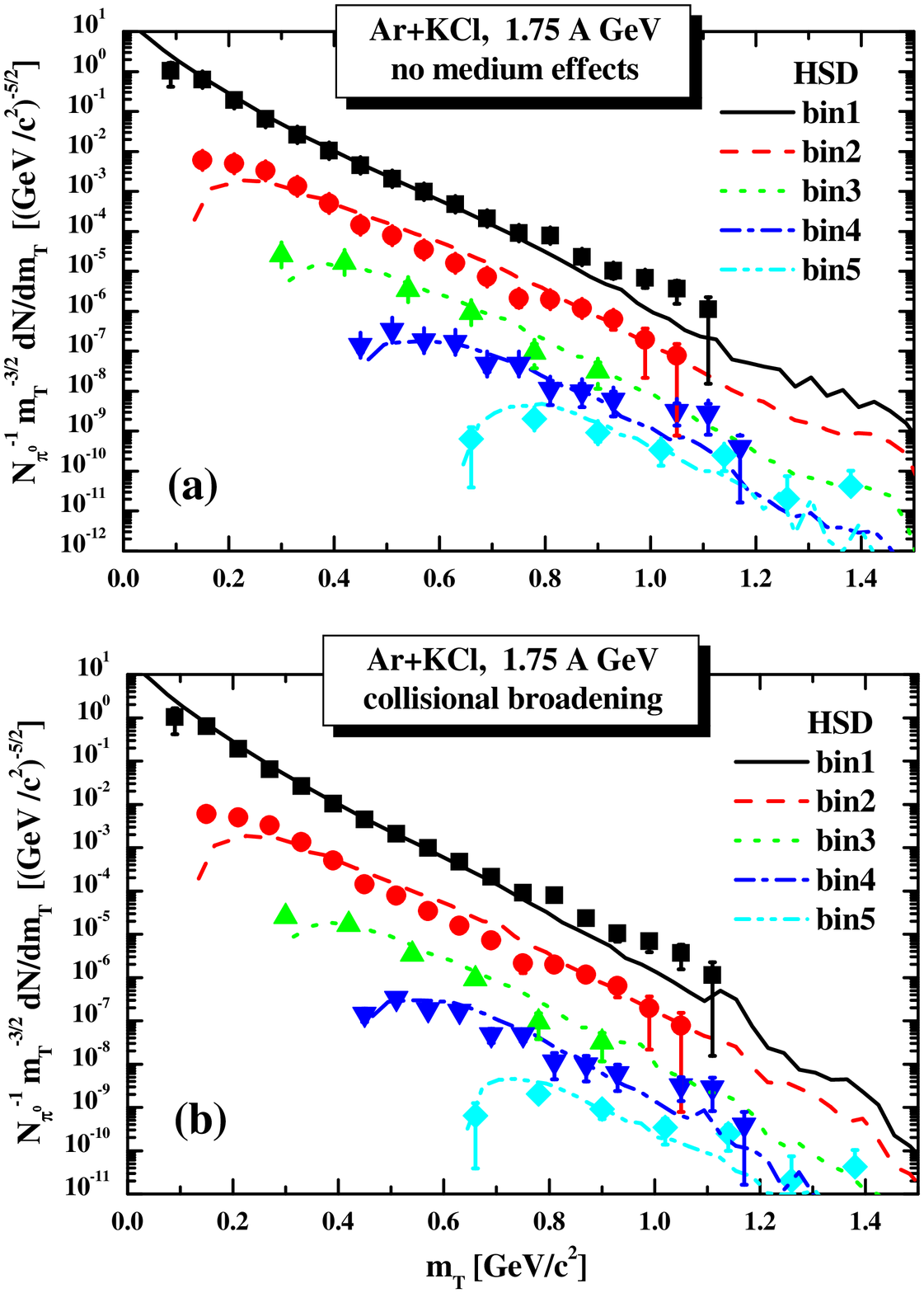}
\caption{(Color online) The HSD results for the  transverse momentum spectra
- normalized to the $\pi^0$ multiplicity - for Ar+KCl at 1.75 $A$GeV
for 5 different mass bins:
bin 1: $M \leq$ 0.15 GeV,
bin 2: 0.13 $\leq M \leq$ 0.3 GeV,
bin 3: 0.3 $\leq M \leq$ 0.45 GeV
bin 4: 0.45 $\leq M \leq$ 0.65 GeV
and  bin 5: $M \geq$ 0.65 GeV
in comparison to the HADES data \cite{Agakishiev:2011vf}.  The upper
part (a) shows the case of 'free' vector-meson spectral functions while the
lower part (b) gives the result for the 'collisional broadening' scenario.
The individual coloured lines display the contributions from the various
channels in the HSD calculations (see colour coding in the legend).  The
theoretical calculations passed through the corresponding HADES
acceptance filter and mass/momentum resolutions.  }
\label{Fig_mtArKCl} \end{figure}

The transverse momentum spectra - normalized to the $\pi^0$ multiplicity -
for Ar+KCl at 1.75 $A$GeV have been measured by the HADES collaboration
for 5 different mass bins \cite{Agakishiev:2011vf} :
bin 1: $M \leq$ 0.15 GeV,
bin 2: 0.13 $\leq M \leq$ 0.3 GeV,
bin 3: 0.3 $\leq M \leq$ 0.45 GeV
bin 4: 0.45 $\leq M \leq$ 0.65 GeV
and  bin 5: $M \geq$ 0.65 GeV.
Fig. \ref{Fig_mtArKCl} presents the HADES data in comparison with HSD
calculations; on the top without medium effect, on the bottom for the
dropping mass scenario. We see also here a good agreement between theory and
experiment. Thus one can conclude that the agreement between theory and
experiment ( Fig. \ref{Fig_CC10}  - Fig. \ref{Fig_mtArKCl}) up to $ M\approx$
0.5 GeV is of such a quality that we can use the theory to study the physical
processes involved.

 \begin{figure}[h!]
\phantom{a}\vspace*{5mm}
\centerline{\psfig{figure=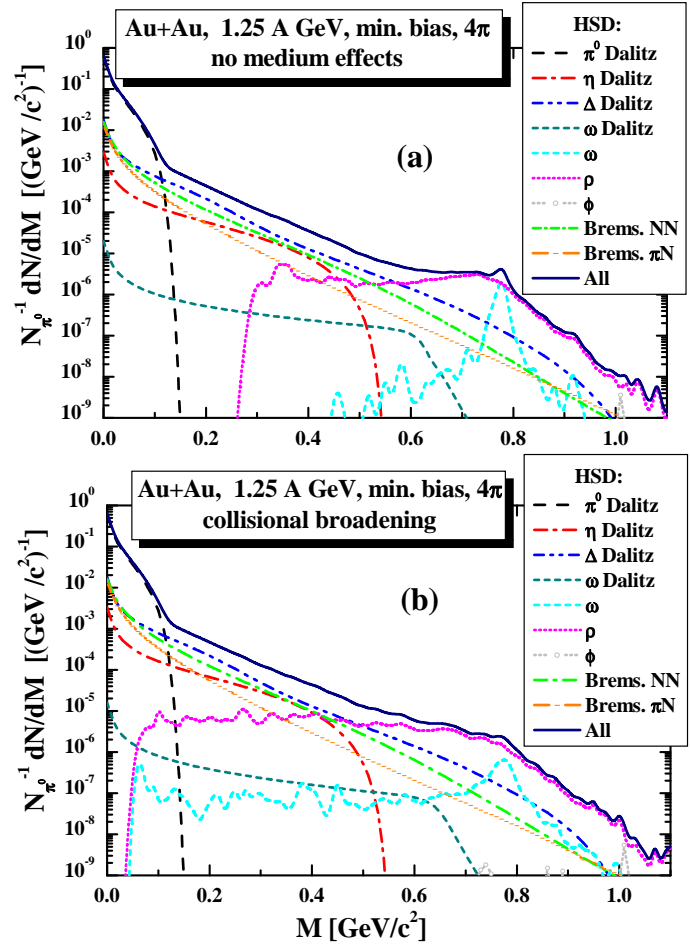,width=9cm}}
\caption{(Color online) The mass differential dilepton spectra - normalized to the
number of $\pi^0$'s - from HSD for minimal bias Au+Au collisions at 1.25
$A$GeV.  The upper part (a) shows the case of 'free' vector-meson
spectral functions while the lower part (b) gives the result for the
'collisional broadening' scenario.  The different color lines display
individual channels in the transport calculation (see legend).
}
\label{Fig_MAu125}
\end{figure}

The HADES collaboration has recently measured also the dilepton invariant mass spectra
for the reaction Au+Au at 1.25 $A$GeV. The analysis is not completed
yet.  Fig. \ref{Fig_MAu125} presents the HSD predictions for the mass
differential dilepton spectra - normalized to the $\pi^0$ multiplicity -
for this reaction.  The upper part shows the case of 'free'
vector-meson spectral functions while the lower part gives the result
for the 'collisional broadening' scenario.

\subsection {Dileptons from the UrQMD model}

In this subsection we present the results from the UrQMD (v. 2.3)
transport model  \cite{UrQMD1,UrQMD2}. In this model the dilepton
afterburner does not contain bremsstrahlung. It is, however,
useful to verify whether it agrees with HSD and IQMD calculations
as far as all hadronic dilepton sources are concerned. For the
details of the dilepton treatment in UrQMD at SIS energies we
refer the reader to Refs. \cite{Schumacher:2006wc,Schmidt:2009}.

\begin{figure}[h!]
\includegraphics[width=9.cm]{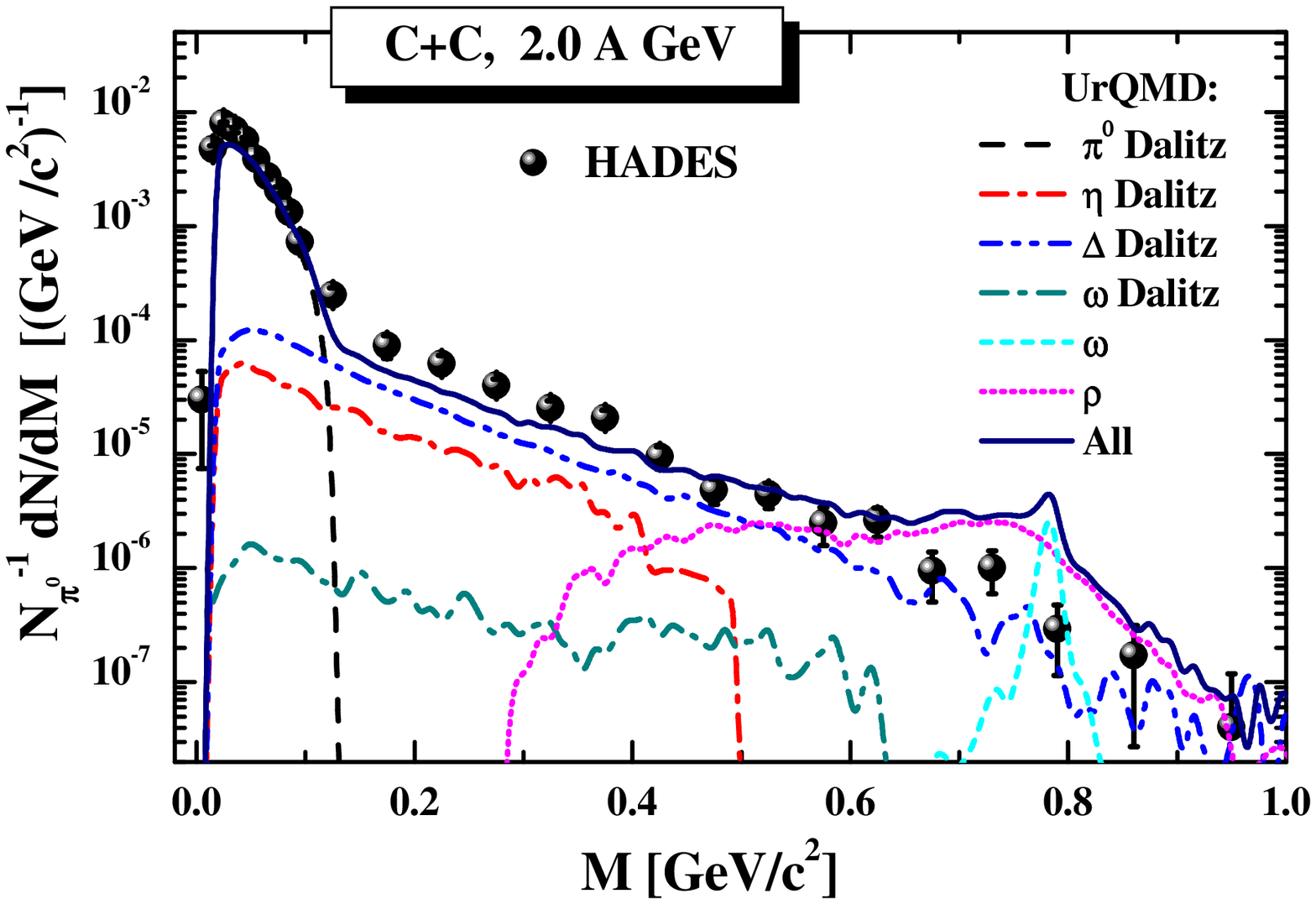}
\caption{(Color online) The mass differential dilepton spectra - normalized to the
number of $\pi^0$'s - from UrQMD for C+C - at 2 $A$GeV in
comparison to the HADES data \cite{Agakishiev:2009yf}.
The different colour lines display individual channels in the transport
calculation (see legend).  The theoretical calculations passed through
the corresponding HADES acceptance filter including mass/momentum
resolutions.
} \label{Fig_UrCC20}
\includegraphics[width=9.cm]{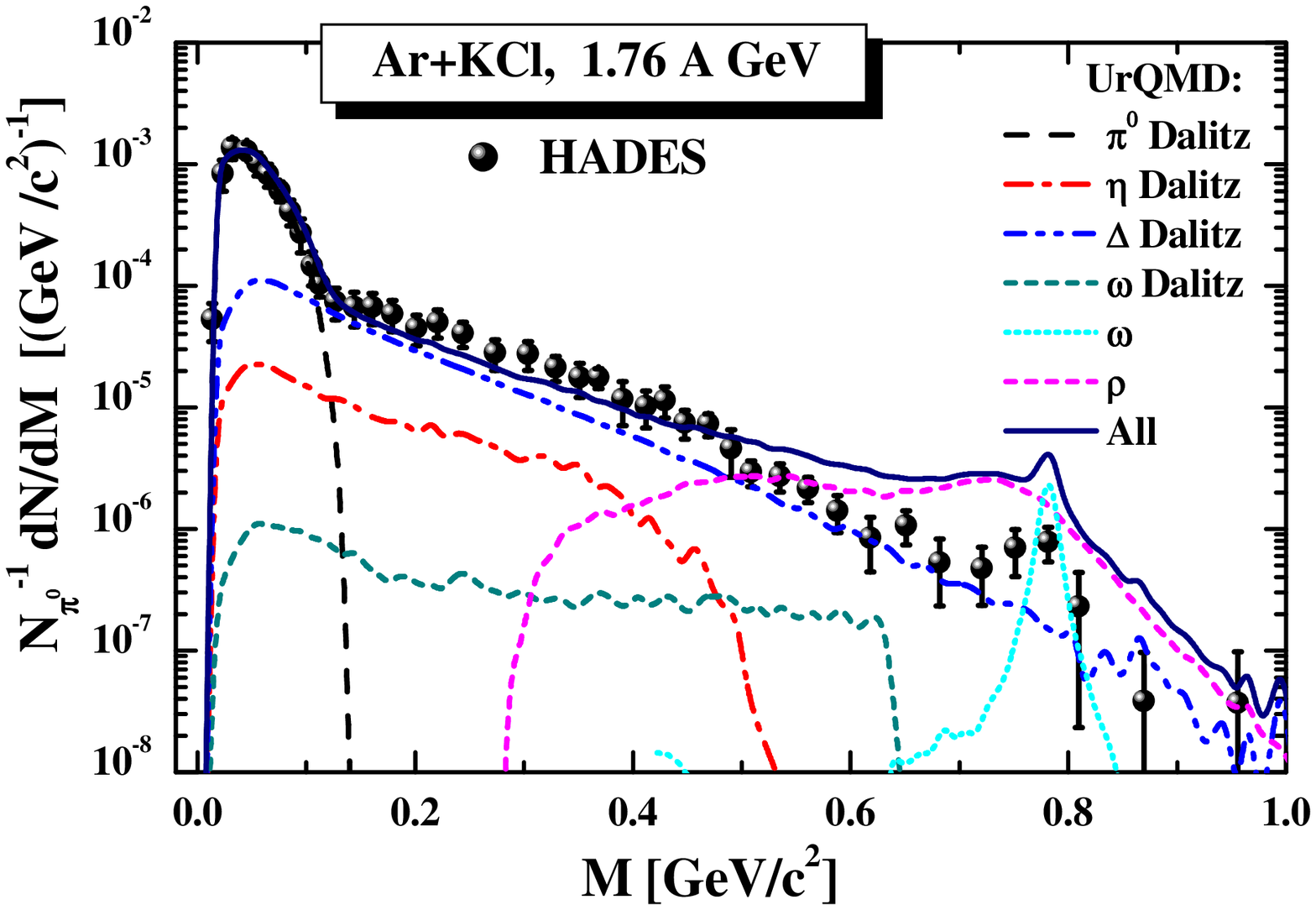}
\caption{(Color online) The mass differential dilepton spectra - normalized to the
number of $\pi^0$'s - from UrQMD for Ar+KCl at 1.76 $A$GeV  in
comparison to the HADES data \cite{Agakishiev:2011vf}.
The individual colored lines display the contributions from the various
channels in the HSD calculations (see color coding in the legend).  The
theoretical calculations passed through the corresponding HADES
acceptance filter including mass/momentum resolutions.
}
\label{Fig_UrArKCl}
\end{figure}

Fig. \ref{Fig_UrCC20} shows the mass differential dilepton spectra -
normalized to the  $\pi^0$ multiplicity - from UrQMD calculations for C+C
- at 2 $A$GeV in comparison to the HADES data \cite{Agakishiev:2009yf}
and  Fig. \ref{Fig_UrArKCl} -- for Ar+KCl at 1.76 $A$GeV  in
comparison to the HADES data \cite{Agakishiev:2011vf}.
As one can see from Figs. \ref{Fig_UrCC20} and \ref{Fig_UrArKCl} the UrQMD
v. 2.3 substantially overestimates the dilepton yield from the vector
mesons. The problem can be traced back to the description of $\rho$
production in elementary $NN$ collisions which proceeds via an excitation
and decay of heavy baryonic resonances $N(1520), N(1770),...$. Their
coupling to the $\rho$ channel is not well known and may therefore be
overestimated. On the other hand the dilepton yield at low invariant
masses is underestimated for both systems. This is, first of all, due
to the lack of the bremsstrahlung contributions but also due to an
underprediction of the $\eta$ yield in UrQMD.

We note that the UrQMD model is presently under improvement and
extension, updated results for the dileptons at SIS energies are
expected soon \cite{Endres}.

\section{Ratios of dilepton yields R(AA/NN)}

\subsection{Comparison with experimental data}

The primary interest of measuring dilepton production in heavy-ion
collisions is to see whether it is a mere
superposition of the production in elementary $(pp+pn(d))$ collisions. Of
course in this threshold energy regime the Fermi motion of the
nucleons inside a nucleus plays an important role and therefore
the question has to be formulated more precisely: Is there an in medium
enhancement beyond the Fermi motion? This question we will address in this
section.

The HADES collaboration has measured the elementary reactions at
different beam energies than the heavy-ion reactions,
i.e. $pp$ and quasi-free $pn$ reactions at 1.25 GeV whereas the
C+C collisions at 1.0 and 2.0 $A$GeV and Ar+KCl at 1.75
$A$GeV. Thus, a comparison of elementary reaction data with
those of heavy ions at the same energy was not possible experimentally.
Therefore, we also have to calculate the 'reference spectrum' $NN=(pp+pn)/2$
at 1.25 GeV in order to compare with experimental  $AA/NN$ ratios.
Then we show the sensitivity of the ratio $AA/NN$  to the energy selection
of reference spectra $NN$ which finally might influence
the interpretation of in-medium modifications in $A+A$ collisions
relative to the $NN$.
All calculations presented here have been performed with  free vector
meson spectral functions.

\begin{figure}[h!]
\phantom{a}\vspace*{5mm}
\includegraphics[width=8.5cm]{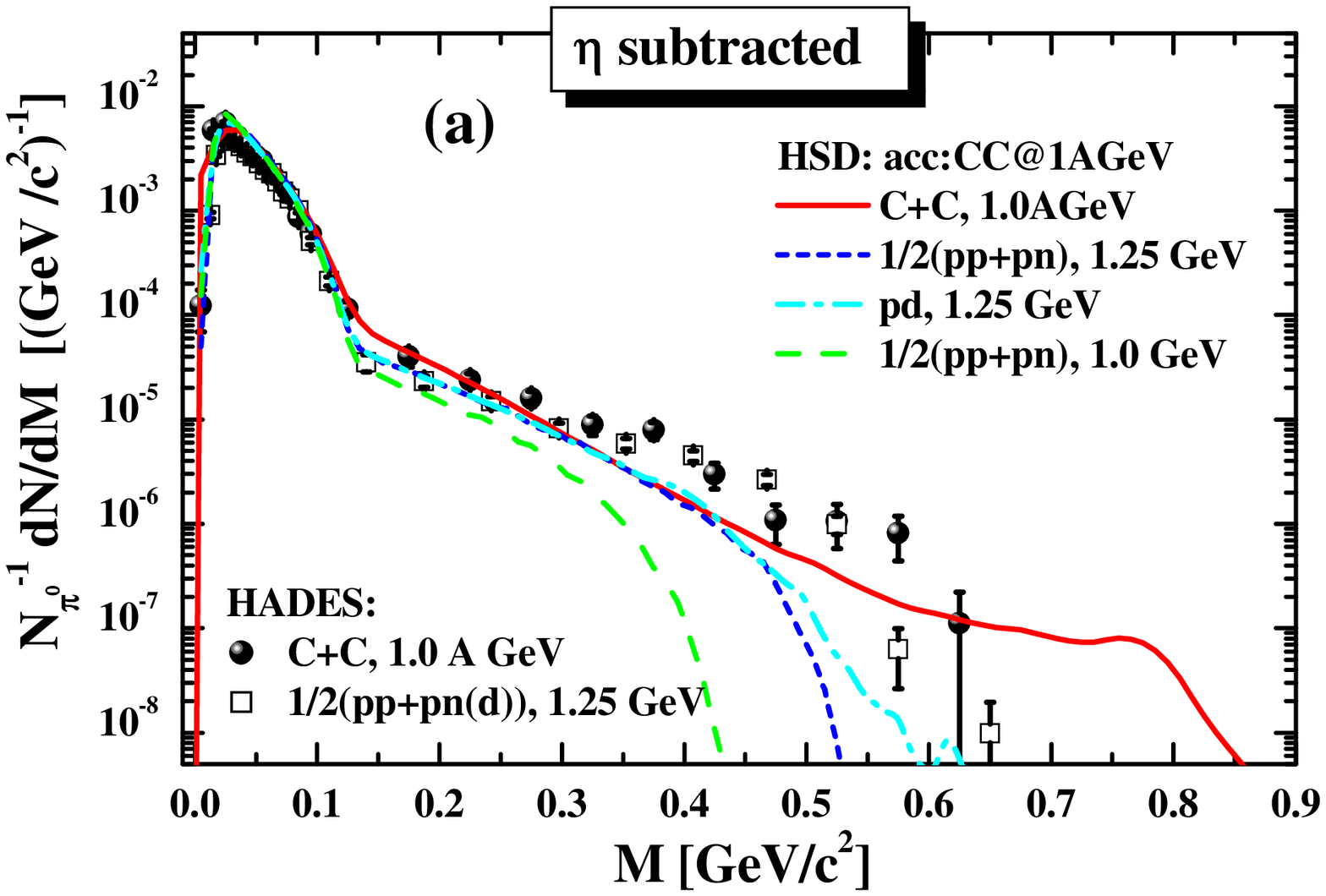}
\hspace*{3mm}
\includegraphics[width=8.5cm]{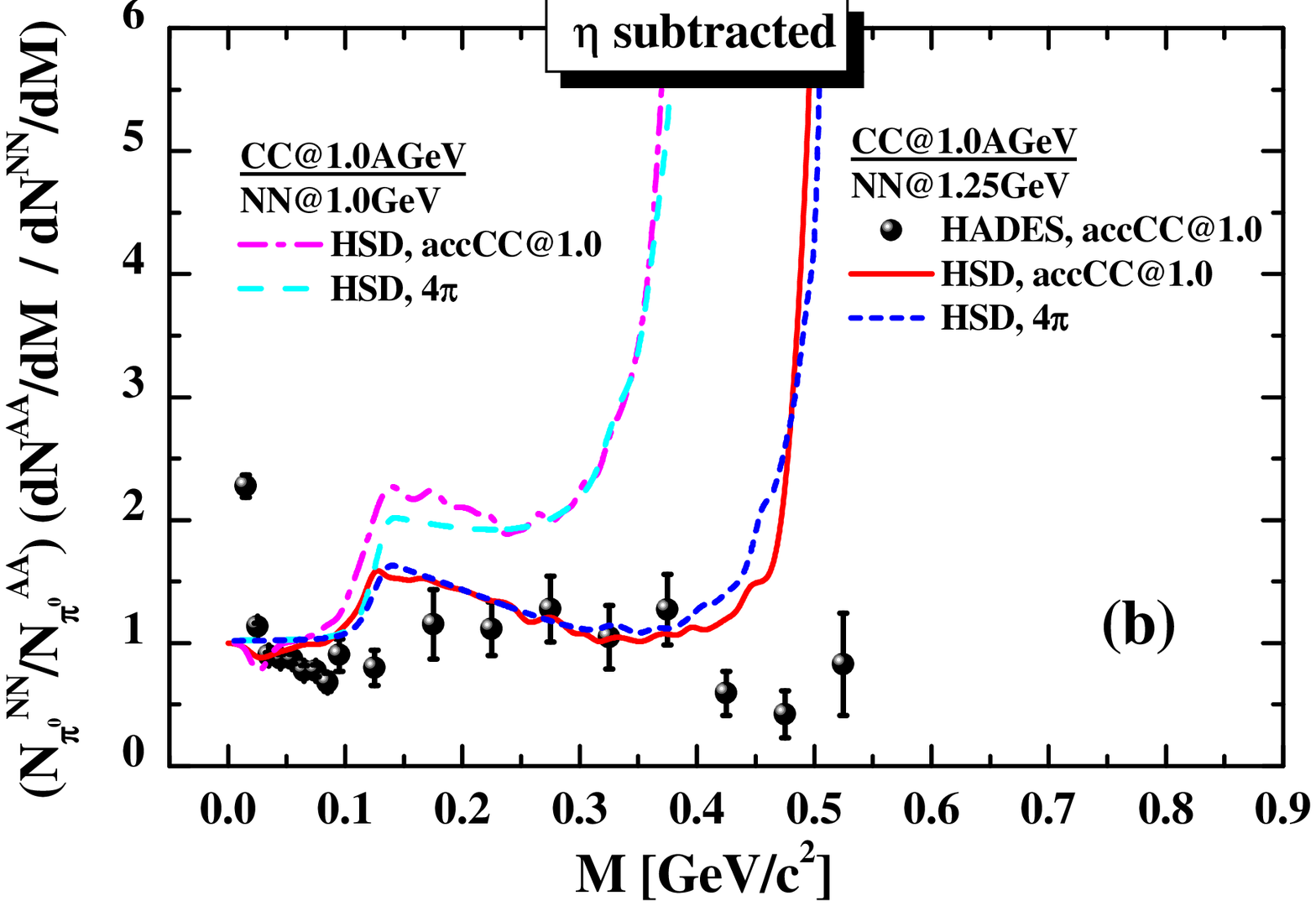}
\caption{(Color online) Left (a): The mass differential dilepton spectra - normalized to
the $\pi^0$ multiplicity and after $\eta$ Dalitz yield subtraction -
from HSD calculations for C+C at 1.0 $A$GeV (solid line),  for the
isospin-averaged reference spectra $NN=(pp+pn)/2$ at 1.25 GeV (short
dashed line) and at 1.0 GeV (dashed line) as well as for $pd$ at 1.25
GeV (dot-dashed line) in comparison to the corresponding HADES
data \cite{Agakishiev:2009yf} - for C+C at 1.0 $A$GeV
and the 'reference' spectra taken as an averaged sum of $pp$ and quasi-free $pn(d)$
(denoted as $(pp+pn(d))/2$) measured at 1.25 GeV.
The theoretical calculations passed through the HADES acceptance filter for
C+C at 1.0 $A$GeV (denoted as "acc:CC@1AGeV") and mass/momentum resolutions.
Right (b): Ratio of the dilepton differential spectra - normalized
to the $\pi^0$ multiplicity and after $\eta$ Dalitz yield subtraction -
to the isospin-averaged reference spectra $NN=(pp+pn)/2$ taken at
1.25$\  GeV$ employing C+C at $1.0\ AGeV$ experimental ("acc:CC@1AGeV") acceptance
(solid line) and in $4\pi$ (short dashed line).
Also the HSD results for the ratio of C+C at 1.0 $A$GeV to
the reference $NN$ spectra at 1.0 GeV are shown with
experimental ("acc:CC@1AGeV") acceptance corrections (dash-dotted line) and in
$4\pi$  (dashed line).
}
\label{Fig_RCC1NN}
\phantom{a}\vspace*{5mm}
\includegraphics[width=8.5cm]{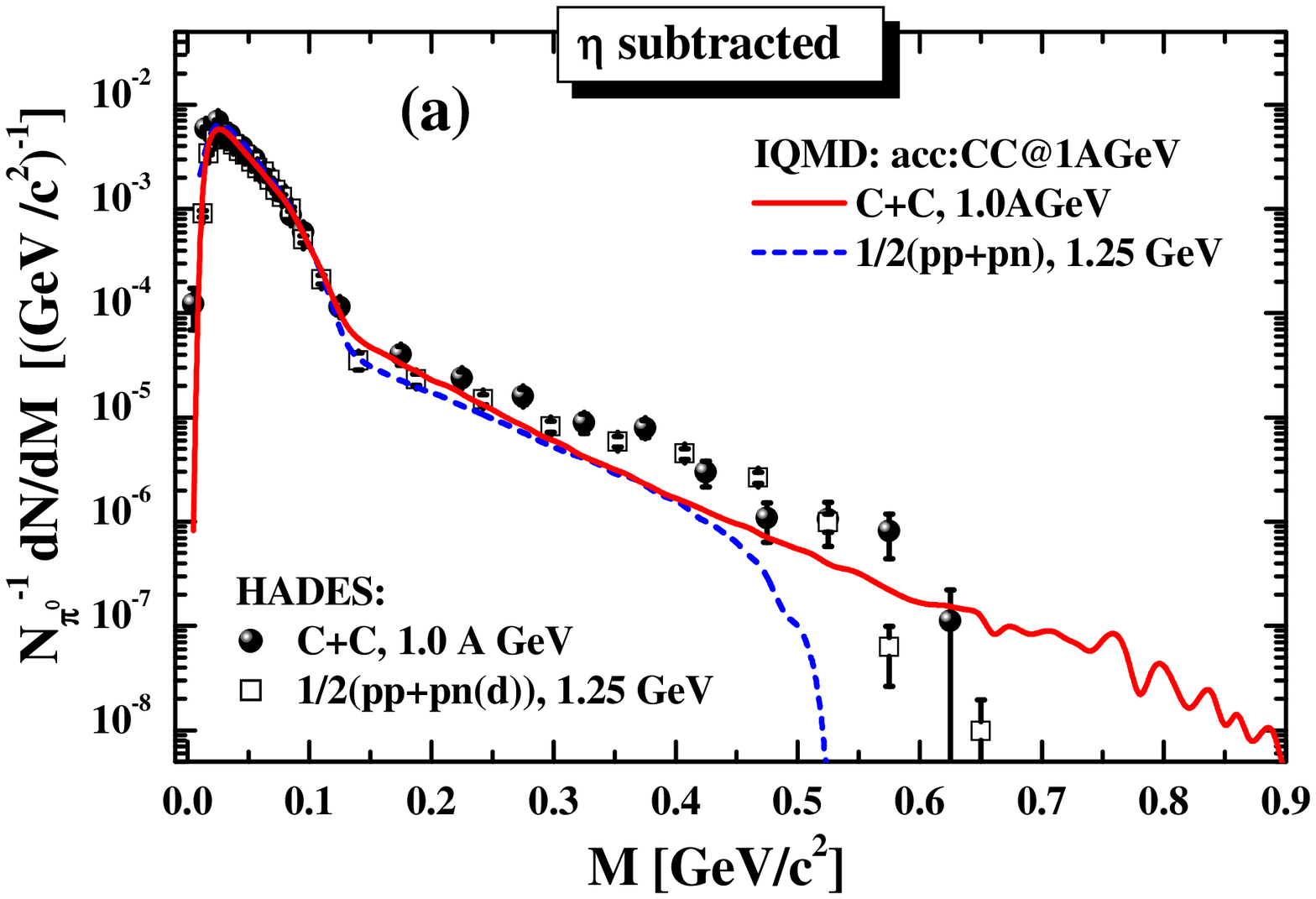}
\hspace*{3mm}
\includegraphics[width=8.5cm]{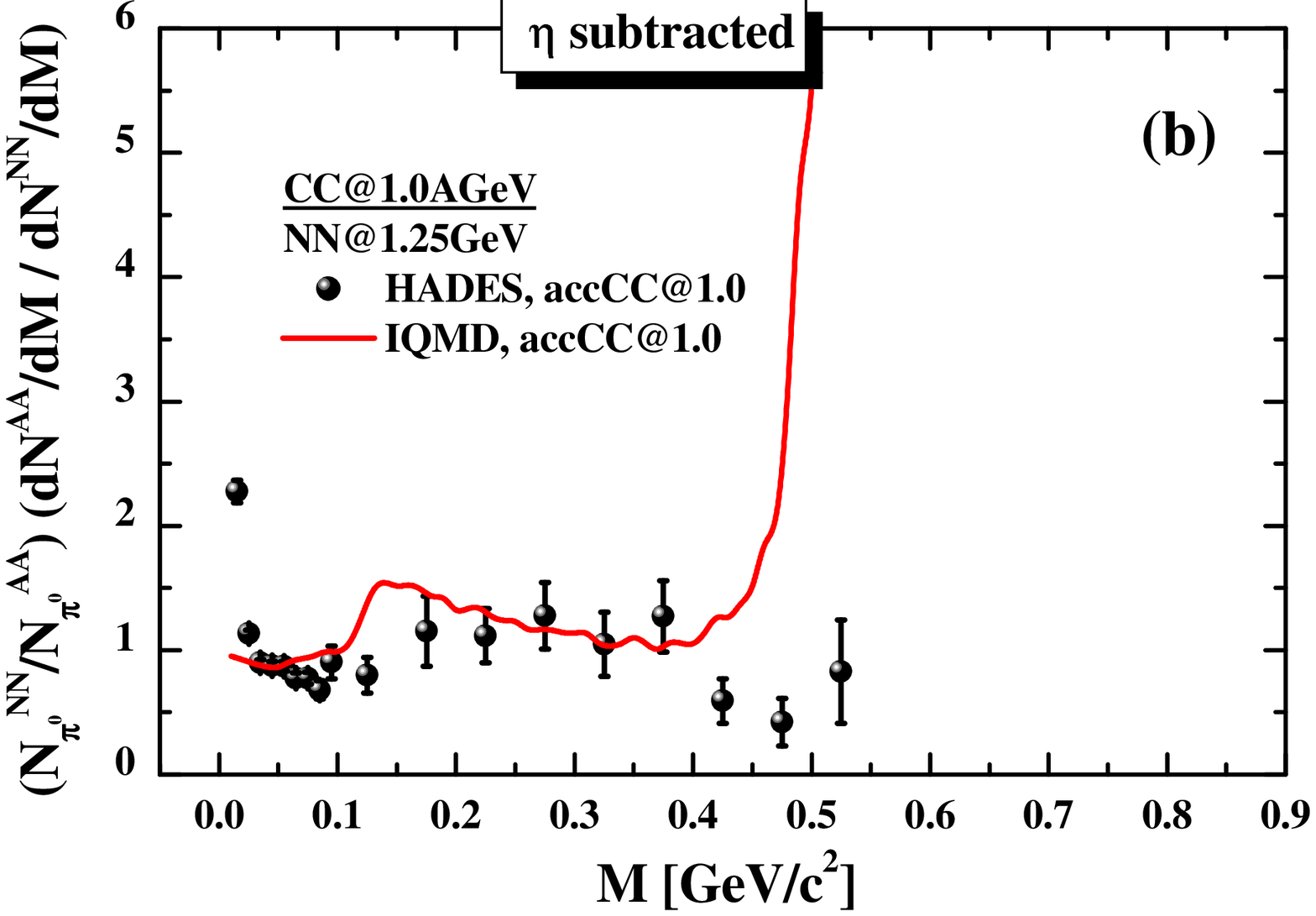}
\caption{(Color online) Left (a): The mass differential dilepton spectra - normalized to
the $\pi^0$ multiplicity and after $\eta$ Dalitz yield subtraction -
from IQMD calculations for C+C at 1.0 $A$GeV (solid line),  for the
isospin-averaged reference spectra $NN=(pp+pn)/2$ at 1.25 GeV (short
dashed line) in comparison to the HADES data \cite{Agakishiev:2009yf}.
The theoretical calculations passed through the HADES acceptance filter for
C+C at 1.0 $A$GeV and mass/momentum resolutions.
Right (b): Ratio of the dilepton differential spectra - normalized to the
$\pi^0$ multiplicity and after $\eta$ Dalitz yield subtraction -
of C+C at 1.0 $A$GeV (employing C+C at $1.0\ AGeV$ experimental
("acc:CC@1AGeV") acceptance) to the isospin-averaged reference spectra
$NN=(pp+pn)/2$ taken at 1.25  GeV. }
\label{Fig_RIQCC1NN}
\end{figure}

Fig. \ref{Fig_RCC1NN}, left, shows the mass differential dilepton
spectra - normalized to the multiplicity of $\pi^0$'s and after $\eta$
Dalitz yield subtraction - from HSD calculations for C+C at 1.0
$A$GeV (solid line), for the isospin-averaged reference spectra
$NN=(pp+pn)/2$ at 1.25 GeV (short dashed line) and at 1.0  GeV
(dashed line) as well as for $pd$ at 1.25  GeV (dot-dashed line).
These calculations are compared to the corresponding HADES data
from Refs. \cite{Agakishiev:2009yf} -  for C+C at 1.0 $A$GeV
and the 'reference' spectra taken as an averaged sum of $pp$ and quasi-free $pn(d)$
(denoted as $(pp+pn(d))/2$) measured at 1.25 GeV.
The theoretical calculations passed through the HADES acceptance filter
for C+C at 1.0 $A$GeV (denoted as "acc:CC@1AGeV") and
mass/momentum resolutions which smears out the high mass region.
The theoretical  reference spectra is taken as the averaged sum of dilepton spectra from
$p+p$ and free $p+n$ collisions. As seen from the figure
there is no essential difference between our theoretical $pd$ and $NN$ spectra up
to $M \approx 0.5$ GeV and only for larger invariant masses the
enhanced 'open' phase space for $pd$ compared to $NN$ becomes
important.

Fig. \ref{Fig_RCC1NN} (r.h.s.)  shows the ratio of the dilepton
differential spectra for C+C at 1.0  $A$GeV  to the
isospin-averaged $NN=(pp+pn)/2$. Both spectra are  normalized to the
$\pi^0$ multiplicity and the $\eta$ Dalitz yield has been subtracted.  The
solid and short dashed line present the ratio of C+C at 1.0 \
$A$GeV to  $NN$ at 1.25 GeV in the acceptance region and in 4$\pi$, respectively.
The dash-dotted  and dashed lines are the corresponding  ratios of C+C
at $1.0 \ AGeV$ to $NN$ at 1.0 $GeV$. If we divide the spectra of
C+C at 1.0 \ $A$GeV by the $NN$ spectra at 1.25  GeV  the ratio is quite flat,
as the experiments show as well. The enhancement in the theory at the upper
end of the $\pi^0$ peak and hence around M=0.15\ GeV comes in about
equal parts from bremsstrahlung and $\Delta$ Dalitz decay. We observe as well that the
acceptance cuts do not change the enhancement. Therefore we can discuss
it in the next section using $4 \pi$ yields. In this figure we display
as well  that the true enhancement,  obtained by comparing C+C and
$NN$ at the same energy, is much larger.  For $0.125 \ GeV \ <M  \ <0.3$
GeV it is about a factor of two.

We note that the HADES collaboration used $pp$
and quasi-free $pn(d)$ spectra at 1.25 GeV as a reference $NN^d=(pp+pn(d))/2$
spectrum for the ratios of the dilepton yields from $AA$ to $NN$. In order to
avoid the additional uncertainties of dilepton production in $pd$
collision, a system which cannot be modeled reasonably good in semi-classical
approaches, we use the reference spectra $NN=(pn+pp)/2$.
As Fig. \ref{Fig_RCC1NN} shows, both methods
are equivalent up to invariant masses of M=0.4 GeV.  Above this
value the ratio increases very fast because in the elementary
reactions the limitation due to phase space is more severe than in
heavy-ion collisions, where the Fermi motion can provide larger
invariant masses.  These HSD results are confirmed by
the IQMD calculations shown in Fig. \ref{Fig_RIQCC1NN} in a form
equivalent to  Fig. \ref{Fig_RCC1NN}.

Now we step to the energy 2.0 $A$GeV.
In order to compare the experimental data for C+C, measured at two different
energies 1.0 and 2.0 $A$GeV, the HADES collaboration transformed the
C+C data measured at 2.0 $A$GeV to the acceptance of C+C at 1.0 $A$GeV
by using - due to lack of statistics - a one-dimensional
transformation (see Ref. \cite{Agakishiev:2009yf}). We denote this transformation
as "$1D-acc:CC@1AGeV$" in order to distinguish it from the standard
three dimensional filtering procedure using the
"3D" (defined above as "acc:CC@1AGeV") experimental acceptance matrix
(which depends on $M, p_T$ and $y$),  provided by the HADES
Collaboration \cite{HADESweb} for the filtering of theoretical $4\pi$ results.

\begin{figure}[h!]
\phantom{a}\vspace*{5mm}
\includegraphics[width=8.5cm]{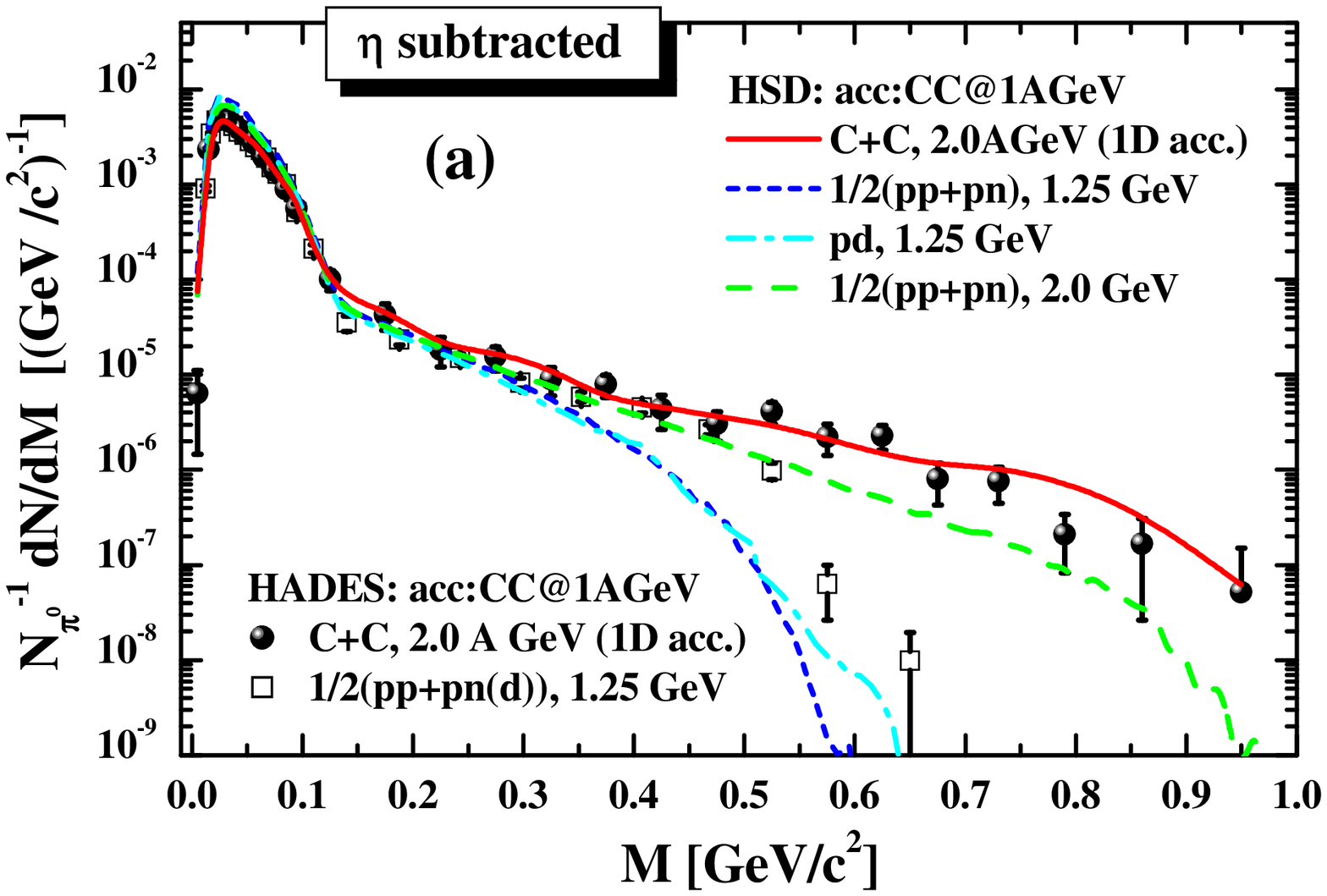}
\hspace*{3mm}
\includegraphics[width=8.5cm]{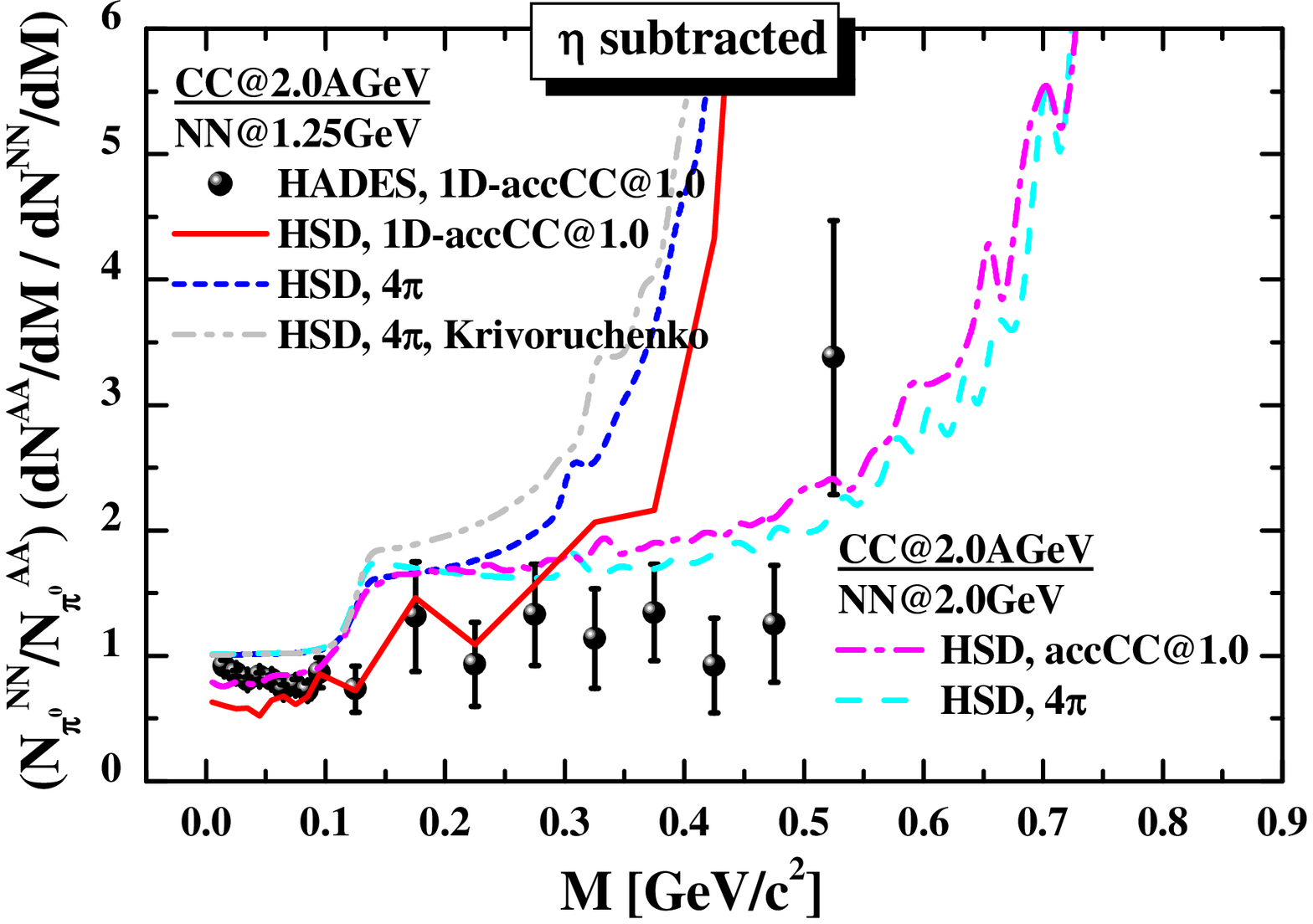}
\caption{(Color online) Left (a): The mass differential dilepton spectra - normalized to
the $\pi^0$ multiplicity and after $\eta$ Dalitz yield subtraction -
from HSD calculations  for C+C at 2.0 $A$GeV  (solid line) and for the
isospin-averaged reference spectra $NN=(pp+pn)/2$ at 1.25 GeV (short
dashed line) and at 2.0 GeV (dashed line) as well as for $pd$ at 1.25
GeV (dot-dashed line) in comparison to the HADES data \cite{Agakishiev:2009yf}
- for C+C measured at 2.0 $A$GeV and $(pp+pn(d))/2$ at 1.25 GeV and transformed
to the acceptance for C+C at 1.0 $A$GeV (see the discussion in the text).
The theoretical calculations passed through the HADES acceptance filter for
C+C at 1.0  $A$GeV ("1D-acc:CC@1AGeV") and mass/momentum resolutions.
Right (b): Ratio of the dilepton differential spectra of C+C at 2.0 $A$GeV
- normalized to the $\pi^0$ multiplicity and after
$\eta$ Dalitz yield subtraction - to the isospin-averaged reference
spectra $NN=(pp+pn)/2$ at 1.25 GeV with experimental ("1D-acc:CC@1AGeV")
acceptance (solid line) and
in $4\pi$ result with the default "Wolf" differential electromagnetic width
for $\Delta$ Dalitz decay (short dashed line) and "Krivoruchenko" width
(dash-dot-dotted line).
Also the HSD results for the ratio of C+C at 2 $A$GeV to the reference
$NN$ spectra at 2.0 GeV are shown: with experimental ("acc:CC@1AGeV")
acceptance for C+C at 1.0  $A$GeV (dash-dotted line) and in $4\pi$ (dashed line).
}
\label{Fig_RCC2NN}
\phantom{a}\vspace*{5mm}
\includegraphics[width=8.5cm]{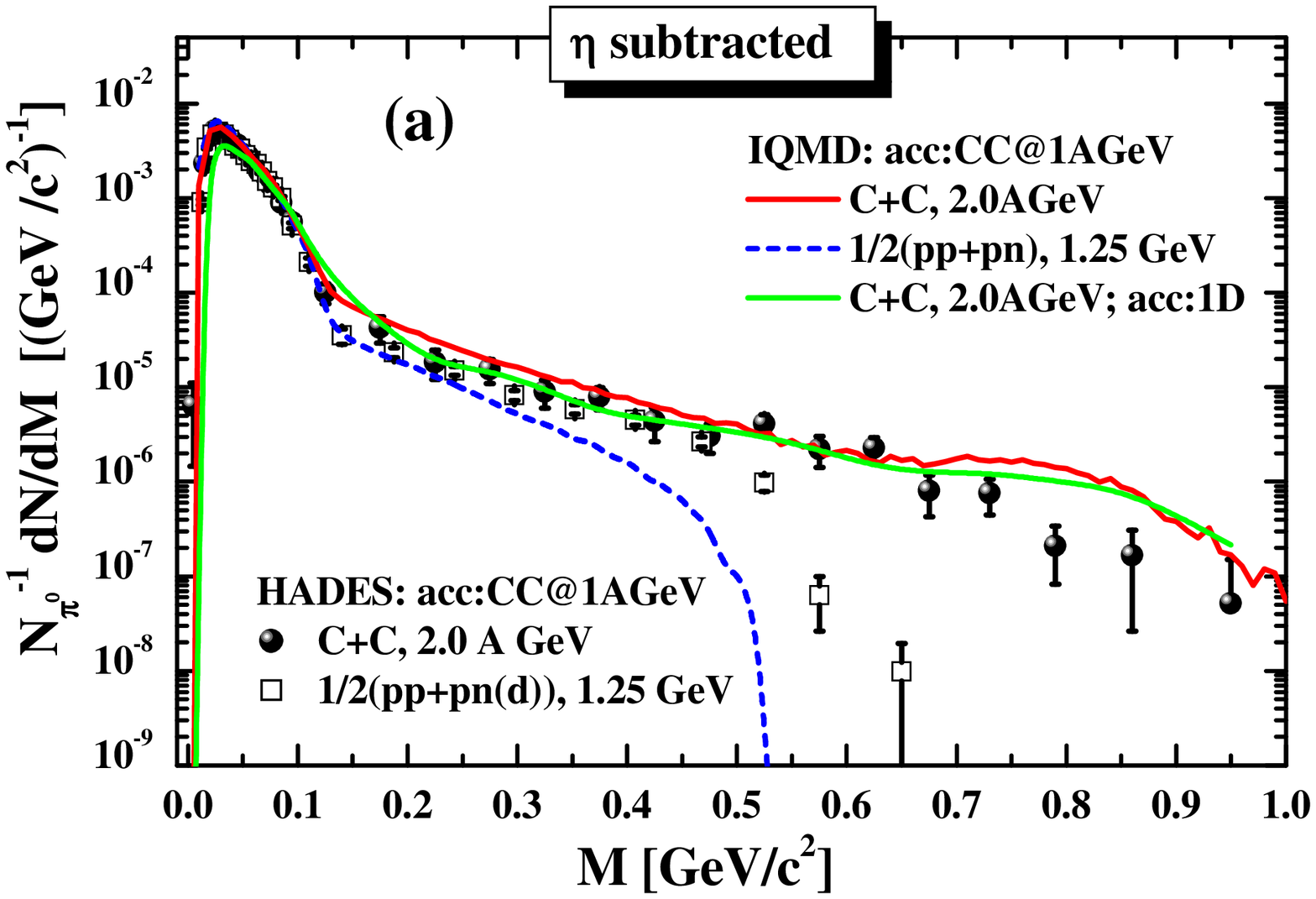}
\hspace*{3mm}
\includegraphics[width=8.5cm]{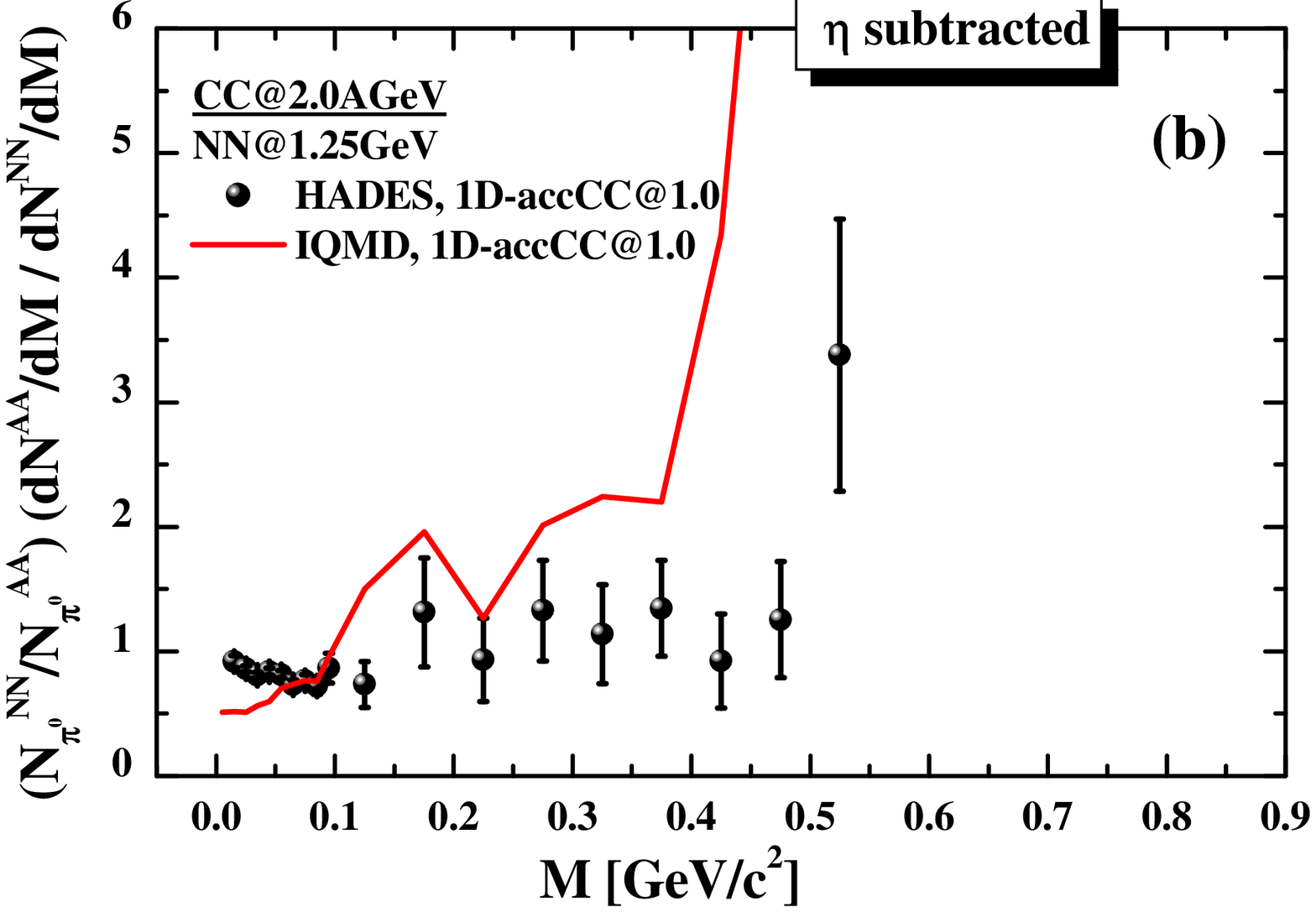}
\caption{(Color online) Left (a): The mass differential dilepton spectra - normalized to
the $\pi^0$ multiplicity and after $\eta$ Dalitz yield subtraction -
from IQMD calculations  for C+C at 2.0 $A$GeV  (solid line) and for the
isospin-averaged reference spectra $NN=(pp+pn)/2$ at 1.25 GeV (short
dashed line)  in comparison to the  HADES data \cite{Agakishiev:2009yf}.
The theoretical calculations passed through the corresponding HADES acceptance filter for
C+C at 1.0 $A$GeV ("1D-acc:CC@1AGeV") and mass/momentum resolutions
(see the discussion in the text).
Right (b): Ratio of the dilepton differential spectra for C+C at 2.0 $A$GeV
 - normalized to the $\pi^0$ multiplicity and after
$\eta$ Dalitz yield subtraction - to the isospin-averaged reference
spectra $NN=(pp+pn)/2$ taken at 1.25 GeV with
experimental ("1D-acc:CC@1AGeV") acceptance for C+C at 1.0  $A$GeV (solid line).
}
\label{Fig_RIQCC2NN}
\end{figure}

Fig. \ref{Fig_RCC2NN} presents for C+C at 2.0  $A$GeV  the same
quantities as Fig.  \ref{Fig_RCC1NN} for C+C at 1.0  $A$GeV.
The  solid line on the left is the result of the HSD calculations, the
short dashed  line and the dashed dotted line are  the
isospin-averaged reference spectra $NN=(pp+pn)/2$ at 1.25 GeV and
$pd$ at $1.25\ GeV $. The dashed line is the reference $NN$ spectrum
at 2.0 GeV; the corresponding HADES data  are taken from Refs.
\cite{Agakishiev:2009yf}.
Note, that the simulated HSD mass distribution for C+C at 2.0 $A$GeV
has been transformed to the corresponding acceptance in the same way
as done for the experimental data using the "$1D-acc:CC@1AGeV$" transformation.
Fluctuations introduced by this procedure result in part from the limited
statistics of the relevant HADES C+C data set and in part from the
necessary re-binning of the latter.

The right part of Fig. \ref{Fig_RCC2NN} shows the ratio
of the dilepton differential spectra - normalized
to the $\pi^0$ multiplicity and after $\eta$ Dalitz yield subtraction -
of C+C at 2.0 $A$GeV to the isospin-averaged reference spectra $NN=(pp+pn)/2$ taken
at 1.25 GeV,  applying the C+C at 2.0 $A$GeV "$1D-acc:CC@1AGeV$" experimental
acceptance (solid line), and
in $4\pi$ result with the default "Wolf" differential electromagnetic width
for $\Delta$ Dalitz decay (short dashed line) and "Krivoruchenko" width
(dash-dot-dotted line) in order to demonstrate the model uncertainties
(cf. discussion in Section VI).
Also the HSD results for the ratio of C+C at 2.0 $A$GeV to the reference $NN$ spectra,
taken at 2.0 GeV, are shown, including the full "3D"- experimental acceptance
(dash-dotted line)  and in $4\pi$ (dashed line). These results show that
the experimental data measured up to an invariant mass of $ M\approx
0.5$ GeV are compatible with a ratio of one and hence with no in-medium
enhancement.  The theoretical results are more complicated. Up to an
invariant  mass of $ M\approx \ 0.3$  GeV theory predicts a
enhancement factor of about 1.8 for $4\pi$. The ratio at the same nominal energy
shows this enhancement even up to invariant masses of $ M \approx 0.6$
GeV before the influence of the Fermi motion sets in.

The IQMD calculations for C+C at 2.0 $A$GeV are presented in Fig.
\ref{Fig_RIQCC2NN} which shows the same quantities as  Fig.
\ref{Fig_RCC2NN}. We see that the both model agree quite well and
the form of the ratio is identical in both approaches.

\begin{figure}[h!]
\phantom{a}\vspace*{5mm}
\includegraphics[width=8.5cm]{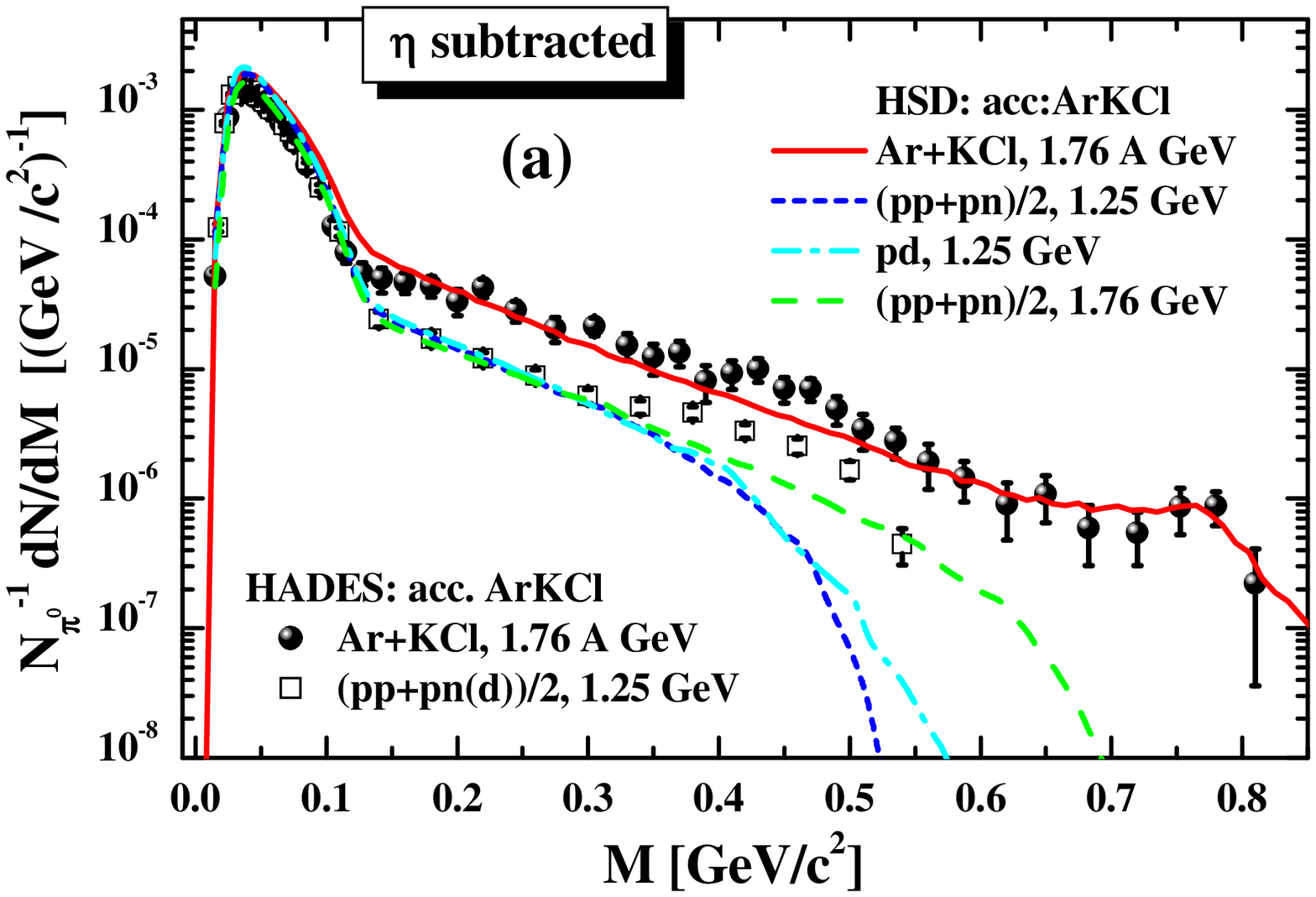}
\hspace*{3mm}
\includegraphics[width=8.5cm]{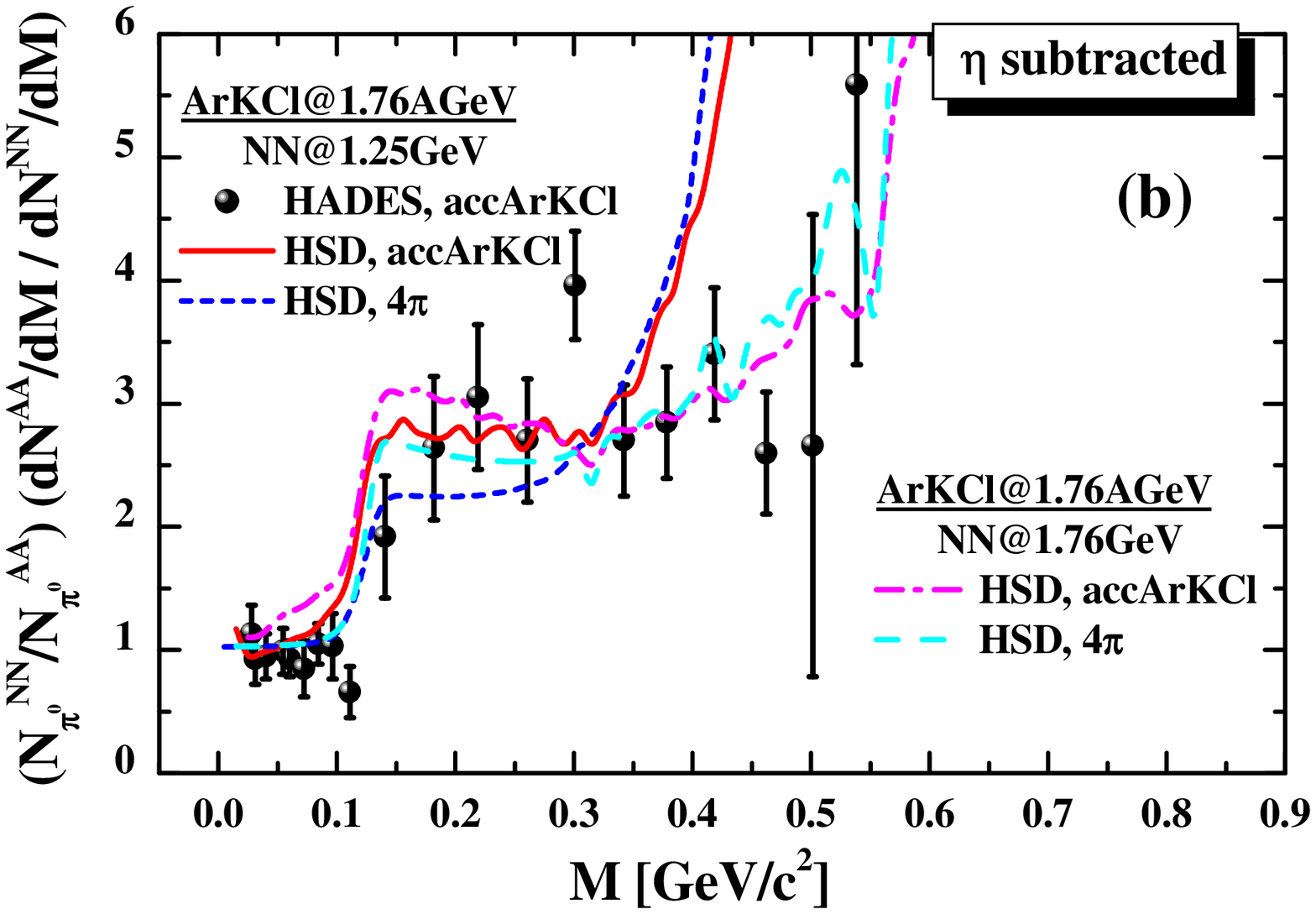}
\caption{(Color online) Left (a): The mass differential dilepton spectra - normalized to
the $\pi^0$ multiplicity and after $\eta$ Dalitz yield subtraction -
from HSD calculations for  Ar+KCl at 1.76 $A$GeV (solid line) and for the
isospin-averaged reference spectra $NN=(pp+pn)/2$ at 1.25 GeV (short
dashed line) and at 1.76 GeV (dashed line) as well as for $pd$ at 1.25
GeV (dot-dashed line) in comparison to the corresponding HADES data
\cite{Agakishiev:2011vf}.  The theoretical calculations for $Ar +
KCl$ and for $NN$ passed through the HADES acceptance filter for Ar+KCl and
mass/momentum resolutions.  Right (b): Ratio of the dilepton differential
spectra - normalized to the $\pi^0$ multiplicity and after $\eta$ Dalitz yield
subtraction - to the isospin-averaged reference spectra $NN=(pp+pn)/2$
taken at 1.25 GeV, involving Ar+KCl experimental acceptance (solid
line) and for $4\pi$ (short dashed line).
Also the HSD results for the ratio to the reference $NN$ spectra taken
at 1.76 GeV are shown, with the Ar+KCl experimental acceptance
(dash-dotted line) and in $4\pi$  (dashed line).
}
\label{Fig_RArKNN}
\phantom{a}\vspace*{5mm}
\includegraphics[width=8.5cm]{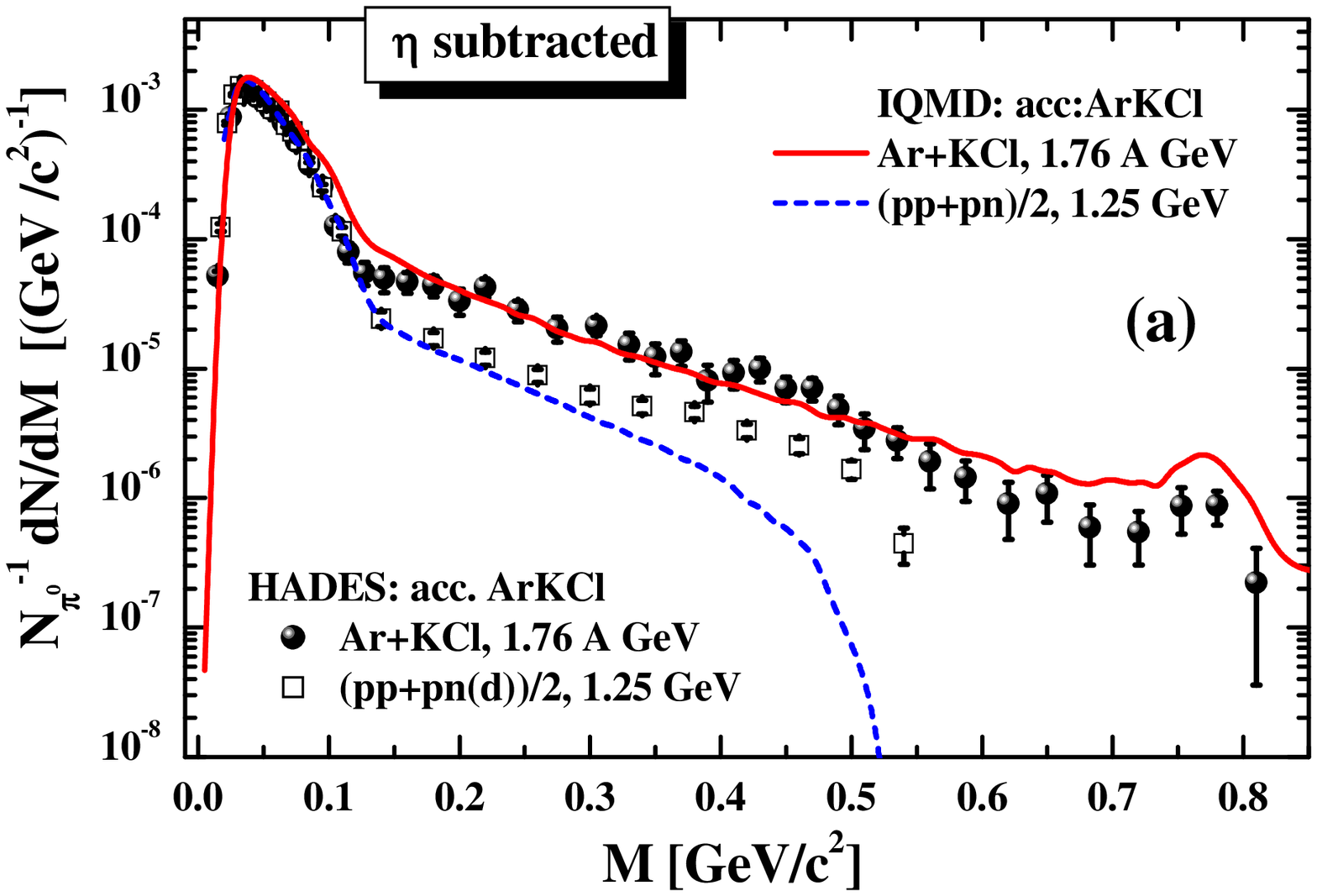}
\hspace*{3mm}
\includegraphics[width=8.5cm]{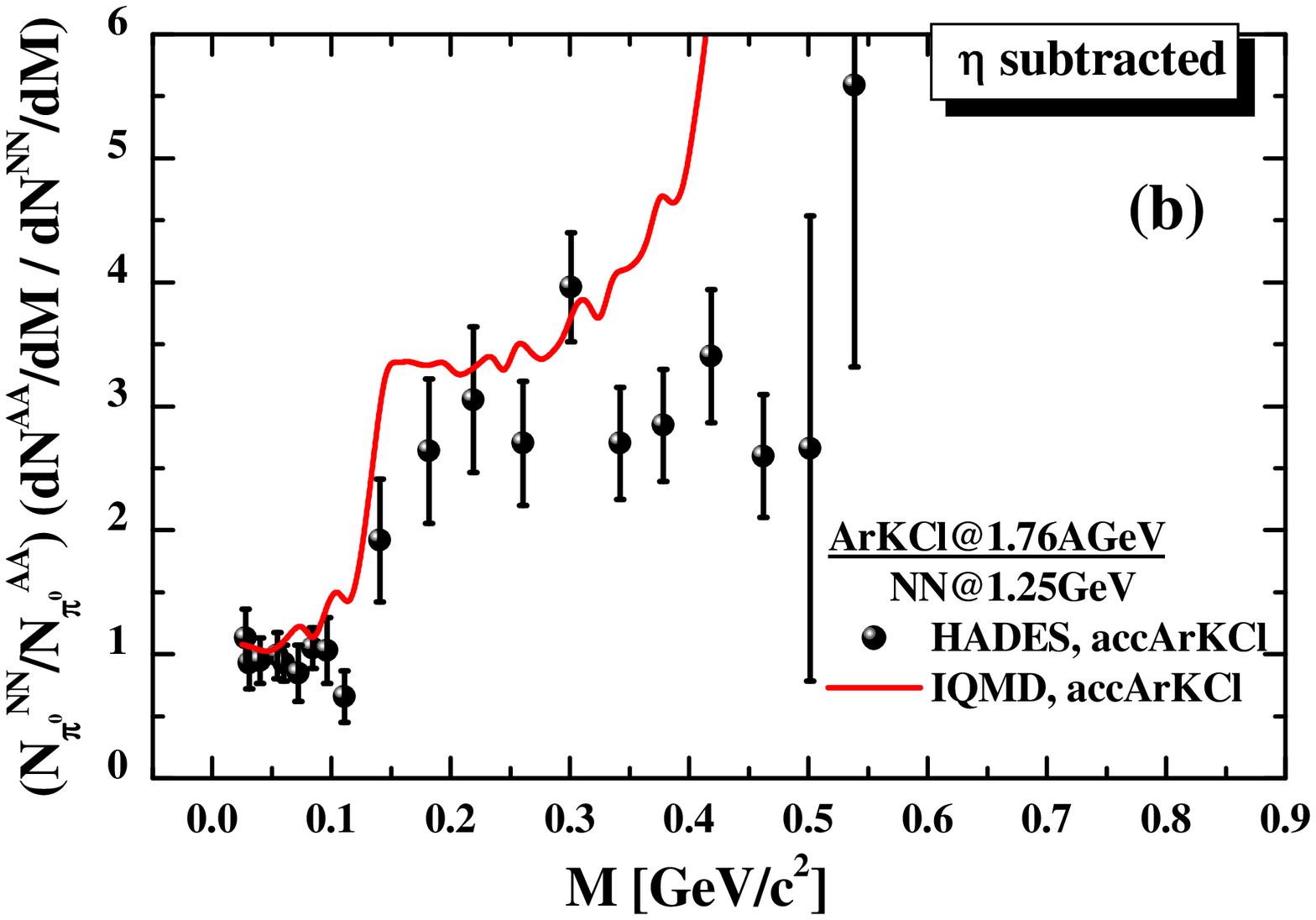}
\caption{(Color online) Left (a): The mass differential dilepton spectra - normalized to
the $\pi^0$ multiplicity and after $\eta$ Dalitz yield subtraction -
from IQMD calculations for of Ar+KCl at 1.76 $A$GeV (solid line) and for the
isospin-averaged reference spectra $NN=(pp+pn)/2$ at 1.25 GeV (short
dashed line) in comparison to the corresponding HADES data \cite{Agakishiev:2011vf}.
The theoretical calculations for $Ar + KCl$ and for $NN$
passed through the HADES acceptance filter for Ar+KCl and
mass/momentum resolutions.  Right (b): Ratio of the dilepton differential
spectra - normalized to the $\pi^0$ multiplicity and after $\eta$ Dalitz yield
subtraction - to the isospin-averaged reference spectra $NN=(pp+pn)/2$,
taken at 1.25 GeV, employing the Ar+KCl experimental acceptance (solid line).
}
\label{Fig_RIQArKNN}
\end{figure}

Fig. \ref{Fig_RArKNN} (l.h.s.) displays the mass differential dilepton
spectra for  Ar+KCl at 1.76 $A$GeV (solid line) - normalized to the
$\pi^0$ multiplicity  and after $\eta$ Dalitz yield subtraction.  We
compare HSD calculations for  Ar+KCl at 1.76 $A$GeV,  for the
isospin-averaged reference spectra $NN=(pp+pn)/2$ at 1.25 GeV (short
dashed line) and at 1.76 GeV (dashed line) as well as for $pd$ at 1.25
 GeV (dot-dashed line)  to the corresponding HADES data,
taken from Ref.  \cite{Agakishiev:2011vf}.  The theoretical
calculations for $Ar + KCl$ and for $NN$ passed through the HADES acceptance
filter for Ar+KCl and mass/momentum resolutions.  The right part of  Fig.
\ref{Fig_RArKNN} shows the ratio of the dilepton differential spectra -
normalized to the $\pi^0$ multiplicity and after $\eta$ Dalitz yield
subtraction - to the isospin-averaged reference spectra $NN=(pp+pn)/2$
taken at 1.25 GeV and employing the Ar+KCl experimental acceptance (solid
line) and in $4\pi$ (short dashed line).  We display as well the HSD
results for the ratio of Ar+KCl at 1.76 $A$GeV to the reference $NN$
spectrum at the same energy, including the experimental $Ar + KCl$ acceptance
(dash-dotted line) and in $4\pi$ (dashed line).  These results show
clearly that for invariant masses of $0 .1 \ GeV < M < 0.35$ GeV the data
as well as theory  are not a mere superposition of the elementary
spectra. The comparison also excludes that this enhancement, observed
in heavy-ion collisions, is due to acceptance since the results
with acceptance and in $4\pi$ are very similar. At larger invariant
masses theory and data do not agree because of the bump at the invariant
masses around   $ M\approx 0.5$ GeV, seen in the experimental $pd$
reactions, is not reproduced by theory. Taking the reference spectra at
the same nominal energy theory predicts that this enhancement is
constant up to energies of $ M\approx 0.5$ GeV. Then the Fermi motion
becomes important and yields a strong increase of the ratio.

Consequently, the experimental ratios of the invariant mass spectra
measured in heavy-ion collisions to the isospin-averaged
reference spectra $NN=(pp+pn)/2$ taken at 1.25 GeV reveals an in-medium
enhancement in Ar+KCl collisions at 1.75 $A$GeV  whereas
in  $ C+C$ collisions at 2 $A$GeV this ratio  is compatible with one and
therefore no in-medium enhancement is seen. The transport models show an enhancement
in all heavy-ion reactions when the reference spectrum is taken at the
same energy. It shows as well that acceptance cuts do not modify this
enhancement. The origin of this enhancement will be discussed in the next
subsection.

In Fig. \ref{Fig_RIQArKNN} we display the same quantities as in Fig. \ref{Fig_RArKNN}
but for IQMD calculations.  The enhancement of the experimental ratio is confirmed by IQMD
calculations,  which are in quantitative agreement with the HSD results.

 \subsection{Energy and system size dependence of the dilepton yield}

In this section we present the energy and system size dependence of the
dilepton yield in $4\pi$ as predicted by the HSD calculations in order
to study the question of a  possible in medium enhancement and to
identify eventually its physical origin.

\begin{figure}[h]
\phantom{a}\hspace*{-25mm}
\includegraphics[width=12.5cm]{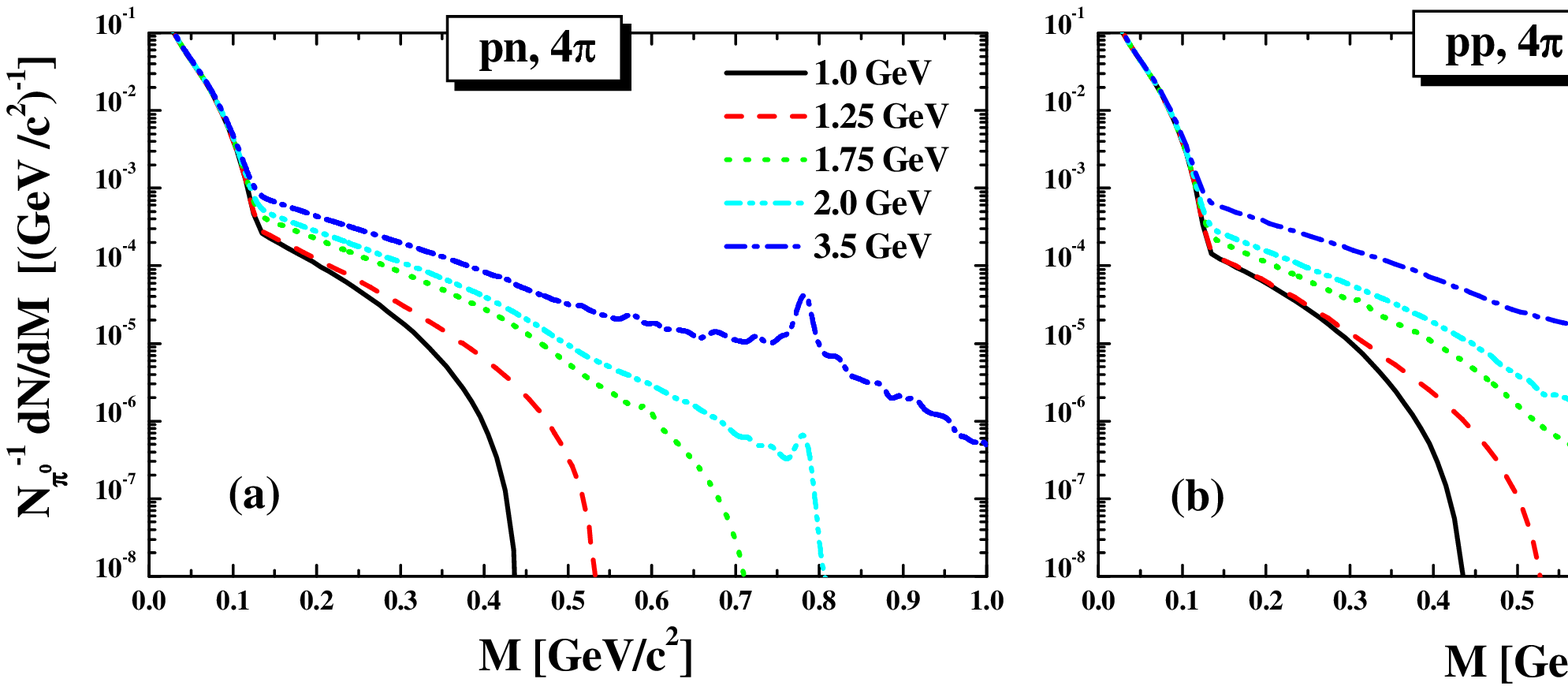}
\caption{(Color online) The $4\pi$  mass differential dilepton spectra - normalized to the
$\pi^0$ multiplicity  - obtained in  HSD calculations for $pn$ (left (a)) and $pp$ (right (b))
collisions at 1.0, 1.25, 1.75, 2.0 and 3.5 GeV.
}
\label{Fig_MNN}
\end{figure}
Fig. \ref{Fig_MNN} shows the HSD calculations for the mass differential
dilepton spectra - normalized to the $\pi^0$ multiplicity - for $pn$
(left) and $pp$ (right) collisions at 1.0, 1.25, 1.75, 2.0 and 3.5 GeV
in $4\pi$ acceptance.
Whereas the normalization renders the low invariant mass part to one,
independent of the beam energy, the spectra at high invariant masses
show a strong beam energy dependence, as expected. Bremsstrahlung
is not coupled to the number of pions (or the number of participants
which is often assumed to be proportional to the number of $\pi$'s) but
to the number of collisions. Also the production of heavier mesons
increases at these energies close to the meson thresholds, either
because it becomes easier to  produce them directly or because the
baryonic resonances which decay into these resonances are more
frequently populated. Last but not least, the phase space limitation of
the invariant  mass changes with energy which makes ratios
between invariant mass spectra at different energies complicated. Due
to the isospin dependence of different processes the $pp$ and $pn$
invariant mass spectra differ in detail but are generally determined by
phase space.  We can conclude from Fig. \ref{Fig_MNN}
that the comparison of dilepton data of heavy-ions and of elementary
reactions suffer substantially if both are measured at different
energies. This renders quantitative conclusions difficult.

\begin{figure}[h!]
\phantom{a}\vspace*{5mm}
\includegraphics[width=9.5cm]{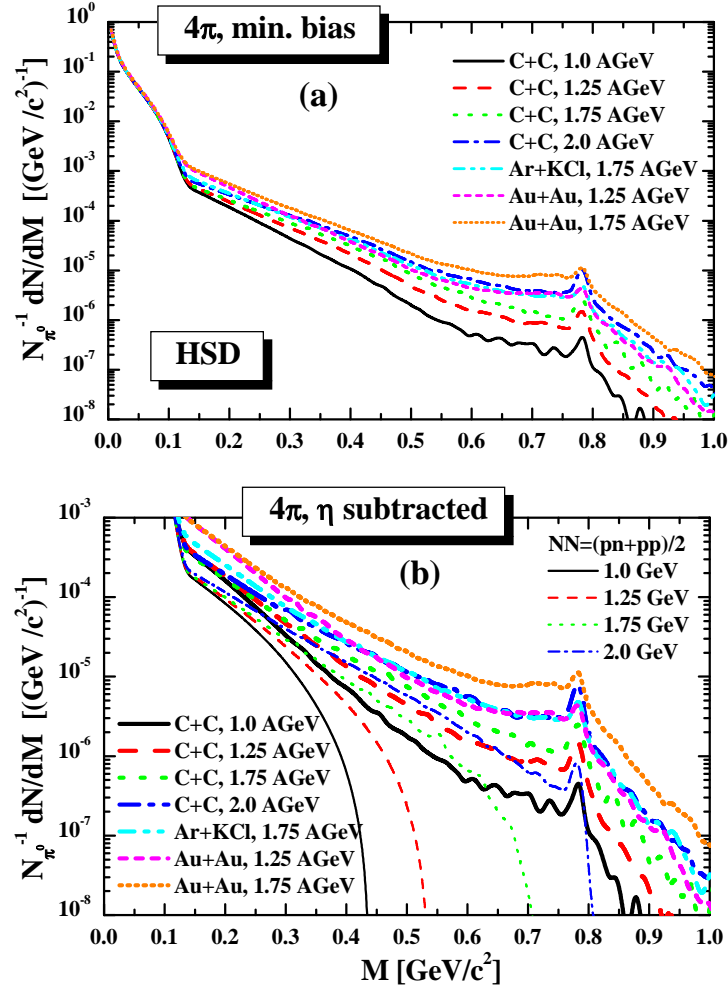}
\caption{(Color online) The invariant mass differential dilepton spectra - normalized to
the $\pi^0$ multiplicity- obtained in HSD calculations for the minimal bias C+C, Ar+KCl,
Au+Au collisions and for the isospin-averaged reference spectra $NN=(pn+pp)/2$  at
1.0, 1.25, 1.75, 2.0 $A$GeV in $4\pi$ acceptance.
The upper plot (a) corresponds to the total dilepton $A+A$ spectra whereas
the lower plot (b) shows the dilepton spectra after $\eta$ Dalitz yield
subtraction.  The thick lines on the lower plot stand for the $A+A$
dilepton yields whereas the thin lines show the $NN$ spectra at the
same energy.
}
 \label{Fig_MAA}
\end{figure}

Fig. \ref{Fig_MAA} displays the results of HSD calculations for the
$4\pi$ mass differential dilepton spectra - normalized to the  $\pi^0$
multiplicity  - for the minimal bias symmetric heavy-ion collisions as
compared to the  isospin-averaged reference spectra $NN=(pn+pp)/2$. We
display calculations for   C+C, Ar+KCl, Au+Au at 1.0, 1.25, 1.75, 2.0
$A$GeV .  The upper plot corresponds to the total dilepton $A+A$ spectra
whereas the lower plot shows the dilepton spectra after $\eta$ Dalitz
yield subtraction.  The thick lines on the lower plot stand for the
$A+A$ dilepton yields whereas the thin lines show the $NN$ spectra at
the same energies.  We see clearly that the dilepton spectra do not
scale with the $\pi^0$ multiplicity for invariant masses $M > 0. 11$
GeV. There is a strong energy and system size dependence of this
invariant mass region due to the complicated dynamics of baryon
resonances and mesons. Generally the invariant mass spectra in $A+A$
collisions are smoother due to the Fermi motion.

\begin{figure}[h!]
\phantom{a}\vspace*{5mm}
\includegraphics[width=9.5cm]{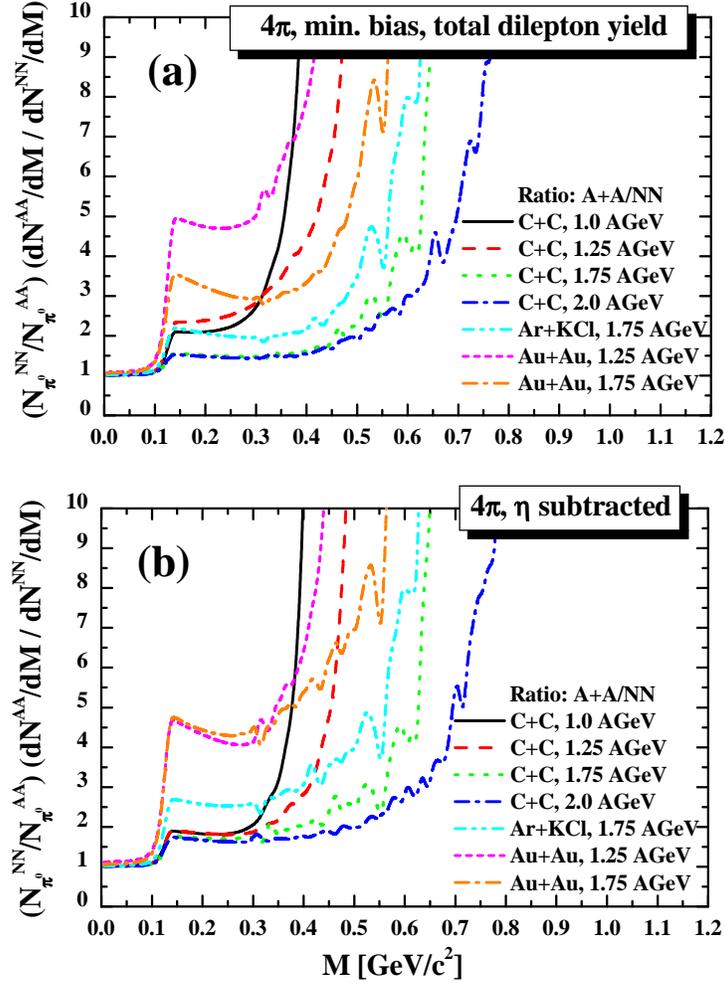}
\caption{(Color online) Upper plot (a): The ratio $(1/N_{\pi^0}^{AA}
dN^{AA}/dM)/(1/N_{\pi^0}^{NN} dN^{NN}/dM)$ of the invariant mass differential
dilepton $4\pi$ spectra - normalized to the $\pi^0$ multiplicity  - from HSD calculations for
minimal bias $A+A$ collisions: We display C+C, Ar+KCl, Au+Au collisions in comparison
to the isospin-averaged reference spectra $NN=(pn+pp)/2$  at 1.0, 1.25, 1.75, 2.0 $A$GeV.
Lower plot (b): the same ratios but for the dilepton spectra after $\eta$ Dalitz yield subtraction.
}
\label{Fig_RAA}
\end{figure}

Fig. \ref{Fig_RAA} presents the ratio $(1/N_{\pi^0}^{AA}
dN^{AA}/dM)/(1/N_{\pi^0}^{NN} dN^{NN}/dM)$ of the mass differential
dilepton spectra - normalized to the $\pi^0$ multiplicities  - obtained
in  HSD calculations. Displayed are the ratios of minimal bias C+C, Ar+KCl, Au+Au
collisions and of  the isospin-averaged reference
spectra $NN=(pn+pp)/2$ at the same energy.  The lower plot depicts the
same ratios but for the dilepton spectra after $\eta$ Dalitz yield
subtraction. Clearly we see a quite complex structure. We start with
the energy dependence of the ratio which decreases with energy.
Including the $\eta$ production this can be clearly seen by comparing
the Au+Au collisions at 1.75 and at 1.25 $A$GeV as well as by
comparing the C+C system at different energies; $\eta$ subtraction
modifies some details but does not change the tendency. It is also
obvious that the ratio increases with the system size. The ratio for
Au+Au at 1.25 $A$GeV is about 4.5, that of C+C at the same energy
around 2.5. We study now the origin of this enhancement in detail.

\begin{figure}[t]
\phantom{a}\vspace*{5mm}
\includegraphics[width=9.5cm]{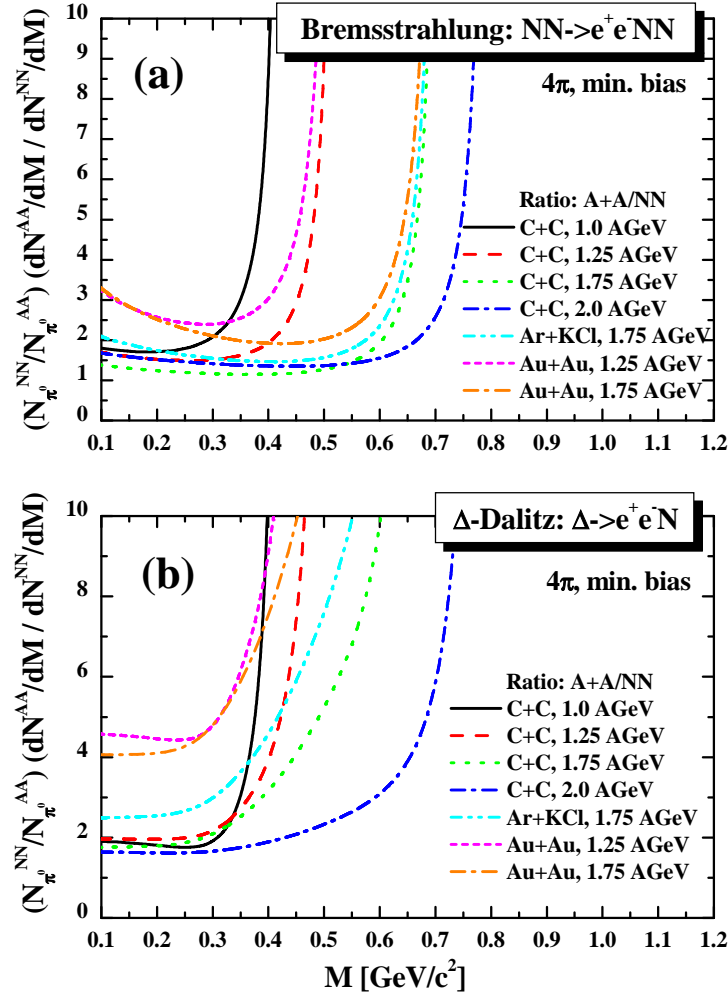}
\caption{(Color online) The ratio $(1/N_{\pi^0}^{AA}
dN^{AA}/dM)/(1/N_{\pi^0}^{NN} dN^{NN}/dM)$ of the dilepton yield from
the bremsstrahlung channel (upper part (a)) and $\Delta$ Dalitz decay (lower
part (b))  - normalized to the multiplicity of $\pi^0$ .
We display HSD calculations for the ratio of the
minimal bias C+C, Ar+KCl, Au+Au collisions and the isospin-averaged
reference spectra $NN=(pn+pp)/2$  at the same energy.
}
\label{Fig_RDBr}
\end{figure}

In Fig. \ref{Fig_RDBr} we display the enhancement factor in heavy-ion
collisions for two different processes: Bremsstrahlung and $ \Delta$
Dalitz decay.   We show the ratio $(1/N_{\pi^0}^{AA}
dN^{AA}/dM)/(1/N_{\pi^0}^{NN} dN^{NN}/dM)$ of the dilepton yield from
HSD calculations of the minimal bias $A+A$ collisions: C+C, Ar+KCl,
Au+Au and of  the isospin-averaged reference spectra $NN=(pn+pp)/2$
at the same energy.  The upper part shows the contribution from
bremsstrahlung, the lower part that from the $\Delta$ Dalitz decay.

We do not expect that bremsstrahlung, one of the dominant sources at
beam energies around 1 $A$GeV, scales with the number of pions; therefore
the ratio should deviate from one. It has to be systematically larger
than one due to multiple collisions of incoming nucleons in heavy-ion
collisions. We see that the ratio depends on the mass but little on the
energy of the system. In Au+Au collisions where the number of
elementary collisions is large the enhancement can reach a factor of 3.
At higher energies the bremsstrahlung contribution is not really
settled because there are no reliable calculations for the
elastic and inelastic elementary channels.

\begin{table}[h!]
\caption{Ratio of $\pi^0$ mesons and the integrated dilepton yield
$(N(\Delta\to e^+e^-)=\int dM \frac{dN(\Delta\to e^+e^-)}{dM}) $
from $\Delta$ Dalitz
decays for C+C and Au+Au  at b=0.5 fm and 1 $A$GeV  and that from the
'elementary' $NN$ reactions for different scenarios: with/without  Fermi motion ('Fermi m.'),
with/without secondary meson-baryon collisions ('mB col.')}
\phantom{a}\vspace*{1mm}
\begin{tabular}{|c|c|c|c|c|c|c|c|c|}\hline
1&2&3&4&5&6&7&8\\
\hline
{Fermi m.}
& mB col. & \ system \ & \ $N(\pi^0)$ \
& $N(\Delta\to e^+e^-)$
& $R(\pi^0)={N^{AA}(\pi^0)\over N^{NN}(\pi^0)}$
&  $ R(e^+e^-)= {N^{AA}(\Delta\to e^+e^-)
     \over N^{NN}(\Delta\to e^+e^-)} $
& ${ R(e^+e^-)\over R(\pi^0)}={(7)\over(6)}$\\
 \hline
- & - & CC& 0.743 &  0.565$\times 10^{-4}$ & 6.74  & 5.56 & 0.83\\
- & - & AuAu& 18.76 & 1.688$\times 10^{-3}$ & 170.08  & 166.3 & 0.98\\
\hline
+ & - & CC& 1.407 &  1.16$\times 10^{-4}$ & 12.76  & 11.42 & 0.89\\
+ & - & AuAu& 31.07 & 2.75$\times 10^{-3}$ & 281.69 & 270.93 & 0.97\\
\hline
- & + & CC& 0.633 &  0.86$\times 10^{-4}$ & 5.74  & 8.47 & 1.47\\
- & + & AuAu& 10.75 & 3.45$\times 10^{-3}$ & 97.46  & 339.8 & 3.49\\
\hline
+ & + & CC& 1.07 &  1.77$\times 10^{-4}$ & 9.70  & 17.44 & 1.80\\
+ & + & AuAu& 16.62 & 6.32$\times 10^{-3}$ & 150.68  & 622.66 & 4.13\\
\hline
\end{tabular}
\end{table}

The other dominant source for dilepton production at beam energies
around 1 $A$GeV is $\Delta$ Dalitz decay. One may assume that the
$\Delta$ Dalitz decay scales with the number of pions because the
relative ratio is given by the branching ratio but this is not the
case. First of all, we are here in a threshold region where the Fermi
momentum only can lead to a substantial enhancement of the production.
Secondly, pions from $\Delta$ decay can be reabsorbed by nucleons and
can form again a $\Delta$ which may later disappear in a $\Delta N\to
NN$ collisions. This process is even important in systems as small as
C+C.  Dileptons, on the contrary, cannot be reabsorbed and are seen
in the detector. Table I shows quantitatively the consequences of
these processes for reactions at 1 $A$GeV. We compare there the pion and
dilepton yield for C+C and Au+Au for different conditions.  If
there is neither a Fermi momentum (Fermi m.) nor meson absorption on
baryons (mB col.) the ratio of $\pi^0$'s to dileptions corresponds to the
branching ratios and the enhancement factor (last column) is one, independent of the
system size of the heavy-ion reaction. The Fermi motion alone increases
the pion yield ($6^{th}$ column) as well as the dilepton yield ($7^{th}$ column) by almost a factor of two.
Because in the ratio displayed in Fig. \ref{Fig_RDBr} one
divides by the number of pions  this ratio remains one for small
invariant masses whereas the Fermi motion makes the ratio explode for
invariant masses close to the phase space boundary.  Meson-baryon interactions (mB coll)
lower the number of pions in heavy-ion collisions,  by  15\% in C+C
collisions and by 47\% in Au+Au collisions because they can lead to a disappearance
of the pions  if the $\pi N \to \Delta$ collision is followed by a $\Delta N \to NN$ collision.
At the same time they enhance the dilepton yield because dileptons do not get reabsorbed
and therefore every $\Delta$ which is produced contributes to the dilepton yield.
The meson-baryon interactions are therefore the reason that dileptons behave
differently than pions. This cycle of $\Delta$ production, $\Delta$ decay and $\pi$
reabsorption in $\pi N \to \Delta$ collisions, which leads in heavy system to the creation of several
generations of $\Delta$'s,  has been studied already 20 years ago as one
of the key elements to the pion dynamics in heavy-ion collision which
allows the pions to equilibrate with the system and to serve as a
measure of the number of participants \cite{Bass:1994ee}. The last two lines of
Table I  show that the pion absorption enhances the dilepton production as
compared to the pions by a factor of about 1.5-1.7 in C+C collisions
and by a factor of 3.5-4.1 for Au+Au collisions, i.e. the
enhancement grows with the size of the system.

\begin{figure}[h!]
\phantom{a}\vspace*{5mm}
\includegraphics[width=8.5cm]{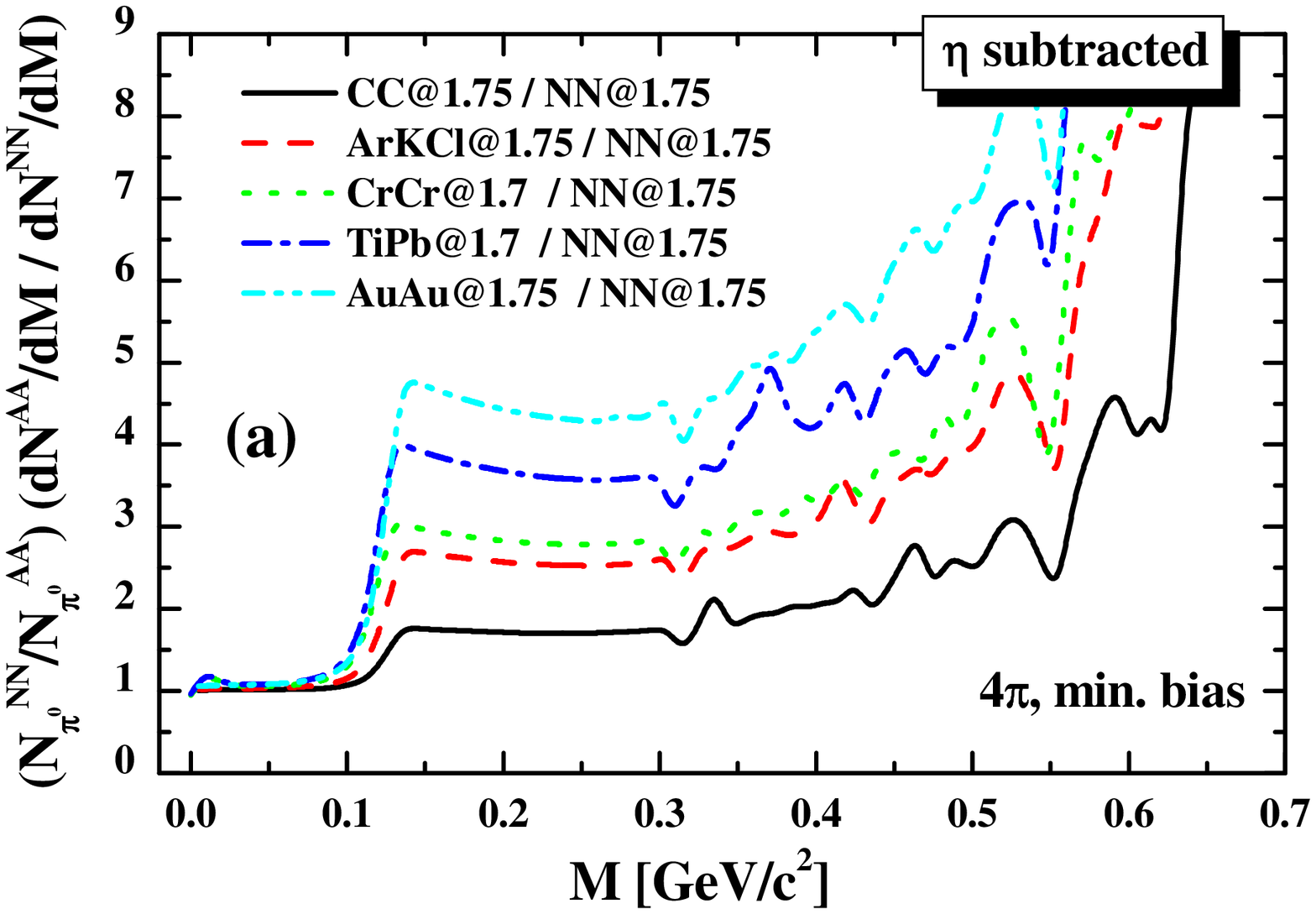}\hspace*{3mm}
\includegraphics[width=8.5cm]{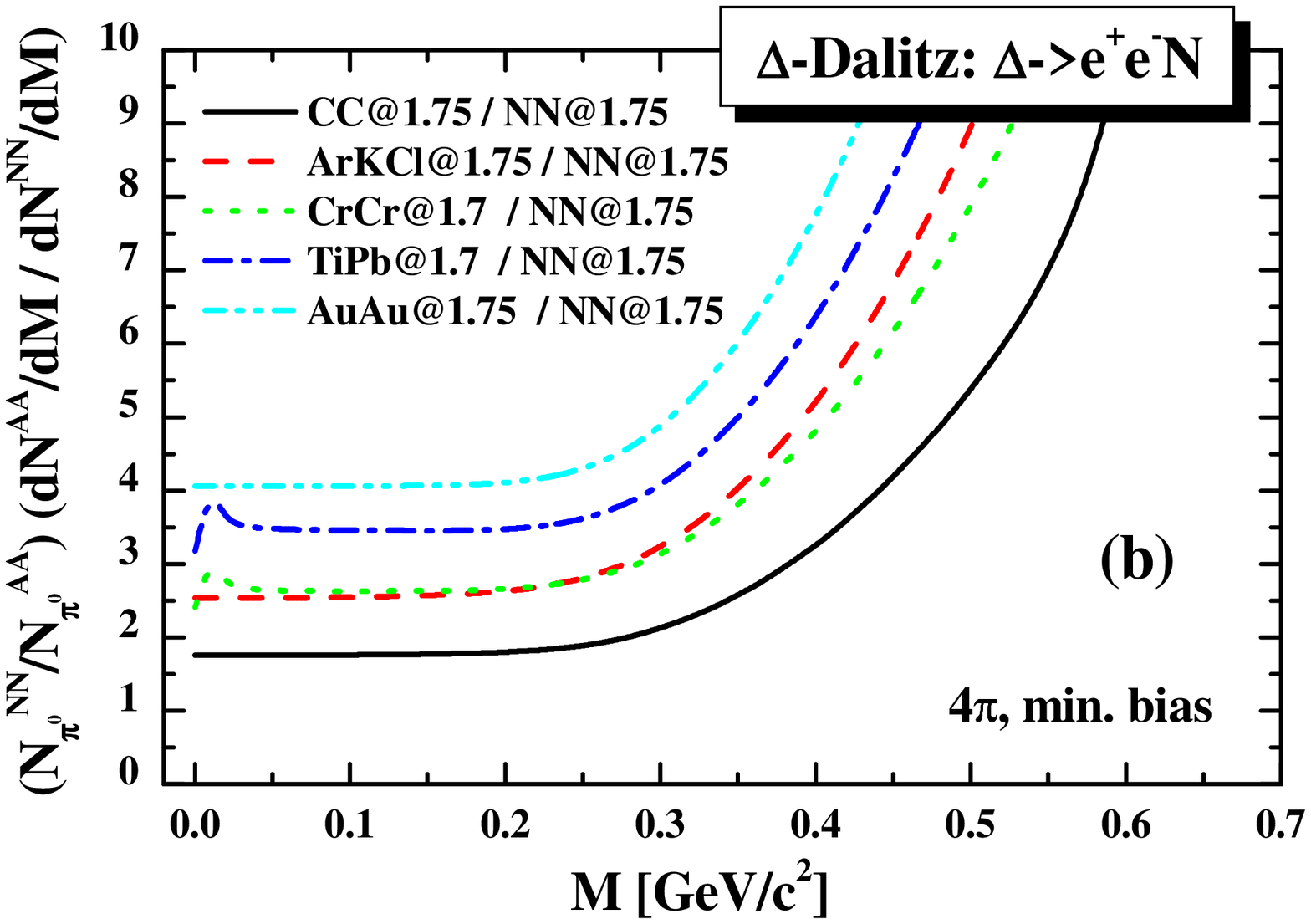}
\caption{(Color online) Left (a): The 4 $\pi$ ratio $(1/N_{\pi^0}^{AA}
dN^{AA}/dM)/(1/N_{\pi^0}^{NN} dN^{NN}/dM)$ of the mass differential
dilepton spectra - normalized to the $\pi^0$ multiplicity and after
$\eta$ Dalitz yield subtraction - from HSD calculations for the minimal bias
C+C, Ar+KCl, Cr+Cr, Ti+Pb and Au+Au collisions to the isospin-averaged reference
spectra $NN=(pn+pp)/2$  at  1.75 $A$GeV.  Right (b): same as the left plot but for the $\Delta$
Dalitz decay contributions, only.
}
\label{Fig_RAA175}
\end{figure}

The system size effect is demonstrated explicitly in Fig. \ref{Fig_RAA175}:
the left plot shows the ratio of the
mass differential dilepton spectra
$(1/N_{\pi^0}^{AA} dN^{AA}/dM)/(1/N_{\pi^0}^{NN} dN^{NN}/dM)$  - normalized to the
$\pi^0$ multiplicity and after $\eta$ Dalitz yield subtraction - from HSD calculations for the
minimal bias C+C, Ar+KCl, Cr+Cr, Ti+Pb, Au+Au collisions and of the
isospin-averaged reference spectra $NN=(pn+pp)/2$  at  1.75 $A$GeV.  The
right plot shows the same but for the $\Delta$ Dalitz decay contributions only.
We see also here that the different ratios are separated by a factor which is
(almost) independent of invariant mass and depends basically on the
size of the colliding nuclei since the effect of multiple $\Delta$ regeneration
increases with the atomic number of the colliding ions.

Thus, the dilepton enhancement observed in Fig. \ref{Fig_RAA} (and hence
also in the experimental spectra) is due to bremsstrahlung and due to
the $\Delta$ dynamics in the medium. Both are not related to
collective effects like the in-medium modifications of
spectral functions but are a mere consequence of the presence of other
nucleons in the nuclei. They also appear if no potential but only
collisional interactions between the nucleons exist. This effect
grows with the nuclear size which is directly related to an
increase of the high baryon density phase from light to heavy-ion collisions.
\begin{figure}[h!]
\phantom{a}\hspace*{-40mm}
\includegraphics[width=13cm]{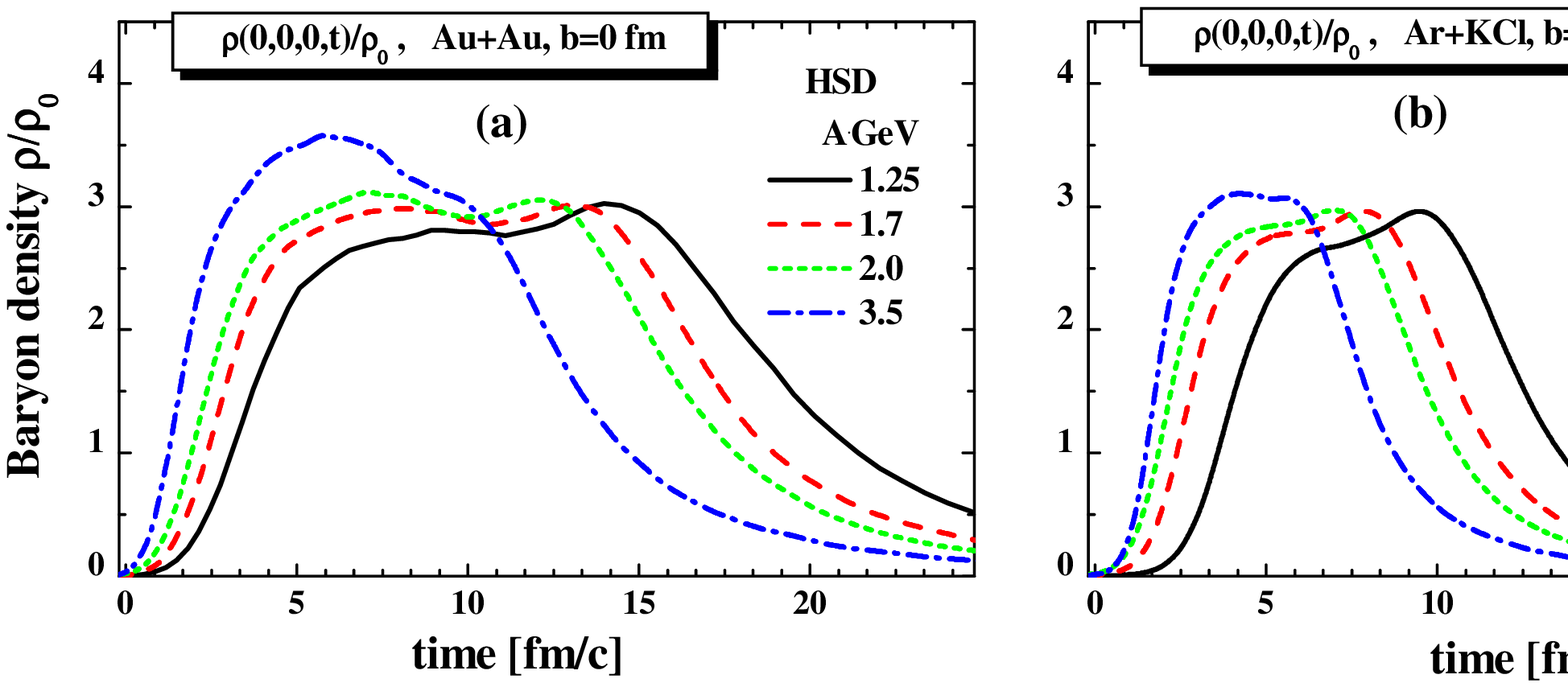}
\caption{(Color online) The time evolution of the baryon density from HSD in the central cell
$\rho(0,0,0,t)$ in units of the normal nuclear density $\rho_0=0.168~fm^{-3}$
for central ($b=0~fm$) Au+Au (left plot (a)) and Ar+KCl (right plot (b))
at different energies - 1.25, 1.7, 2.0 and 3.5 $A$GeV.
}
\label{Fig_denst}
\end{figure}
This is demonstrated in Fig. \ref{Fig_denst} which shows the time evolution
of the baryon density from HSD in the central cell $\rho(0,0,0,t)$ in units of
the normal nuclear density $\rho_0=0.168~fm^{-3}$
for central ($b=0~fm$) Au+Au (left plot) and Ar+KCl (right plot)
at different energies - 1.25, 1.7, 2.0 and 3.5 $A$GeV.
By comparing the Ar+KCl and Au+Au density profiles one sees that the maximum
density reached in the central cell is approximately the same in both cases -
up to $3\rho_0$ and only slightly grows with increasing energy.
However, the high baryon density phase for the heavy Au+Au nuclei collisions is much
longer than for the intermediate Ar+KCl system which implies a
longer reaction time and a stronger influence of secondary reactions
on observables as discussed above.


\subsection{In-medium effects in vector meson production}

Now we come to the question - how the in-medium effects in vector
meson production can influence the ratios.
The dilepton spectra for $p+Nb$ at 3.5 $A$GeV, C+C at 1.0, 2.0 $A$GeV and for Ar+KCl at
1.75 $A$GeV within the collisional broadening scenario for the
vector meson spectral functions have been already presented in
Sections III and IV (cf. Figs.
\ref{Fig_MpNb35},\ref{Fig_CC10},\ref{Fig_CC20},\ref{Fig_MArKCl})
in comparison to the HADES data as well as our predictions for
Au+Au at 1.25 $A$GeV (cf. Fig. \ref{Fig_MAu125}).

\begin{figure}[h!]
\includegraphics[width=9.cm]{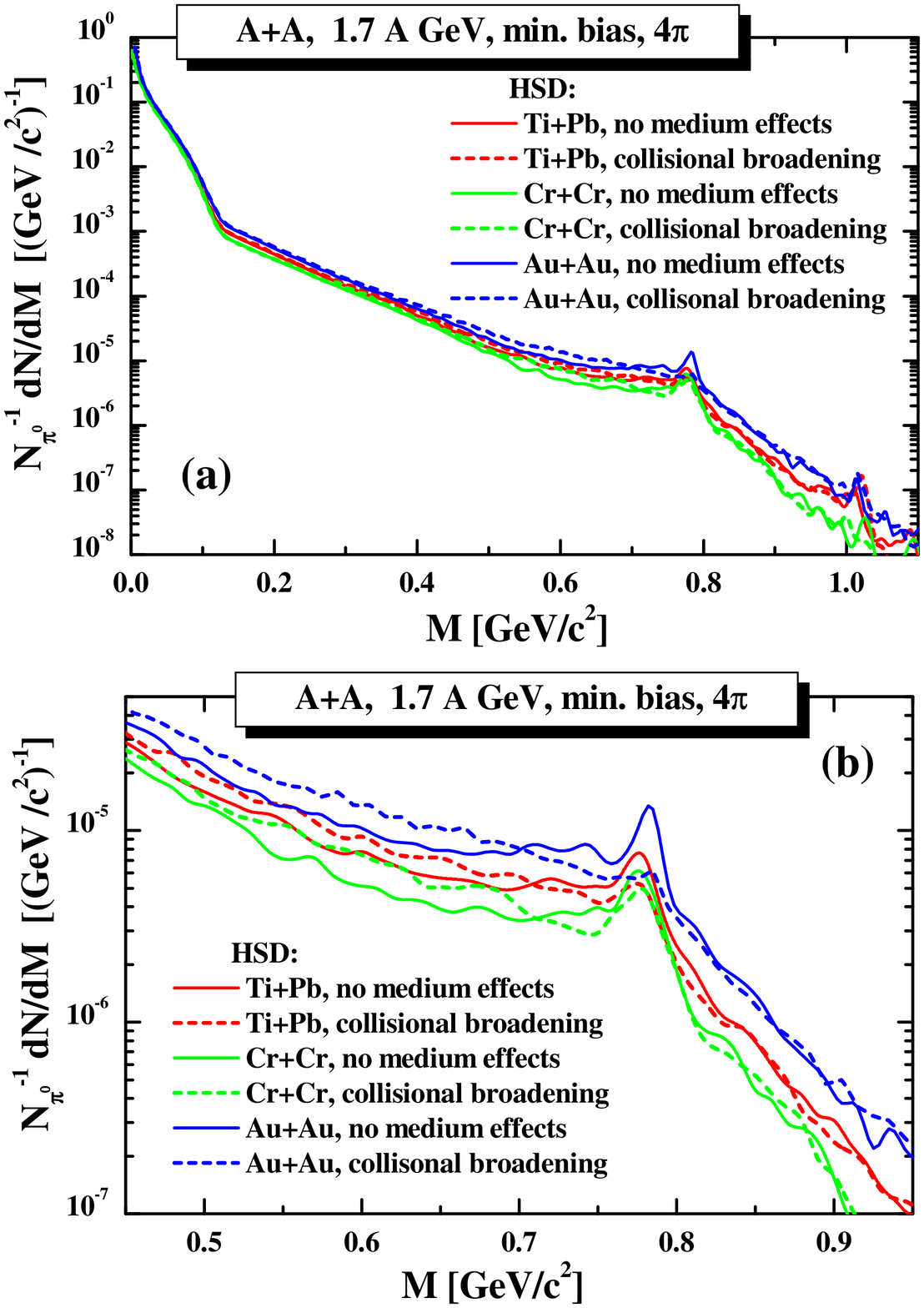}
\caption{(Color online) The 4 $\pi$ mass differential dilepton spectra - normalized to the
$\pi^0$ multiplicity - from HSD calculations for minimal bias Ti+Pb,
Cr+Cr and Au+Au collisions at 1.7 $A$GeV.  The solid
lines stand for the 'no medium effects' scenario whereas the dashed
lines show the dilepton yield for the 'collisional broadening'
scenario.  The lower plot (b) is a 'zoom' of the upper plot (a) for
the mass range $0.4<M<1.0 $ GeV.
}
 \label{Fig_MAA175}
\phantom{a}\vspace*{5mm}
\includegraphics[width=9cm]{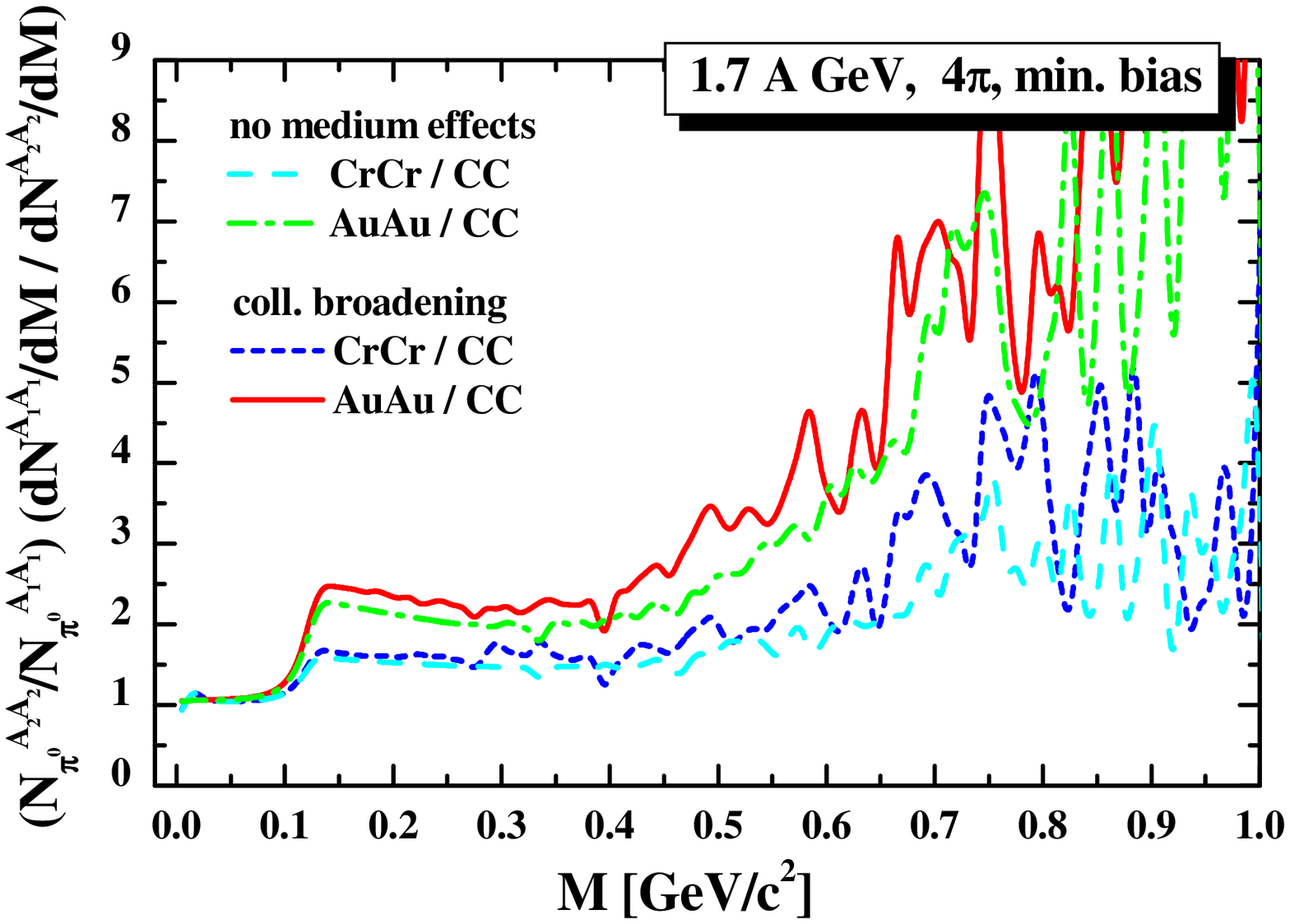}
\caption{(Color online) The 4 $\pi$ ratio $(1/N_{\pi^0}^{A-2A_2}
dN^{A_2A-2}/dM)/(1/N_{\pi^0}^{A_1A_1} dN^{A_1A_1}/dM)$ of the mass differential
dilepton spectra - normalized to the $\pi^0$ multiplicity  - from HSD
calculations for minimal bias Au+Au (Cr+Cr) collisions and C+C collisions.
This ratio is displaed for the 'no medium effects' and for
the 'collisional broadening' scenario. }
\label{Fig_R17}
\end{figure}

In Fig. \ref{Fig_MAA175} we display for reactions at 1.70 $A$GeV
the system size dependence of  the 4$\pi$ mass differential
dilepton spectra - normalized to the $\pi^0$ multiplicity - from HSD
calculations for minimal bias $A+A$ reactions. We display the result
for the symmetric Cr+Cr and  Au+Au systems as well as for the
asymmetric  Ti+Pb system.
The solid lines stand for the 'no medium effects' scenario whereas the
dashed lines show the dilepton yield for the 'collisional broadening'
scenario.  The lower plot is a 'zoom' of the upper one for the mass
range $0.4<M<1.0$  GeV.
First of all we note the growth of the dilepton yield for  $0.15 \le
M \le 0.6$ GeV when going from the intermediate Cr+Cr to the heavy
system Au+Au. The larger the system mass, the more
important is the aforementioned $\Delta$ reaction cycle and the more the dilepton
production is enhanced as compared to pion production.
As we have discussed already in Sections III and IV, for the
collisional broadening scenario one sees clearly the influence of the
larger width of the vector meson resonances (the peaks get smaller and broader).

What would be the consequence of this in-medium effect on the dilepton
ratio of $AA$ spectra to the 'reference spectrum'?  Would this
observable yield information on the underlying dynamical processes?
Previously we concentrated on the ratio $R(AA/NN)$ where the 'reference
spectrum' is constructed as an average of $pp$ and $pn$ yields:
$NN=(pp+pn)/2$. However, such a ratio would not be well suited for
studying in-medium effects in the vector meson mass region due to the
limited open phase space in $NN$ collisions relative to $AA$ collisions
- taken at the same energies - since the Fermi motion in $AA$ extends
the kinematical limits, which leads to a fast rise of  $R(AA/NN)$ at
larger invariant masses $M$.  Moreover, as has been discussed in
Section III.A, there is a general problem with $NN$ as a 'reference
spectrum' since experimentally  $pn$ are usually quasi-free $pd$
reactions.  For the beam energies discussed here, in the interesting
invariant mass region, $M>0.5$ GeV there are no 'quasi-free' $pn$
collisions anymore but genuine three-body $pd$ collisions.

Alternatively, the in-medium enhancement can be studied by comparing
the yield of a heavy system to that of a light system.  Fig
\ref{Fig_R17} displays  for a beam energy of 1.7 $A$GeV the ratio of
the invariant mass differential dilepton spectra for intermediate
Cr+Cr and heavy Au+Au nuclei  and of the light nuclei C+C, which
is chosen as a 'reference spectrum'. We study two scenarios -- the 'no
medium effects' and the 'collisional broading' scenario. One clearly
sees that the enhancement for $M\le 0.5$ GeV due to the multiple
$\Delta$ production and bremsstrahlung persists when one compares
collisions of heavy and light nuclei and can become as large as a
factor of two. Thus, C+C collisions can also be used as  'reference
spectra' to study such nuclear effects. Moreover, we observe as well
that the difference between the two scenarios is small for low
invariant masses and becomes only noticeable at invariant masses close
to the $\rho$ mass. However, even there the differences remain
moderate. Therefore high precision data are required to study the
question whether vector mesons are modified by the strongly interacting
medium in this energy region. On the other hand the ratio for AuAu/CC
grows much faster (for both - the no medium and the in-medium scenario)
than for CrCr/CC. This is due to the enhancement of the vector meson
productions by secondary meson-baryon and meson-meson interactions in
heavy system relative to the light system. This effect is hence easy to
observe experimentally.

\section {Uncertainties due to  different assumptions in the transport models}

In this section we discuss the different assumptions in different transport approaches
due to the lack of experimental information and theoretical knowledge
and  the consequences for the prediction
of these approaches. The uncertainties related to the production cross sections
in elementary reactions have been addressed already in Section II.
There are, however,  other sources of uncertainties, in particular for the
dilepton production by $\Delta$ Dalitz decay  - the lack of knowledge
of the electromagnetic decay width of the $\Delta$ resonance,  of the
mass distribution of the $\Delta$ resonance in elementary $NN$ collisions and of its
total decay width as well as different assumptions on how the total decay width is related
to the $\Delta$ life time.

\subsection {Electromagnetic decay width of $\Delta$ resonance}

The differential electromagnetic decay width of a $\Delta$ resonance into dileptons  of an invariant mass M,
$\Delta \to N  e^+e^-$, can be related to the Delta decay into a nucleon and a virtual photon,
$\Delta \to N\gamma^*$, by (cf. \cite{Wolf90}):
\begin{eqnarray}
 {d\Gamma\over dM}^{\Delta\to N \ l+l-}(M) = {2\alpha\over 3\pi}
{\Gamma^{\Delta\to N \gamma^*}(M,M_\Delta)\over M}, \label{DalDel}
\end{eqnarray}
where $\alpha=1/137$ and  $M_\Delta$ is the current mass of the
$\Delta$-resonance.
Unfortunately there is no direct measurement of the $\Delta \to N\gamma^*$
width and starting from the pioneering work of Jones and  Scadron \cite{Jones73}
there is a series of different models \cite{Wolf90,Ernst:1997yy,Krivor02,ZatWolf03}.
In the  present versions of the HSD, IQMD and UrQMD transport approaches
the "Wolf" model is employed for the electromagnetic decay width \cite{Wolf90} :
\begin{eqnarray}
&& \Gamma^{\Delta\to N \gamma^*}(M,M_\Delta) = {\lambda^{1/2}(M^2,m_N^2,M_\Delta^2) \over
       16 \pi M_\Delta^2} \cdot m_N
       \cdot [ 2m_T(M,M_\Delta) +m_L(M,M_\Delta) ] \nonumber\\
&& m_L(M, M_\Delta) = (e f g)^2 {M_\Delta^2 \over 9 m_N} M^2 \cdot
       4 (M_\Delta -m_N -q_0), \ \ e^2=4\pi\alpha, \ \ g=5.44 \nonumber\\
&& m_T(M, M_\Delta) = (e f g)^2 {M_\Delta^2 \over 9 m_N}
   \left[q_0^2 (5M_\Delta -3(q_0+m_N)) -M^2(M_\Delta+m_N+q_0)\right] \nonumber \\
&& f = -1.5 {M_\Delta +m_N \over m_N ((m_N+M_\Delta)^2-M^2)} \nonumber \\
&& q_0 = (M^2+p_f^2)^{1/2} \label{DalDel1} \\
&& p_f^2={(M_\Delta^2 -(m_N+M)^2)(M_\Delta^2 -(m_N-M)^2) \over 4 M_\Delta^2}
       \nonumber\\
&& \lambda (M^2,m_N^2,M_\Delta^2) = M^4+m_N^4+M_\Delta^4 - 2 (M^2 m_N^2
   +M^2 M_\Delta^2 + m_N^2 M_\Delta^2). \nonumber
\end{eqnarray}

\begin{figure}[t!]
\phantom{a}\hspace*{-45mm}
\includegraphics[width=12.5cm]{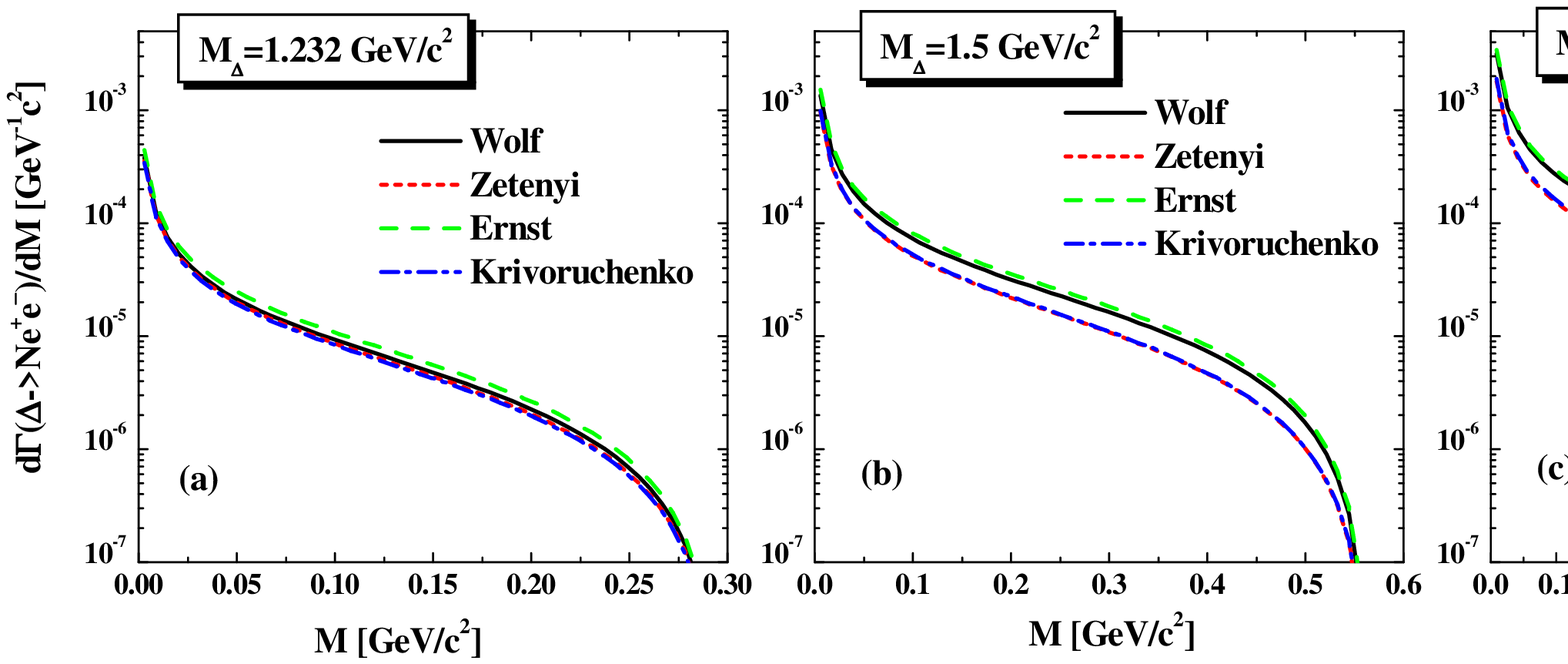}
\caption{(Color online) The electromagnetic decay width of $\Delta$ resonance
to dileptons $\Delta \to N  e^+e^-$ for different models denoted
as "Wolf" \cite{Wolf90}, "Zetenyi" \cite{ZatWolf03}, "Ernst" \cite{Ernst:1997yy}
and "Krivoruchenko" \cite{Krivor02} for different Delta masses of 1.232 GeV (a),
1.5 GeV (b) and 1.8 GeV (c).
}
\label{Fig_Gamee}
\end{figure}

There is a variety of models for the  electromagnetic decay width of $\Delta$ resonance
to dileptons $\Delta \to N  e^+e^-$  - cf. \cite{Wolf90,ZatWolf03,Ernst:1997yy,Krivor02}.
Figure \ref{Fig_Gamee} shows $\Gamma^{\Delta\to N \ e+e-}(M,M_\Delta)$ for different models
denoted as "Wolf" \cite{Wolf90}, "Zetenyi" \cite{ZatWolf03}, "Ernst" \cite{Ernst:1997yy}
and "Krivoruchenko" \cite{Krivor02} and for three different Delta masses --  1.232 GeV (a),
1.5  GeV (b) and 1.8 GeV (c).
One can see that in the low mass region, i.e. around the
$\Delta$ pole mass 1.232, all approaches give similar results whereas with
increasing $M_\Delta$ the differences grow. The models "Wolf" and "Ernst"
lead to a similar dilepton yield which is, however,  up to a factor of 3 higher then
that from the models "Krivoruchenko" and "Zatenyi". This introduces a systematic
error for the prediction of the dilepton yield for large mass dileptons.

\subsection {Total decay width and the life-time of $\Delta$ resonance }

The population of high mass Delta's in NN reactions depends on the shape of
the differential mass distribution which is given by the $\Delta$ spectral
function. The spectral function of a $\Delta$ resonance of mass $M_\Delta$
is usually assumed to be of the relativistic Breit-Wigner form:
\begin{eqnarray}
A_\Delta(M_\Delta) = C_1\cdot {2\over \pi} \ {M_\Delta^2 \Gamma_\Delta^{tot}(M_\Delta)
\over (M_\Delta^2-M_{\Delta 0}^2)^2 + (M_\Delta {\Gamma_\Delta^{tot}(M_\Delta)})^2}.
\label{spfunD}
\end{eqnarray}
with $M_{\Delta 0}$ being the pole mass of the $\Delta$.
The factor $C_1$ is fixed by the normalization condition:
\begin{eqnarray}
\int_{M_{min}}^{M_{lim}} A_\Delta(M_\Delta) dM_\Delta =1,
\label{SFnorma}\end{eqnarray} where $M_{lim}=2$~GeV is chosen as
an upper limit for the numerical integration. The lower limit for
the vacuum spectral function corresponds to the nucleon-pion decay,
$M_{min}=m_\pi+m_N$. In  $NN$ collisions the
Deltas can be populated up to the $M_{max}=\sqrt{s}-m_N$ and hence
the available part of spectral function is defined by the
beam energy.

The shape of spectral function (and correspondingly the production
of high mass Delta's)  depends strongly on the total width $\Gamma_\Delta^{tot}$.
Due to the lack of experimental information this total width has to be assumed
and different parametrizations exist.

For the present HSD calculations we adopt the "Monitz" model \cite{Monitz}
(cf. also Ref. \cite{Wolf90}) :
\begin{eqnarray}
\Gamma_\Delta^{tot} (M_\Delta)&=& \Gamma_R {M_{\Delta 0} \over M_\Delta}
       \cdot \left(q\over q_r\right)^3 \cdot F^2(q), \label{WidthDel}\\
&& q^2={(M_\Delta^2 -(m_N+m_\pi)^2)(M_\Delta^2 -(m_N-m_\pi)^2)
       \over 4 M_\Delta^2}, \nonumber \\
&&  \Gamma_R = 0.11 {\ \rm GeV}, \ \ M_{\Delta 0} = 1.232 {\ \rm GeV}; \nonumber \\
&& F(q) ={\beta_r^2 +q_r^2 \over \beta_r^2 +q^2}, \label{WidthDel1} \\
&&  q_r^2 = 0.051936, \ \ \beta_r^2 = 0.09.\nonumber
\end{eqnarray}
In the UrQMD model one employs  the "Bass" parametrization \cite{UrQMD1}
which differs from the "Monitz" model (\ref{WidthDel}) by the
formfactor (\ref{WidthDel1}):
\begin{eqnarray}
 F_{B} (q) = 1.2 \ {\tilde\beta_r^2 \over \tilde\beta_r^2 +q^2}, \ \ \ \ \
 \tilde\beta_r^2=q_r^2/0.2  \label{FqUrQMD}.
\end{eqnarray}

\begin{figure}[h!]
\includegraphics[width=9.2cm]{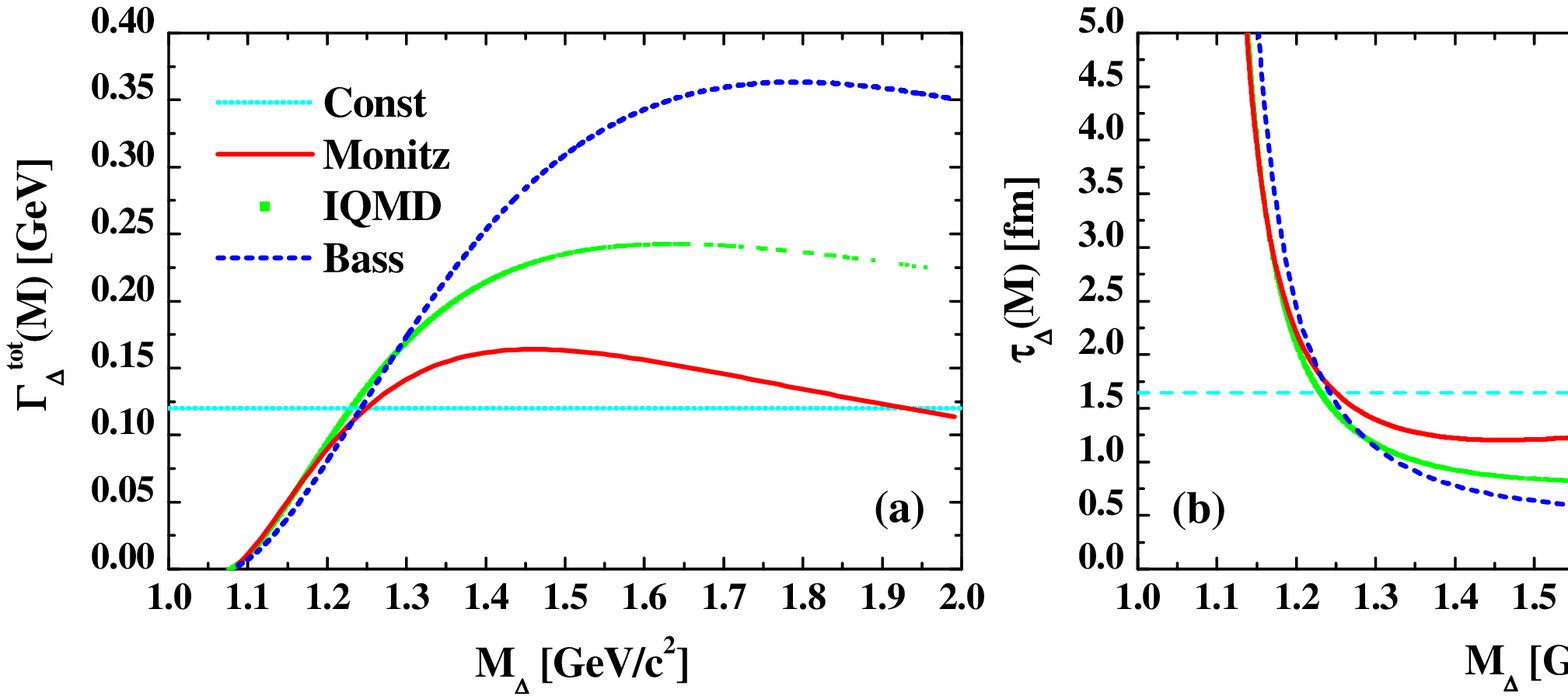}\hspace*{25mm}
\includegraphics[width=6cm]{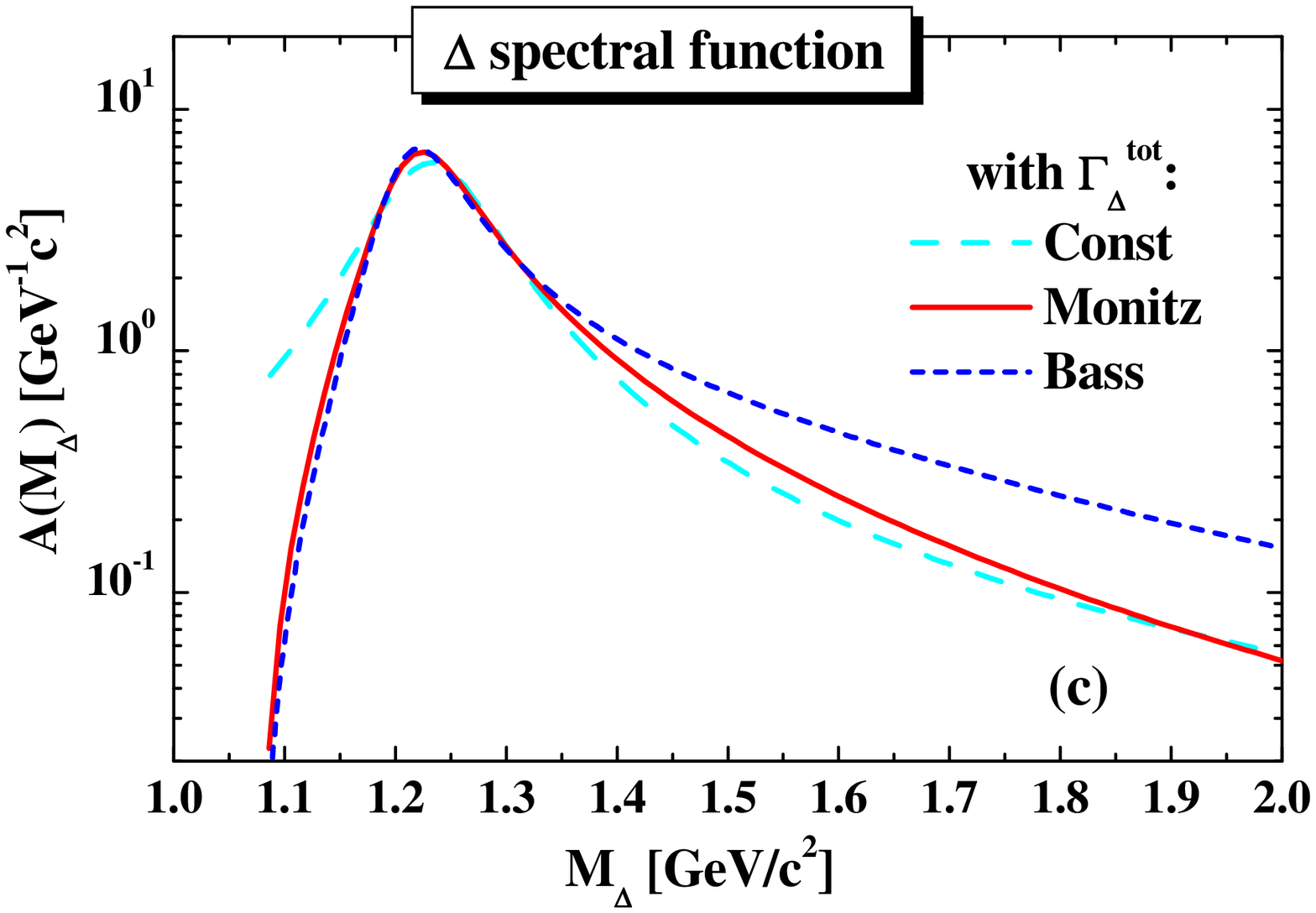}
\caption{(Color online)  The mass dependence of the total width $\Gamma_\Delta^{tot}
(M_\Delta)$ (left plot (a)), life time (middle plot (b)) and the
spectral function (right plot (c))
from different models: "Const" - the constant width $\Gamma_{\Delta 0}^{tot}=0.12$
GeV, "Monitz" - from  Eq. \ref{WidthDel} (cf. \cite{Monitz}) , "Bass" -
from \ref{FqUrQMD} (cf. \cite{UrQMD1}), "IQMD" - from
\cite{Hartnack:1997ez,Hartnack:2011cn}.
}
\label{Fig_Gamt}
\end{figure}

In the left part (a) of Fig. \ref{Fig_Gamt} we show the  mass dependence of the
total width $\Gamma_\Delta^{tot} (M_\Delta)$
from different models: "Const" - a constant width $\Gamma_\Delta^{tot}=0.12$
GeV, "Monitz" - from  Eq. \ref{WidthDel} (cf. \cite{Monitz}) , "Bass" -
from \ref{FqUrQMD} (cf. \cite{UrQMD1}) as well as the parametrization used in the
IQMD model \cite{Hartnack:1997ez,Hartnack:2011cn} denoted as "IQMD".
We observe substantial differences between the models, especially
for large mass $\Delta$. These differences become more important at higher energies.
For lower energies, especially for the 1 $A$GeV data, phase space limits the
$\Delta$ masses to $M_\Delta < 1.4$~GeV.

The total decay width is related to the life time of the resonances
 - an other important quantity for the transport approaches -
by
\begin{eqnarray}
\tau_\Delta(M_\Delta) ={\hbar c \over \Gamma_\Delta^{tot} (M_\Delta) }. \label{tau}
\end{eqnarray}
The lifetime  as a function of $M_\Delta$ is illustrated in the middle part (b)
of Fig. \ref{Fig_Gamt}. The lifetime of
large mass $\Delta$ is in the  "Bass" parametrization up to three times lower than in the
Monitz parametrization. $\Delta$'s of such a high mass
are rare, however, as can be see from the right part (c) of Fig. \ref{Fig_Gamt}
which shows the mass dependence of the spectral
function for different parametrizations of the width.

\subsection {Consequences for the dilepton yield}

Now we show how the uncertainties in the modelling of the total $\Delta$ width
and of the electromagnetic decay width affect the final results for the dilepton yield.

\subsubsection{Convolution model}

We start out with a simple example: the dilepton yield from the $\Delta$ Dalitz decay
is a convolution of the mass distribution of the $\Delta$ resonances
- which we take for our model study to be defined by the spectral function
$A_\Delta(M_\Delta)$ (Eq.(\ref{spfunD})) - and the $\Delta$ mass
dependent branching ratio for the electromagnetic decay into dileptons
which is defined as a ratio of electromagnetic partial width
${d\Gamma\over dM}^{\Delta\to Ne^+e^-}\!\!\!(M,M_\Delta)$
and  the total width $\Gamma_\Delta^{tot}(M_\Delta)$:
\begin{eqnarray}
{dN\over dM}^{e^+e^-}\!\!\!\!\!\!\!\!\!\!(M)
&& =\int dM_\Delta \cdot A_\Delta(M_\Delta) \cdot
{d\Gamma\over dM}^{\Delta\to Ne^+e^-}\!\!\!\!\!\!\!\!\!\!\!\!\!\!\!\!\!\!\!\!(M,M_\Delta)
\cdot {1 \over \Gamma_\Delta^{tot}
(M_\Delta)}  \label{dNdMee} \\
&& =\int dM_\Delta \cdot A_\Delta(M_\Delta) \cdot
{d\Gamma\over dM}^{\Delta\to Ne^+e^-}\!\!\!\!\!\!\!\!\!\!\!\!\!\!\!\!\!\!\!\!(M,M_\Delta)
\cdot {\tau_\Delta(M_\Delta)}. \label{dNdMee1}
\end{eqnarray}
where the  expression  (\ref{dNdMee}) has been re-written in terms of the
$\Delta$ life time using relation (\ref{tau}).

\begin{figure}[h!]
\includegraphics[width=8.5cm]{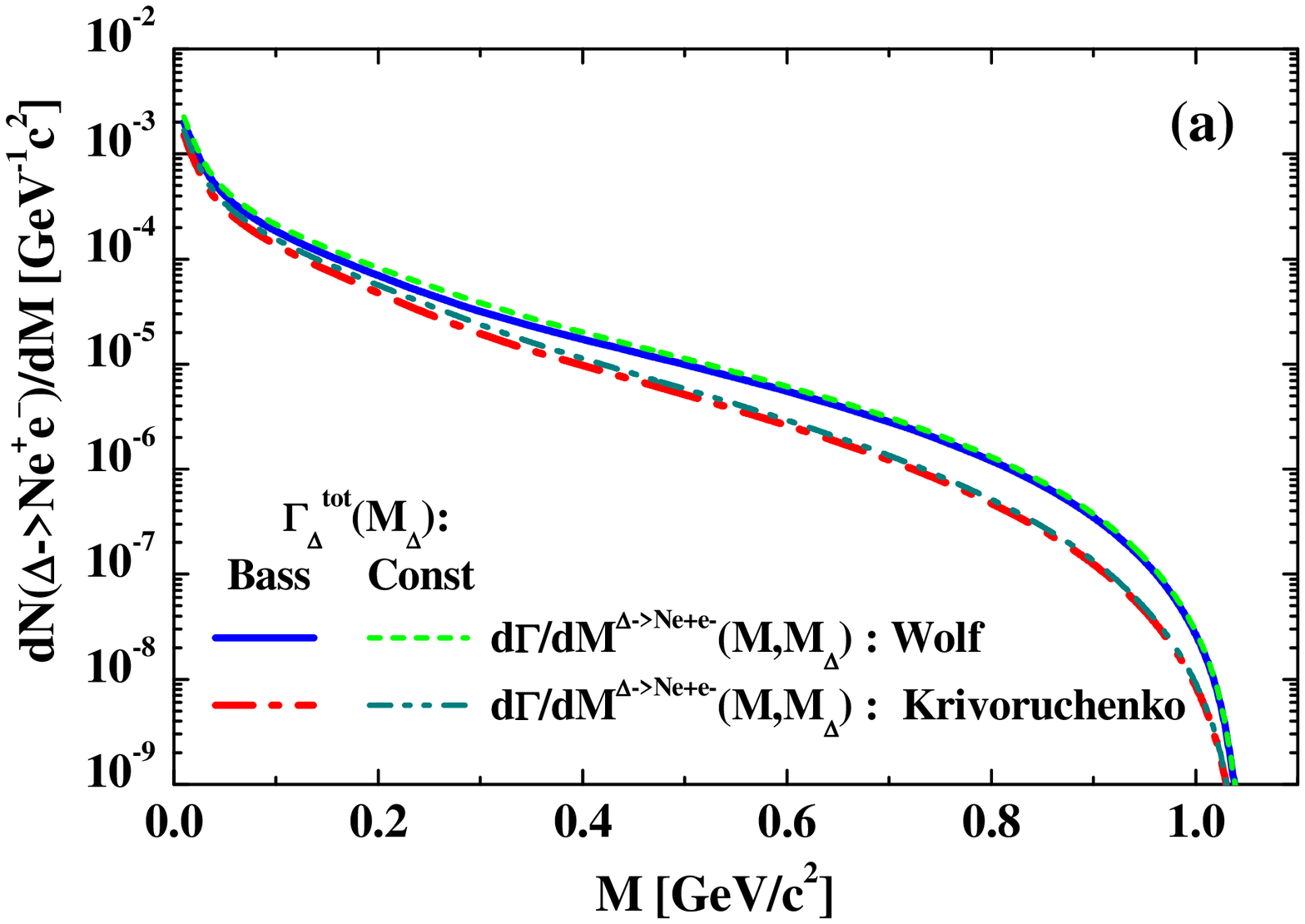}\hspace*{5mm}
\includegraphics[width=8.5cm]{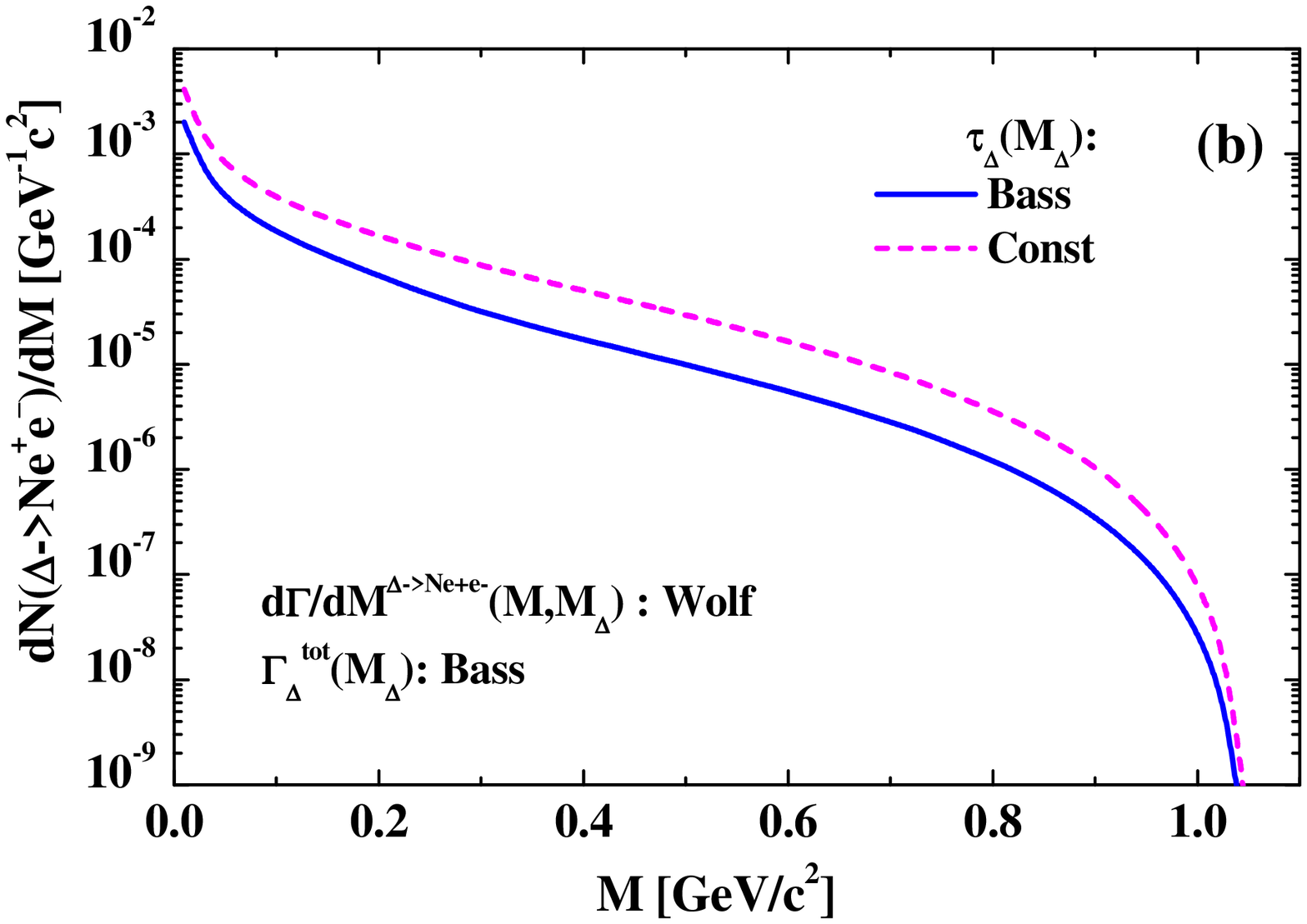}
\caption{(Color online) Left(a): the dilepton yield as a function of invariant dilepton mass
for the 2 parametrization of the total width $\Gamma_\Delta^{tot}(M_\Delta)$
2 models for the partial electromagnetic decay width
 ${d\Gamma\over dM}^{\Delta\to Ne^+e^-}\!\!\!(M,M_\Delta)$:
1) solid line: total width - "Bass", electromagnetic - "Wolf";
2) dot-dashed line: total width - "Bass", electromagnetic - "Krivoruchenko"
3) dashed line: total width - "Const", electromagnetic - "Wolf"
4) dot-dot-dashed line: total width - "Const", electromagnetic - "Krivoruchenko".
Right(b): the dilepton yield as a function of invariant dilepton mass
for the 2 assumptions of the life time $\tau_\Delta(M_\Delta)$:
solid line - "Bass", dashed line - "Const" life time,
while using the "Bass" total width  for spectral function
and the "Wolf" model for the partial electromagnetic decay width in both cases.}
\label{Fig_dNdM}
\end{figure}

In the left side (a) of Fig. \ref{Fig_dNdM} we show the dilepton yield
for two different assumptions for $\Gamma_\Delta^{tot}(M_\Delta)$ and for
 ${d\Gamma\over dM}^{\Delta\to Ne^+e^-}\!\!\!(M,M_\Delta)$:
1) solid line: total width - "Bass", electromagnetic - "Wolf";
2) dot-dashed line: total width - "Bass", electromagnetic -
"Krivoruchenko"; 3) dashed line: total width - "Const", electromagnetic -
"Wolf"; 4) dot-dot-dashed line: total width - "Const", electromagnetic -
"Krivoruchenko".
The variation of $\Gamma_\Delta^{tot}(M_\Delta)$ changes
the dilepton yield only marginally as long as the same electromagnetic decay
width is used - cases 1),3) and 2),4). The reason can easily be  seen from Eq.
(\ref{dNdMee}) : the total width $\Gamma_\Delta^{tot}(M_\Delta)$
enters in the numerator of spectral function   (Eq. \ref{spfunD})
and in the denominator of the branching ratio and thus cancels.
The only remaining dependence comes from the denominator of
Eq. (\ref{spfunD}) but far from the pole mass $\Gamma_\Delta^{tot}(M_\Delta)$
this term is small as compared to the other part of the denominator.
Oppositely, for a fixed total width $\Gamma_\Delta^{tot}(M_\Delta)$
the variation of ${d\Gamma\over dM}^{\Delta\to Ne^+e^-}\!\!\!(M,M_\Delta)$
leads to differences of the  dilepton yield up to the factor of 3
for high invariant masses - cases 1),2) and 3),4).

Thus we can conclude that different assumptions on the total width
$\Gamma_\Delta^{tot}(M_\Delta)$  have little influence on the
invariant mass distribution of dileptons whereas the lack of
knowledge of  ${d\Gamma\over dM}^{\Delta\to Ne^+e^-}\!\!\!(M,M_\Delta)$
introduces an uncertainty of up to a factor of three for the dilepton
yield from $\Delta$ decay at large invariant masses.

Different assumptions have been made of how to relate $\Gamma_\Delta^{tot}(M_\Delta)$
to the lifetime of the $\Delta$. We do not discuss here the rational behind
the different approaches. Rather we concentrate on the consequences for the dilepton yield.
In HSD the total width for the $\Delta$ production (i.e. that which enters
the spectral function $A_\Delta(M_\Delta)$) is the same as the width used to
determine the life time  Eq. (\ref{tau}). In this case, we have a cancellation
of the total width in Eq. (\ref{dNdMee}) as discussed above, which leads to the
low sensitivity of dilepton spectra to different $\Gamma_\Delta^{tot}(M_\Delta)$ .
In the UrQMD model (cf. e.g. the corresponding discussion in  Ref. \cite{UrQMD1},
Section 3.3.4), on the contrary, the width used for the $\Delta$ production
differs from that in the life time definition (\ref{tau}), so there is no
cancellation of the widths any more, rather the ratios of the two  widths
enters the  Eq. (\ref{dNdMee}).

\begin{figure}[h!]
\includegraphics[width=8.5cm]{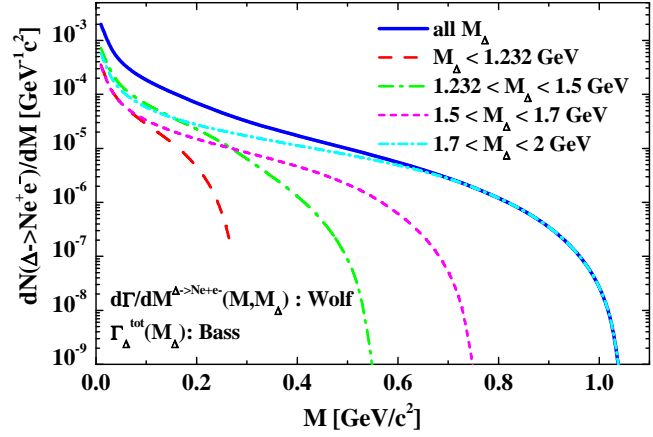}\hspace*{5mm}
\caption{(Color online) The contribution of $\Delta$'s from the 4 mass bins
to the dilepton yield  $dN/dM(\Delta\to Ne^+e^-)$:
1) $M_\Delta \le 1.232$ GeV,
2) $  1.232 \le M_\Delta \le 1.5$ GeV,
3) $  1.5 \le M_\Delta \le 1.7$ GeV,
4) $  1.7 \le M_\Delta \le 2.0$ GeV.
Calculations are done using the "Bass" total width for spectral function
and "Wolf" model for the partial electromagnetic decay width.
}
\label{Fig_Dal_MD}
\end{figure}

In the right part (b) of Fig. \ref{Fig_dNdM} we demonstrate the
consequences of the different life time definitions. We employ
in all cases the "Wolf" parametrization of  ${d\Gamma\over
dM}^{\Delta\to Ne^+e^-}\!\!\!(M,M_\Delta)$ and
$\Gamma_\Delta^{tot}(M_\Delta)$  of "Bass" but vary the
description of the life time. The full blue line shows the
dilepton yield under the assumption that
$\Gamma_\Delta^{tot}(M_\Delta)$ of "Bass" determines the lifetime
(Eq. (\ref{tau})) whereas the dashed red line shows the result
assuming that for the calculation of the life time (Eq. (\ref{tau}))
a constant width of 120 MeV is employed. For a constant width we observe a strong enhancement which
is mainly related to the large contribution of the high mass
$\Delta$'s to the  dilepton yield.
This is illustrated in
Fig. \ref{Fig_Dal_MD} where we show the contribution of
$\Delta$'s from different mass ranges to the dilepton yield.
The sum of all 4 bins gives the solid curve of fig.
\ref{Fig_dNdM}. One has to keep in mind, however, that in real
$NN$ collisions at low energies the high mass tail of the $\Delta$
distribution is strongly suppressed due to the limitation of the
phase space.

\subsubsection{$pp$ and heavy-ion collisions}

Now we extend our study  of systematic errors to $pp$ and heavy-ion
calculations. For this purpose we use the HSD model.

\begin{figure}[h!]
\includegraphics[width=8.5cm]{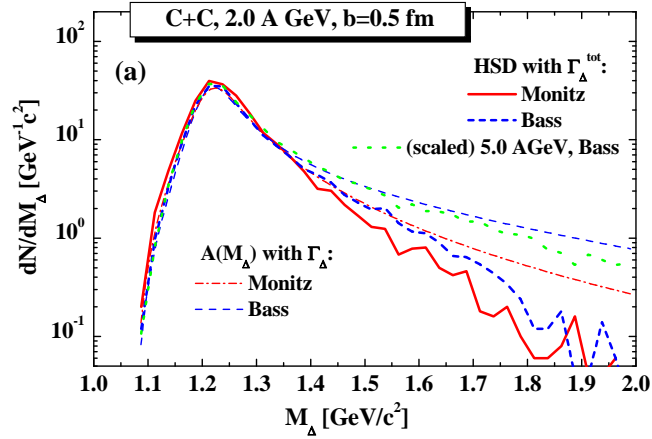}\hspace*{5mm}
\caption{(Color online) The $\Delta$ mass distribution from HSD for the central
C+C collisions at 2 $A$GeV for the 2 model cases for the total
width:
solid line - "Monitz", short dashed - "Bass" width. The thin
dash-dotted and dashed lines show the spectral function
$A_\Delta(M_\Delta)$ calculated with "Monitz" and "Bass" widths,
correspondingly. The dotted line stands for $dN/dM_\Delta$
from  HSD for C+C collisions at 5.0 $A$GeV with "Bass" width
(scaled to the maximum of $dN/dM_\Delta$  at
2 $A$GeV for easy comparison of the shape of mass distributions)}.
\label{Fig_dNdMD}
\end{figure}

The $\Delta$ resonances can be produced dominantly in $NN$ or $\pi N$
collisions. The mass distribution of the produced $\Delta$'s
-- $dN/dM_\Delta(s,M_\Delta)$ is defined by the spectral function
$A_\Delta(M_\Delta)$ (Eq.(\ref{spfunD})) integrated over the
corresponding phase space which depends on the
invariant energy $\sqrt{s}$ of the $NN$ or $\pi N$ collisions and the masses
of the final associated particles $M_X$ (e.g. $NN\to \Delta+X$).
At low energies the phase space leads to the suppression of high mass $\Delta$'s.

We start with the time integrated $\Delta$ mass distribution $dN/dM_\Delta$.
It is shown in Fig. \ref{Fig_dNdMD} for central C+C collisions at 2$A$GeV.
We display the mass distribution  for 2 choices  for the total width:
The solid line displays the calculation for the  "Monitz" width, the
short dashed line that for the  "Bass" width.
For comparison we also show the spectral function
$A_\Delta(M_\Delta)$ (scaled to the maximum of $dN/dM_\Delta$)
for the both widths.
Due to the limited available energy in low energy heavy-ion collisions
only a part of the full spectral function can be explored
(the absorption and rescattering effects for C+C
collisions do not distort the initial production shape of
$\Delta$ mass distribution too much). This lowers
the uncertainties of the predicted dilepton yields related to the very
high mass tail of the distribution.
Going to higher energies the phase space opens more and more
and  the high mass tail of spectral function  can be populated. This is shown
by the dotted line which displays  $dN/dM_\Delta$
for C+C collisions at 5.0 $A$GeV employing the  "Bass" width
which we scaled to the maximum of  $dN/dM_\Delta$ for C+C at 2 $A$GeV
for easy comparison of the shape of corresponding mass distributions.

\begin{figure}[h!]
\centerline{\includegraphics[width=8.6cm]{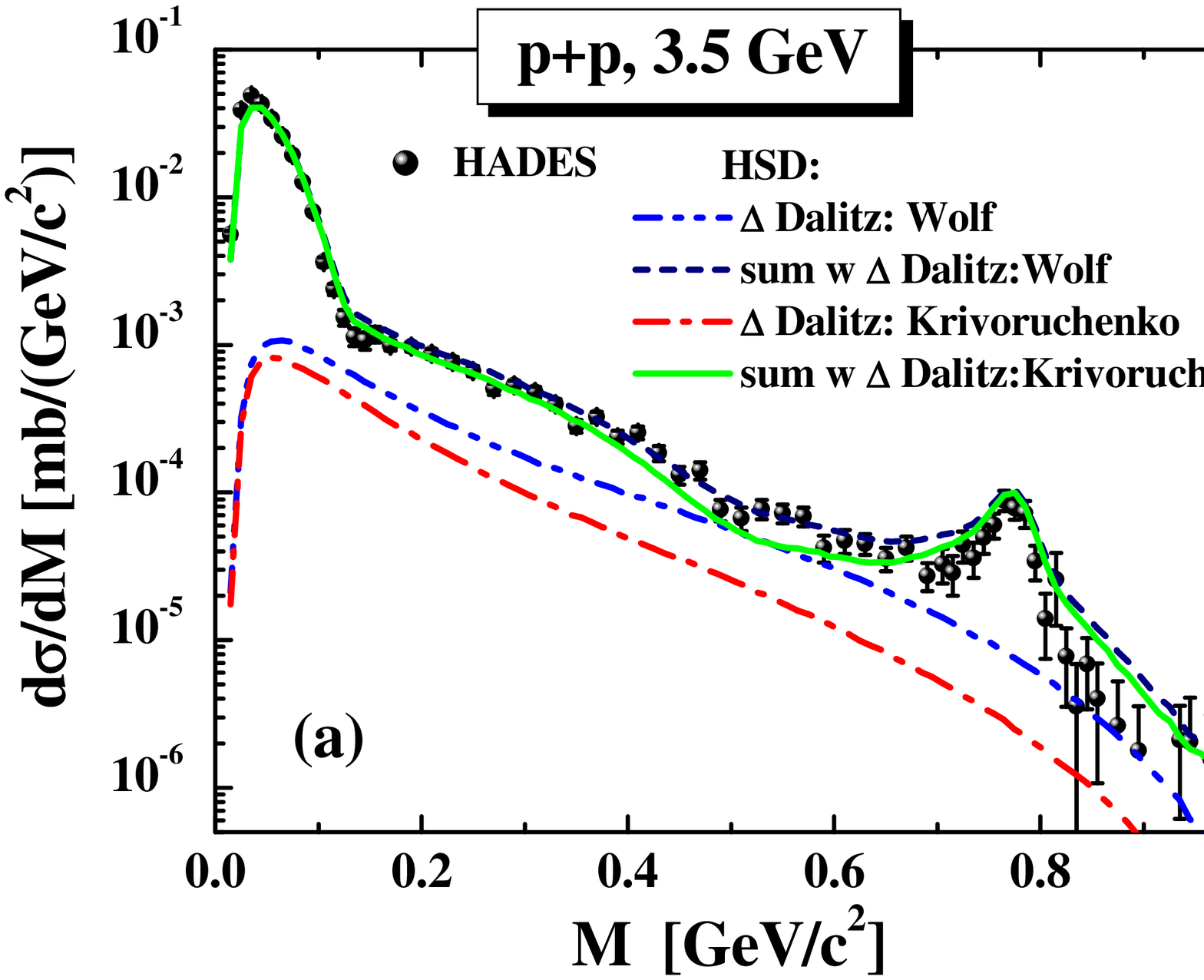}}
\vspace*{2mm} \includegraphics[width=8.6cm]{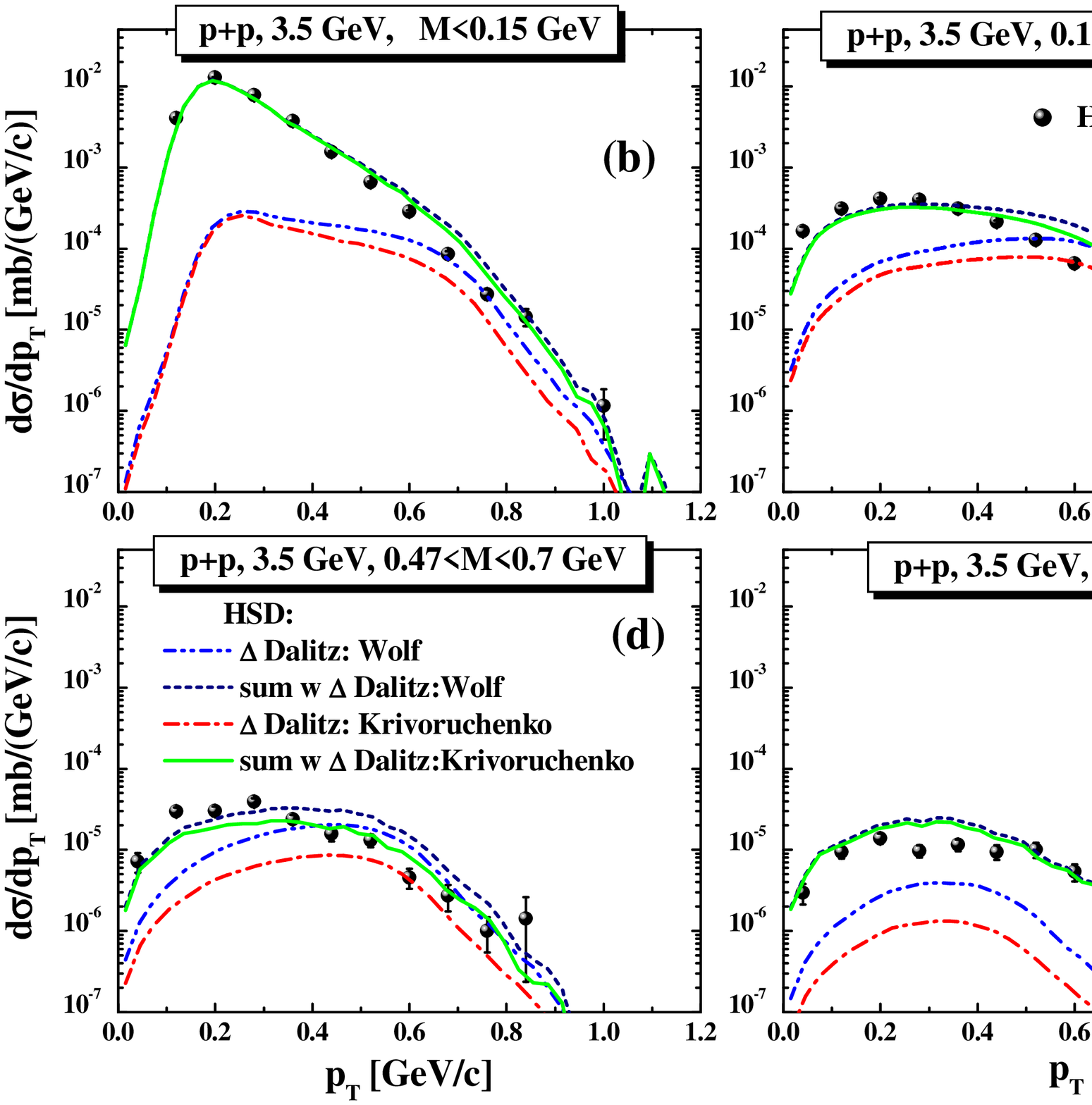}
\caption{(Color online) Left (a):
The differential  cross section $d\sigma/dM$ from HSD calculations for
$e^+e^-$ production in $pp$ reactions at a bombarding energy of 3.5 GeV
in comparison to the HADES data \cite{HADES_pp35}.
Right(b-e):
The HSD results for the transverse momentum spectra  for $pp$ at 3.5 GeV and for 4
different mass bins:  $M \leq$ 0.15 GeV,  0.15 $\leq M \leq$ 0.47 GeV,
0.47 $\leq M \leq$ 0.7 GeV and  $M \geq$ 0.7 GeV in comparison to the
HADES data \cite{HADES_pp35}. The individual lines  similar to the left part.
The dash-dot-dotted and the dashed lines shows the Delta Dalitz contribution
and the corresponding total sum  of all channels without $pp$ Bremsstrahlung as in Figs.
\ref{Fig_M35},\ref{Fig_y35} - for the "Wolf" electromagnetic decay width.
The dash-dotted line stands for the parametrization using
"Krivoruchenko" width, the solid line is the corresponding sum.
}
\label{Fig_pp35Kriv}
\label{Fig_GameeHSD}
\end{figure}

\begin{figure}[h!]
\includegraphics[width=8.5cm]{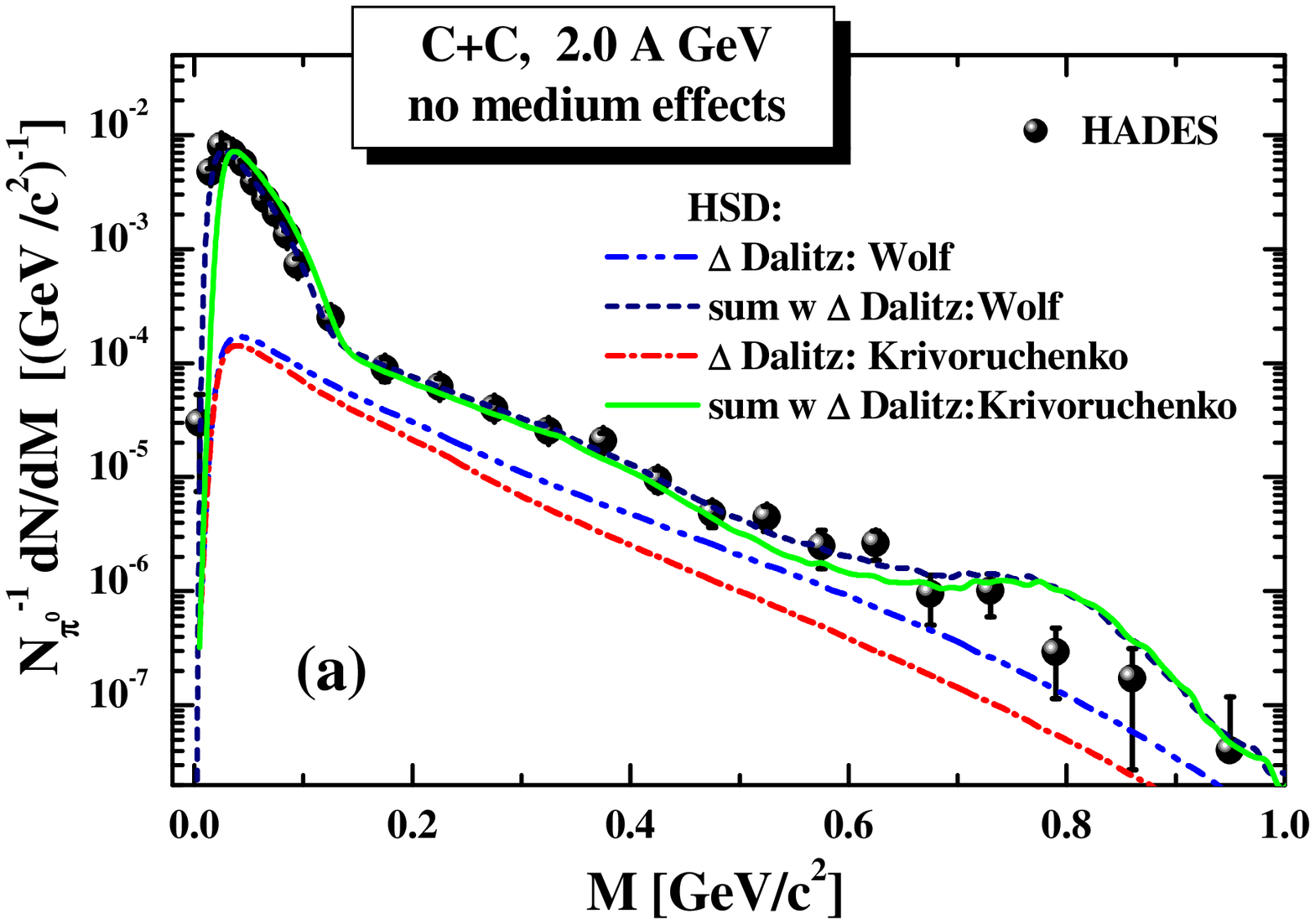}
\includegraphics[width=8.5cm]{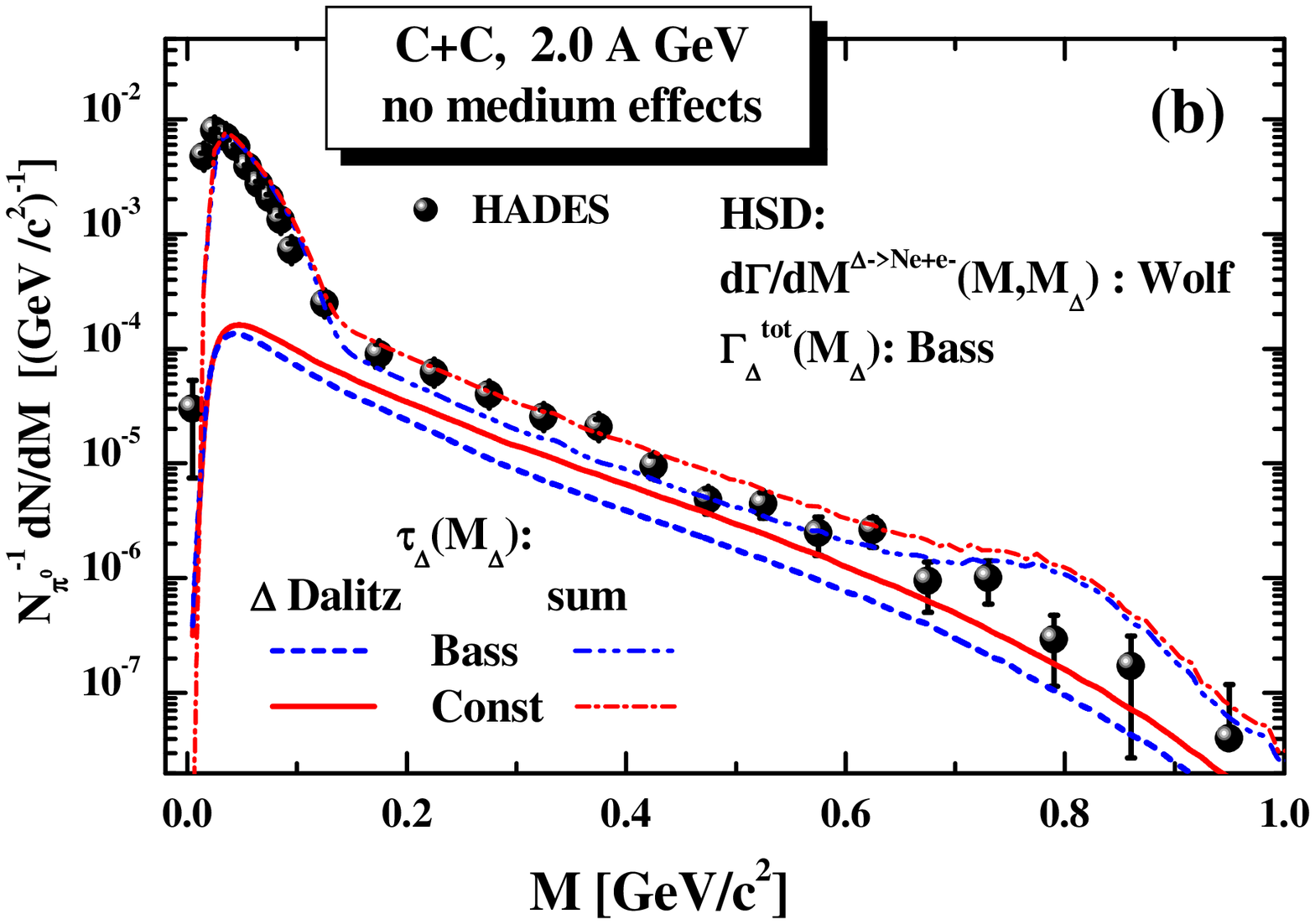}
\caption{(Color online) Left (a): The mass differential dilepton spectra -
normalized to the $\pi^0$ multiplicity  - from HSD calculations
for central (b=0.5 fm) C+C collisions at 2 $A$GeV in comparison
to the HADES data \cite{Agakishiev:2009yf}.
The dash-dot-dotted and the dashed lines shows the Delta Dalitz contribution
and the corresponding total sum  of all channels as in Fig.
\ref{Fig_CC20}- for the "Wolf" electromagnetic decay width.
The dash-dotted line stands for the parametrization using
"Krivoruchenko" width, the solid line is the corresponding sum.
Right(b): similar to the left part - the mass differential dilepton spectra
as a function of invariant dilepton mass for the 2 model cases of
life time $\tau_\Delta(M_\Delta)$:
solid line - "Const", dashed line - "Bass" life time,
while using the "Bass" total width for the $\Delta$ production and dynamics
and "Wolf" model for the partial electromagnetic decay width in both cases.
}
\label{Fig_GameeHSD}
\end{figure}

Similar to the left (a) part of  Fig.
\ref{Fig_dNdM} we demonstrate in Fig. \ref{Fig_pp35Kriv}  the
consequences of the variation of the electromagnetic decay width
on the differential  cross section $d\sigma/dM$ (left (a)) and the
transverse momentum spectra  (right (b-e)) for 4 different mass
bins  ($M \leq$ 0.15 GeV,  0.15 $\leq M \leq$ 0.47 GeV, 0.47
$\leq M \leq$ 0.7 GeV and  $M \geq$ 0.7 GeV) from HSD
calculations for $e^+e^-$ production in $pp$ reactions at a
bombarding energy of 3.5 GeV. The dash-dot-dotted and the dashed
lines show the Delta Dalitz contribution and the corresponding
total sum  of all channels without $pp$ Bremsstrahlung as in Figs.
\ref{Fig_M35},\ref{Fig_y35} - for the "Wolf" electromagnetic decay
width. The dash-dotted line stands for the parametrization using
"Krivoruchenko" width, the solid line is the corresponding sum. We
point out that we have selected the $pp$ reaction at 3.5 GeV here
since at this high energy the open phase space is large enough to
populate the high mass $\Delta$'s. Thus, one expects a large
deviation in the dilepton mass spectra coming from the high mass tail
of the $\Delta$ spectral function - as follows from Figs.
\ref{Fig_Gamee} and \ref{Fig_dNdM}(left) -  compared to the
low energy reactions where the available energy limits the
production of heavy $\Delta$'s. Furthermore, Figure  \ref{Fig_pp35Kriv}
(right) demonstrates the sensitivity of $p_T$ distribution to the
form of the electromagnetic decay width. In spite that the deviation
is bigger for the bin with the largest dilepton masses ($M>0.7$ GeV),
this effect is not visible in the final $p_T$ spectra due to the
dominant contributions from the direct decay of vector
mesons. For the lower mass bins (0.15 $\leq M \leq$ 0.47 GeV,
and 0.47 $\leq M \leq$ 0.7 GeV) the difference is better
observed in the final $p_T$ spectra. Thus, the measurement of the
$p_T$ distributions at various mass bins can help in distinguishing
of different models.

We continue with the comparison of the final  mass differential dilepton
spectra for central C+C collision at 2 $A$GeV -- Fig. \ref{Fig_GameeHSD} (left (a)).
The legend for the individual lines is the same as in Fig. \ref{Fig_pp35Kriv}.
We observe  similar deviations as obtained within
the "convolution" model - see the left (a) part of  Fig. \ref{Fig_dNdM}.

Thus, we conclude that the uncertainty of the electromagnetic decay width
of the $\Delta$ resonance translates to an uncertainty of about a factor of 1.5
in the dilepton yield from $\Delta$ decays in heavy ion collisions.
For large invariant masses of the $\Delta$ this uncertainty reaches even
a factor of 3.
However, these large invariant masses are only populated at beam energies
at which $\eta$ production becomes important with the consequence
that the $\eta$ Dalitz decay and bremsstrahlung
are the dominant sources for dilepton production. Therefore the uncertainty
of the dilepton yield from large mass $\Delta$ decay has little influence
on the measured total dilepton yield at large invariant masses
as well as on the ratio $R(AA/NN)$ - cf. dash-dot-dotted line in Fig. \ref{Fig_RCC2NN} (b).

The simulations for the different assumptions about the life time for the central C+C
at 2 $A$GeV are presented in the right part (b) of Fig. \ref{Fig_GameeHSD}.
The assumptions correspond to that of the right (b) part of  Fig. \ref{Fig_dNdM}.
The solid line - "Const" - displays the results assuming a constant life
time whereas the dashed line show the result assuming the  "Bass" life time.
In both cases the "Bass" total width has been employed for the $\Delta$
spectral function and  the "Wolf" model has been used for the partial
electromagnetic decay width.
One can see that the two assumptions about the lifetime yield an uncertainty
of a factor of 2, slightly less than the factor we obtained for elementary reactions.
(Fig. \ref{Fig_dNdMD}).

We would like to stress here that the uncertainties in the electromagnetic decay width
of $\Delta$ resonance as well as that of  total width/life time of the $\Delta$
can be reduced by measuring the dilepton yield in  $\pi N$ reactions
at different energies. Such a measurement would allow for preciser predictions
than presently possible.

\subsection {Electromagnetic $\Delta-N$ transition  formfactor}

The introduction of the electromagnetic $\Delta-N$ transition
formfactor $F_{\Delta N}$ for the $\Delta$ Dalitz decay
has been studied  in Ref.  \cite{GiBUU} within the GiBUU transport model
for $NN$ and $pNb$ reactions. The model from Refs. \cite{Iachello}
has been choosen for the  $\Delta-N$ transition formfactor which
is based on the Vector Dominance Model (VDM) model assuming that
the virtual photon is converted first to a $\rho_0$ meson, i.e.
the transition $\Delta\to \gamma* N \to N e^+e^-$ can be considered as
$\Delta \to \gamma^* N \to \rho_0 N \to e^+e^-N$. For that one needs
to extrapolate the $\Delta-N$ transition formfactor  from the
space-like region to the time-like region where its strength is unknown
experimentally.

\begin{figure}[h!]
\includegraphics[width=7.5cm]{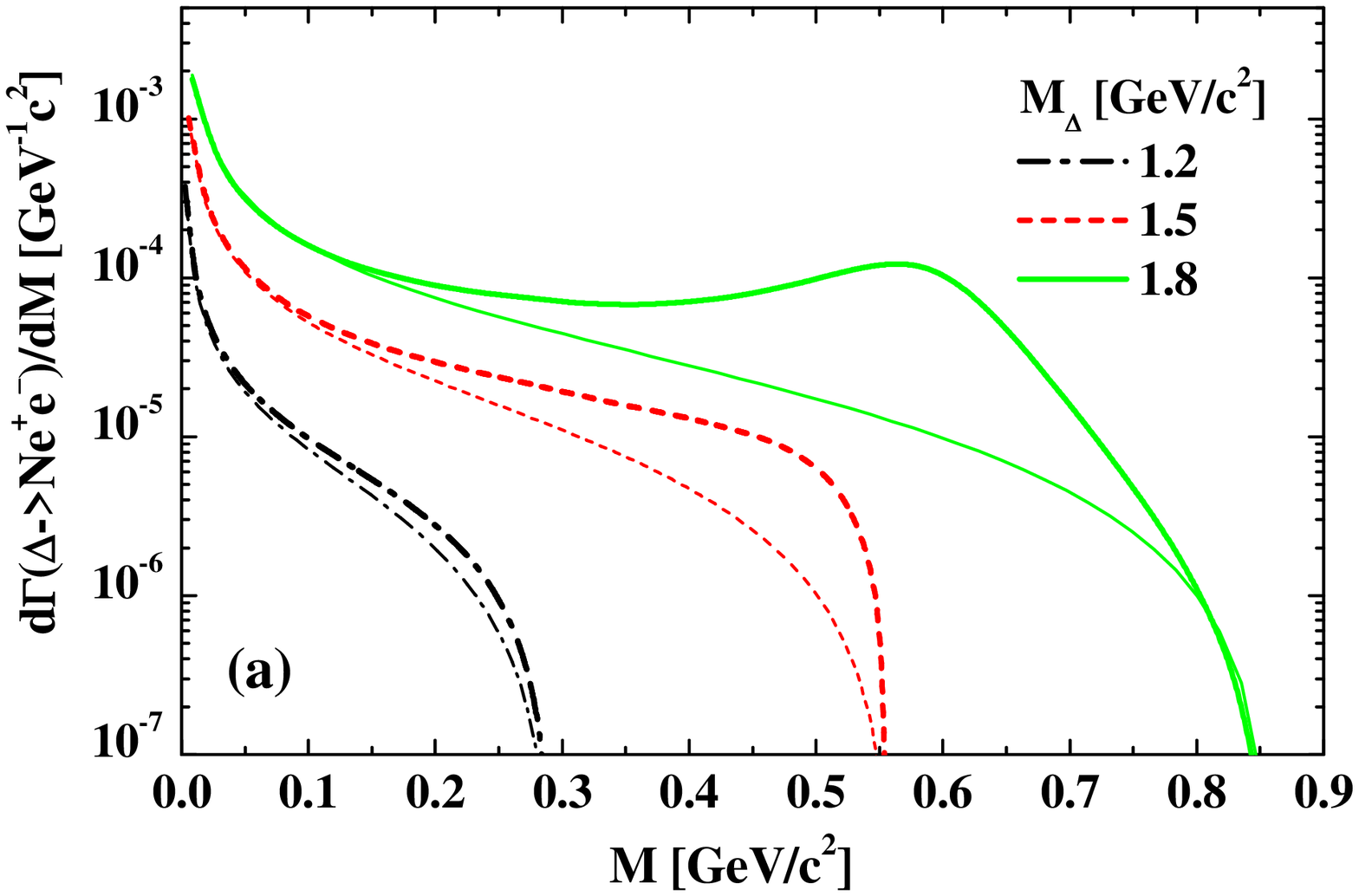}
\includegraphics[width=7.5cm]{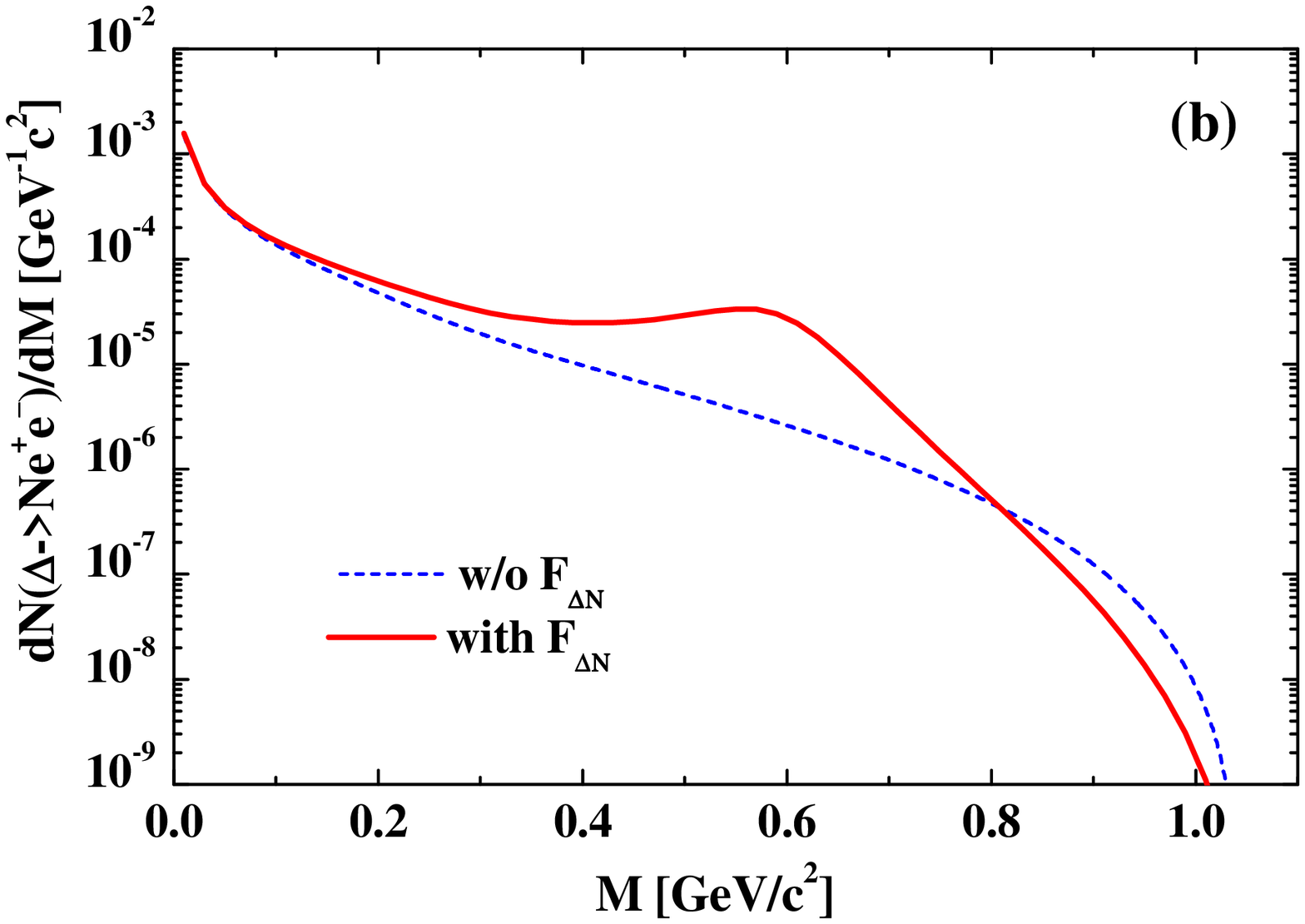}
\caption{(Color online) Left (a): the electromagnetic decay width of the $\Delta$ resonance
to dileptons $\Delta \to N  e^+e^-$ using the "Krivoruchenko" model
for  different $\Delta$ masses of 1.232 GeV (a),
1.5 (b) and 1.8 GeV (c): the thick lines - with the $\Delta-N$ formfactor
$F_{\Delta N}$  from Ref. \cite{Iachello}, the thin lines - without the $\Delta-N$ formfactor.
Right (b):  the dilepton yield as a function of the invariant dilepton
mass using the "Krivoruchenko" model for the electromagnetic decay
width and the "Bass" model for the total width:
the solid lines - with including the $\Delta-N$ formfactor $F_{\Delta N}$,
the dashed line - without the $\Delta-N$ formfactor.
}
\label{Fig_Iachel}
\end{figure}
\begin{figure}[h!]
\includegraphics[width=8.5cm]{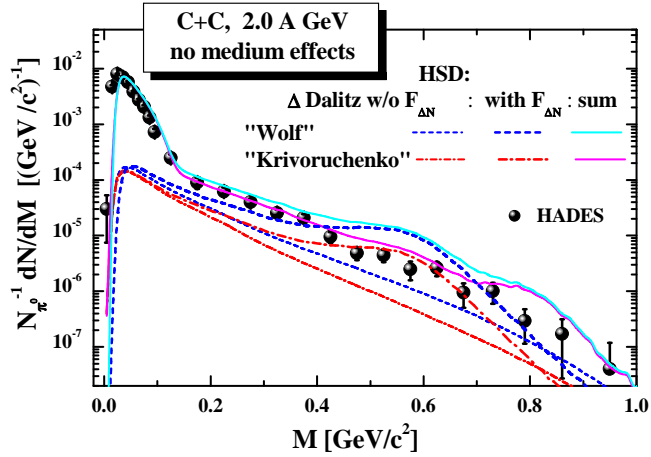}
\caption{(Color online) The results of the HSD transport calculation  for the mass
differential dilepton spectra - normalized to the
$\pi^0$ multiplicity  - for C+C at 2.0 $A$GeV:
the thick lines - with the $\Delta-N$ formfactor
$F_{\Delta N}$  from Ref. \cite{Iachello}, the thin lines -
without the $\Delta-N$ formfactor.
Here the dashed lines corresponds to the "Wolf" model for the
electromagnetic decay width, the dash-dotted lines - for the "Krivoruchenko" model.
}
\label{Fig_CC20Ich}
\end{figure}

In Fig. \ref{Fig_Iachel} we demonstrate the effect of the electromagnetic
$\Delta-N$ transition formfactor: the left plot (a) shows
the electromagnetic decay width of the $\Delta$ resonance into dileptons
$\Delta \to N  e^+e^-$ using the "Krivoruchenko" model  for  different
$\Delta$ masses of 1.232 GeV (a), 1.5 GeV (b) and 1.8 GeV (c):
the thick (thin) lines - with (without) the $\Delta-N$ formfactor
$F_{\Delta N}$  from Ref. \cite{Iachello}. The right plot (b)
displays the dilepton yield using the "convolution" model
(Eq. (\ref{dNdMee})) with the "Bass" total width. One can see that
the inclusion of the formfactor leads to an enhancement  of the dilepton
yield up to a factor of 10 at $M\sim 0.6$~GeV.

We have investigated the consequences of the electromagnetic
$\Delta-N$ transition  formfactor for heavy-ion collisions
for C+C reactions at 2 $A$GeV where the $\Delta$ channel is
one of the dominant channels. Fig. \ref{Fig_CC20Ich} shows the results
of the HSD calculations for 2 different models (in line with our discussion
above on the model uncertainties) for the electromagnetic decay width
- "Krivoruchenko" (dashed lines) and "Wolf" (dash-dotted lines)
which provide the lower and upper limit for the effects of the
form factor for the final spectra (solid lower and upper lines).
One can conclude that the introduction of the  $\Delta-N$
transition formfactor \cite{Iachello} leads to overestimation
of the dilepton yield in heavy-ion collisions, i.e. is not
in line with the HADES data.

\subsection {Electromagnetic pion formfactor for the $pn$ bremsstrahlung }

Here we discuss the uncertainties related to the implementation of
the electromagnetic pion formfactor $F_\pi(M)$ motivated by the Vector Dominance Model
(VDM) model for the $pn$ bremsstrahlung contribution as advocated in the OBE model
in Ref. \cite{Shyam10}. With the help of this formfactor
one hopes to account for the dilepton radiation from the internal charged pion exchange line
in $pn \to pn e^+e^-$ processes assuming vector dominance, i.e. that the photon couples
to dileptons via a $\rho^0$ meson. This diagramm doesn't exist for
the $pp$ reaction, so the enhancement should be seen only in $pn$ dilepton yield.
There is a debate if the virtual photon converts fully (i.e. by 100\%) to $\rho$
meson ($\gamma^* \to \rho\to e^+e^-$)
\cite{eri88} or by 50\% only \cite{bro86} and the rest decays
directly into dileptons ($\gamma^* \to e^+e^-$).
In Ref. \cite{bro86} the electromagnetic pion formfactor $F_\pi(M)$
has been parametrized as
\begin{eqnarray}
F_\pi (M^2) & = & \frac{0.4}{1-M^2/\lambda^2} + \frac{0.6}{1-M^2/2m_\rho^2}
      \frac{m_\rho^2}{m_\rho^2 - M^2 - im_\rho \Gamma_\rho(M^2)},
      \label{eqFpi}
\end{eqnarray}
where $\lambda^2$ = 1.9 $GeV^2$. The width $\Gamma_\rho(M^2)$ is given in
Ref.~\cite{bro86} as
\begin{eqnarray}
\Gamma_\rho^{tot}(M) & = & \Gamma_{\rho^0 \to \pi \pi} \frac{r_C^2 k^3}
                         {M(1+r_C^2k^2)},
\end{eqnarray}
with $k^2 = M^2/4 - m_\pi^2$. The parameter $r_C=2$ fm is an interaction
radius, $\Gamma_{\rho^0 \to \pi \pi}= 0.150$~GeV.

\begin{figure}[t!]
\includegraphics[width=14.cm]{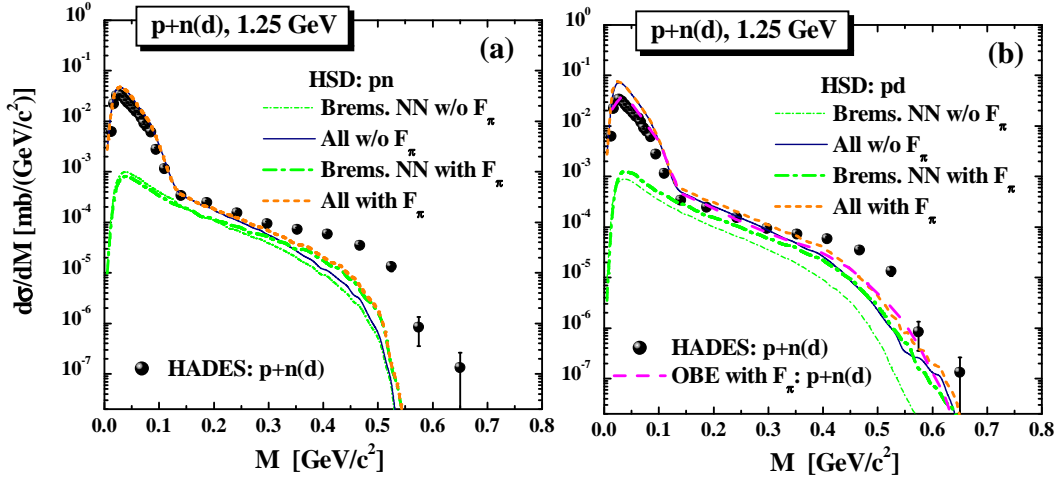}
\caption{(Color online) The HSD results for the dilepton differential cross section
$d\sigma/dM$ for $pn$ (left plot (a)) and $pd$ (right
plot (b)) reactions at 1.25 GeV including the electromagnetic pion formfactor
for the bremsstrahlung channel (denoted as "Brems. NN with $F_\pi$")
in comparison to the "standard" HSD calculations without formfactor
(denoted as "Brems. NN w/o $F_\pi$") as in Fig. \ref{Fig_pp125}.
The dashed line (denoted as "OBE with $F_\pi: p+n(d)$") shows the OBE
results for $p+n(d)$ from Ref. \cite{Shyam10}.
}
\label{Fig_pdFpi}
\end{figure}

According to the OBE calculations of  \cite{Shyam10} for the $pn$ reaction at 1.25 GeV
the incorporation of the formfactor $F_\pi(M)$ leads to the enhancement
of the bremsstrahlung contribution and a better agreement with the HADES
data for quasi-free $pn$ scattering.
Following Ref. \cite{Shyam10} we have performed a model study by including
the formfactor from \cite{bro86} in our calculations of $pn$
bremsstrahlung by simply multiplication of the parametrized OBE results
from \cite{Kaptari:2005qz} used in HSD by the formfactor of eq.
(\ref{eqFpi}). Indeed, this provides an upper limit since we can not
distinguish the individual diagramms in our parametrization of the bremsstrahlung
cross section. We obtain a good agreement with the full OBE calculations
from \cite{Shyam10},as shown in Fig. \ref{Fig_pdFpi}, which
presents the HSD results for the dilepton differential cross section $d\sigma/dM$ for $pn$
(left plot (a)) and $pd$ (right plot (b)) reactions
at 1.25 GeV including the electromagnetic pion formfactor
for the bremsstrahlung channel (denoted as "Brems. NN with $F_\pi$")
in comparison to the "standard" HSD calculations without formfactor
(denoted as "Brems. NN w/o $F_\pi$") as in Fig. \ref{Fig_pp125}.
The dashed line (denoted as "OBE with $F_\pi: p+n(d)$") shows the OBE
results for $p+n(d)$ from Ref. \cite{Shyam10}.

\begin{figure}[h!]
\phantom{a}\vspace*{5mm}
\includegraphics[width=6cm]{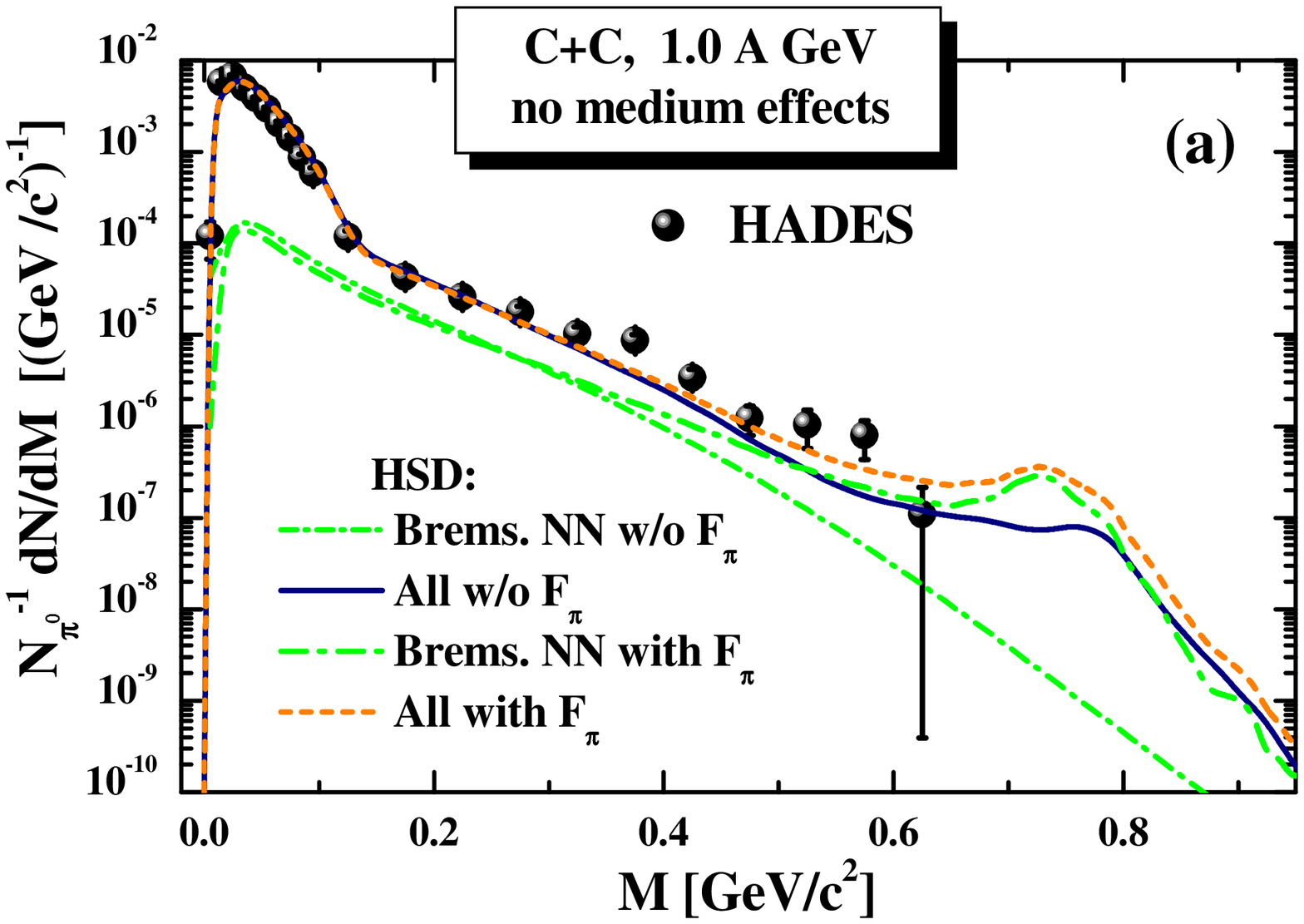}
\includegraphics[width=6cm]{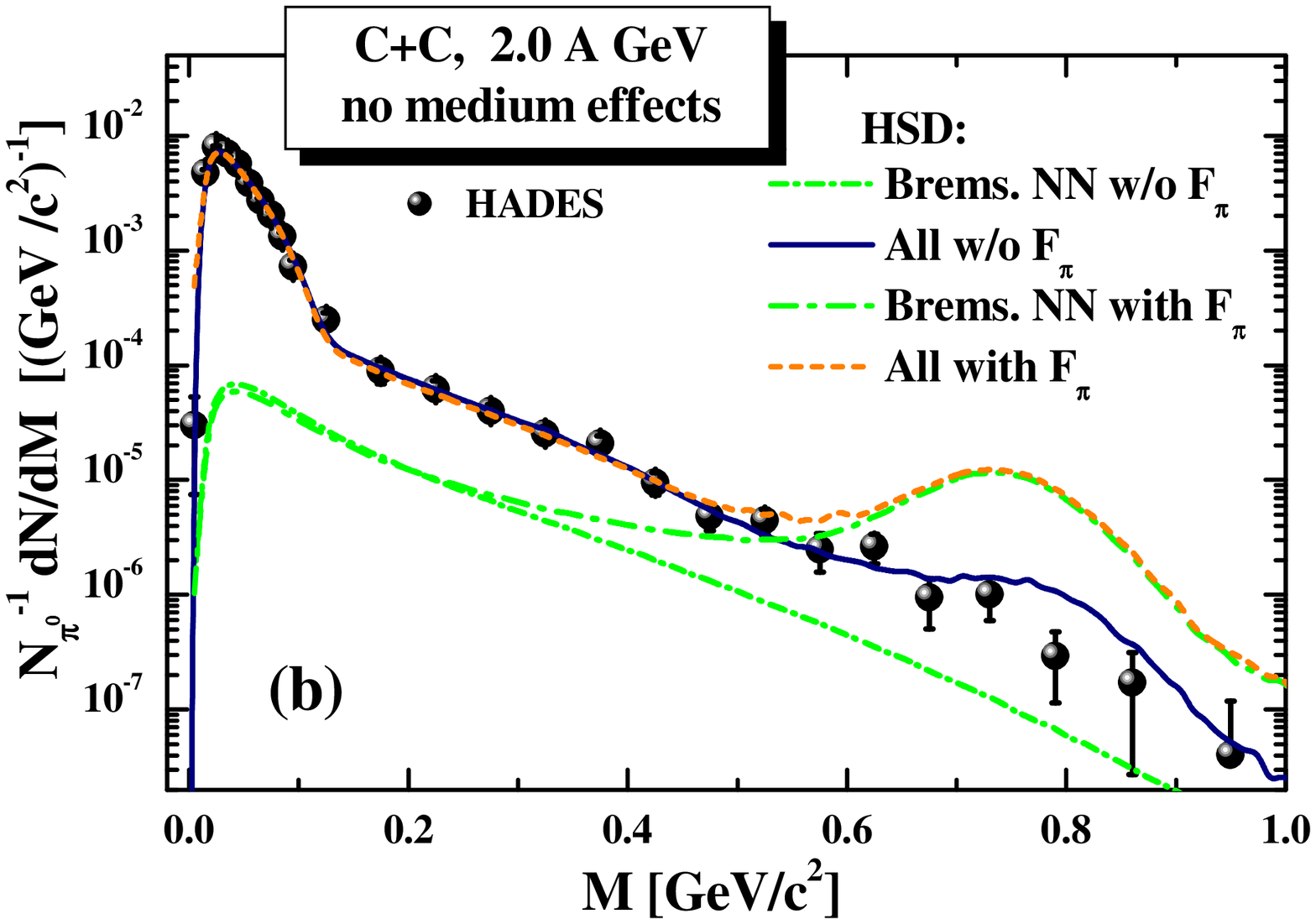}
\includegraphics[width=6cm]{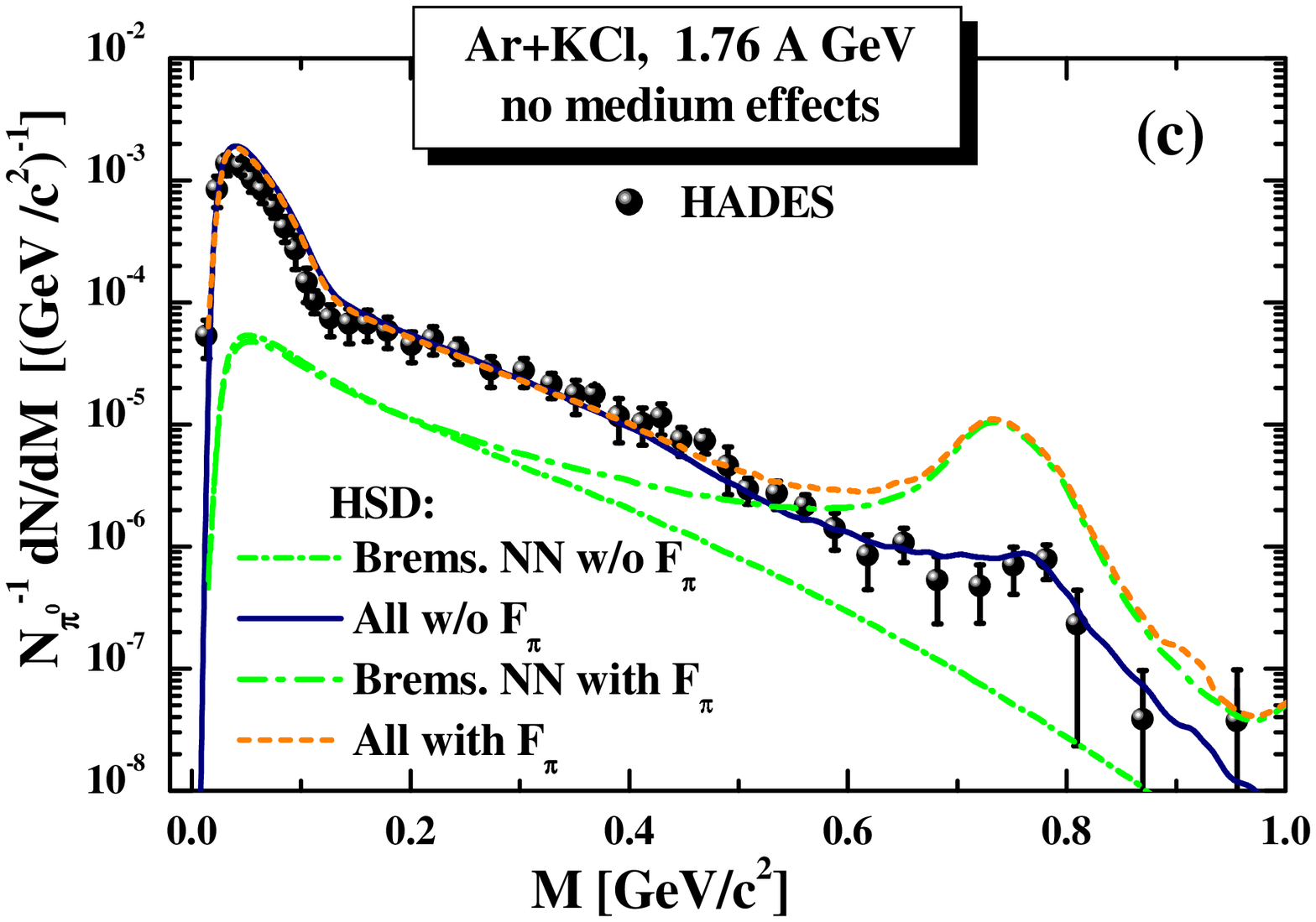}
\caption{(Color online) The HSD results for the dilepton differential cross section
$d\sigma/dM$ for C+C at 1.0 $A$GeV (left (a)) 2.0 $A$GeV (middle
(b)) and Ar+KCl (right (c)) reactions at 1.75 $A$GeV.
The lines description as in Fig. \ref{Fig_pdFpi}.
}
\label{Fig_AAFpi}
\end{figure}

As seen form Fig. \ref{Fig_pdFpi} the inclusion of the formfactor doesn't explain the
experimental HADES quasi-free $p+n(d)$ data. We check now how the formfactor
will change the heavy-ion results where we have  reliable
experimental constraints from the HADES measurements.
The HSD results for the dilepton differential cross section
$d\sigma/dM$ for C+C at 1.0 $A$GeV (left (a)) 2.0 $A$GeV (middle
(b)) and Ar+KCl (right (c)) reactions at 1.75 $A$GeV are shown
in Fig. \ref{Fig_AAFpi}. One can clearly see a sizeable
overestimation of the dilepton yields for all systems which brings us
to the conclusion that to include a form factor is not supported by the experimental data on
heavy-ion collisions.

\section {Conclusions}

We have studied the production of dileptions in $pp$, $pn$, $pA$ and
$AA$ collisions at energies between  1 and 3.5  $A$GeV by comparing
the results of three  independent transport approaches - HSD, IQMD and
UrQMD - with all existing data in this energy domain.  These data allowed for
the first time to study the cycle of creation and absorption of the
$\Delta$ resonance in heavy ion reaction which has been theoretically
predicted since long.

Despite of common general ideas of transport approaches which are based on the
modeling/parameterizations of baryon-baryon, meson-baryon and
meson-meson elementary reactions with further dynamical evolution
including the propagation in a self-generated mean-field potential and
explicit interactions, the models differ in the actual realization and
underlying assumptions where no control from experimental data is
available.

Especially for one of the dominant channels for the dilepton
production in this energy domain, the $\Delta$ Dalitz decay, the
experimental results do not allow for a robust parametrization of the
input for these transport theories.  Neither  the spectral function of the $\Delta$
nor the differential decay width of the $\Delta$ into dileptons
are well known. At energies around 1 $A$GeV the $\Delta$ production
is in agreement with the isospin model which assumes that the difference
$\Delta$ states are produced according to the isospin Clebsch Gordon coefficients
at higher energies, where 2 pion channels contribute substantially, little information
is available on population of the different $\Delta$ states. This situation
will hopefully change with the planned experiments on dilepton production
in $\pi N$ collisions. Such an information would substantially improve the predictive power
of transport theories for heavy ion results.

Similar is the situation for the bremsstrahlung
contribution which turns out to be the dominant channel for the low energy
collisions at 1 $A$GeV.  The present OBE models provide different
predictions as compared to the soft-photon-approximation and do not
agree among each other.  More precise data, especially on dilepton
production in elementary $pp$, $pn$ and especially $\pi N$ collisions
for different energies are need to reduce this systematic error.
They would allow for a more reliable predictions of bremsstrahlung in transport approaches.

We stress the importance of providing such constraint form the
experimental side since the transport models are the only reliable tool
to study the physics of heavy ion collisions at those energies where neither thermal
models nor hydrodynamic models are applicable because the created matter
is far from equilibrium.

Despite of these uncertainties the results of different  transport
approaches for the final dilepton yield agree quite well among each other
even if there are the deviations in the channel decompositions.  The
data are in between the systematical error of the transport
predictions.

We have started our investigation with the dilepton spectra from elementary reactions
which can be described as a superposition of the emission from known
dilepton sources. In $pp$ collisions at energies of around 1 GeV
dileptons stem dominantly from the $\Delta$ Dalitz decay whereas in
$pn$ collisions the bremsstrahlung radiation becomes equally important.
At higher energies the $\eta$ production sets in and contributes to the
invariant mass range $M < 0.55$ GeV. At higher invariant masses the
vector meson decays dominate but  the data are presently not precise
enough to allow for firm conclusions. New experimental differential
data would be very useful to check the underlying model assumptions.

Our study demonstrates that  in heavy-ion reactions the dilepton
production for invariant masses below  $M < 0.6$ GeV cannot be
interpreted as a simple convolution of the average dilepton yield from
$pn$ and $pp$ collisions times the number of elementary collisions. The
presence of a nuclear medium manifests itself in several ways:  First
of all the Fermi motion of nucleons in nuclei smears out the energy
distribution of primary $NN$ collisions substantially. This has a big
influence on the particle production at (sub-)threshold energies. The
Fermi motion enhances the pion as well as the dilepton yields in $AA$
collisions at threshold energies by up to a factor of two.  The
enhancement is, however, identical for pions and dileptons. Therefore,
if the Fermi motion would be the only difference between $AA$ and $NN$
collisions, one could expect that if one normalizes the dilepton yield
to the pion multiplicity, as done in the experimental analysis, no
enhancement would be  observed.

The real situation is quite different:  an enhancement of the dilepton
yield in $AA$ relative to $NN$ is observed experimentally even if one
normalizes the dilepton yield by the $\pi^0$ multiplicity.  The
experimental enhancement is, however, plagued partly by the use of $pd$
collisions instead of $pn$ collisions because in the interesting
kinematical regime the $pd$ collisions are true three-body collisions
and cannot be interpreted as quasi-free $pn$ reactions. So the 'true'
enhancement -- as compared to elementary collisions -- cannot be
inferred from present data for invariant masses above 0.5 GeV.

We have analysed this enhancement in detail and found two origins: The
first reason is the bremsstrahlung radiation from $pn$ and $pp$
reactions which does not scale with the pion number (i.e. the number of
participants) rather with the number of elementary elastic collisions.
The second reason is the shining of dileptons from the 'intermediate'
$\Delta$'s, which take part in the $\Delta \to \pi N$ and $ \pi N \to
\Delta$ reaction cycle.  This cycle produces a number of generations of
$\Delta$'s during the reaction which increases with the size of the
system. At the end only one pion is produced but each intermediate
$\Delta$ has contributed to the dilepton yield because emitted
dileptons do not get absorbed. This leads to an enhancement of the
dilepton yield as compared to the final number of pions.  Thus, the
enhancement confirms the predictions of transport theories that in
heavy-ion collisions several generations of $\Delta$'s are formed which
decay and are recreated by $\pi N \to \Delta$ reactions.
Accordingly, the dilepton data from $AA$ reactions shed light on the $\Delta$
dynamics in the medium.

In the investigated invariant mass range, $M<0.5$ GeV, we do not find
evidence that the observed enhancement of the dilepton yield in
heavy-ion collisions over the elementary reactions requires the
assumption of 'conventional' in-medium effects like a modification of
the spectral functions of the involved hadrons. Theory predicts such a
modification for vector mesons and therefore for invariant masses
$M>0.5$ GeV.  More precise data are needed to draw robust conclusions
on the in-medium modifications in this invariant mass range.

We summarize with the final remark that the ratio of dilepton yields
$AA/NN$ is a sensitive observable which allows to penetrate  the
intermediate phase of the heavy-ion reaction and sheds light on the
$\Delta$ dynamics which is not accessible by the hadronic observables.
Thus, the HADES data  provide the first experimental constraint on this
issue in heavy-ion collisions at SIS energies.  Moreover, by measuring
this ratio at a low bombarding energy one can get access to the
bremsstrahlung radiation since it becomes the dominant process there.
One expects to observe in this case the scaling of the ratio with the
number of binary collisions rather than with the number of pions.  A
precise measurement of the ratio of dilepton yields in heavy-nuclei
collisions (as Au+Au or Pb+Pb) and of that in the light systems (as
C+C) for invariant masses $M>0.5$ GeV will help to obtain information
on the in-medium modification of the spectral function of vector
mesons.


\section*{Acknowledgements}
The authors are grateful for fruitful discussions with
W. Cassing, C. Hartnack, S. Endres, L. Fabbietti, M. Gumberidze, F. Krizek, Y. Leifels,
O. Linnyk, A. Rustamov, J. Stroth, P. Salabura, H. van Hees, M. Weber and J.
Weil.
Our special thanks go to T. Galatyuk and R. Holzmann for the
continuous help and many useful advices concerning the experimental data and
filtering procedure.
E.B. and S.V. acknowledge financial support through the ``HIC for
FAIR" framework of the ``LOEWE" program.


\end{document}